\documentclass[10pt]{amsart}

\usepackage[colorlinks]{hyperref}
\usepackage{color,graphicx,shortvrb}
\usepackage[latin 1]{inputenc}
\usepackage[dvipsnames]{xcolor}

\usepackage[active]{srcltx} 

\usepackage{enumerate}
\usepackage{amssymb}
\usepackage{tikz-cd}

\usepackage [ all ]{xy}

\newtheorem{theorem}{Theorem}[section]

\newtheorem{corollary}[theorem]{Corollary}

\newtheorem{lemma}[theorem]{Lemma}

\newtheorem{proposition}[theorem]{Proposition}
\newtheorem{remark}[theorem]{Remark}

\def\J#1#2#3{ \left\{ #1,#2,#3 \right\} }

\def\11{\textbf{$1$}}
\def\11b#1{\mathbf{1}_{_{#1}}}
\def\CC{{\mathbb{C}}}

\DeclareGraphicsExtensions{.jpg,.pdf,.png,.eps}

\begin{document}

\title[Representation of symmetry transformations on sets of tripotents]{Representation of symmetry transformations on the sets of tripotents of spin and Cartan factors}

\author[Y. Friedman]{Yaakov Friedman}
\author[A.M. Peralta]{Antonio M. Peralta}

\address{Jerusalem College of Technology,
Departments of Mathematics and Physics, Extended Relativity Research Center,
P.O.B. 16031 Jerusalem 91160, Israel.}

\address{Departamento de An{\'a}lisis Matem{\'a}tico, Facultad de
Ciencias, Universidad de Granada, 18071 Granada, Spain. \smallskip\smallskip\smallskip}

\email{friedman@g.jct.ac.il}

\email{aperalta@ugr.es}




\date{}

\begin{abstract} There are six different mathematical formulations of the symmetry group in quantum mechanics, among them the set of pure states $\mathbf{P}$--i.e., the set of one-dimensional projections on a complex Hilbert space $H$-- and the orthomodular lattice $\mathbf{L}$ of closed subspaces of $H$. These six groups are isomorphic when the dimension of $H$ is $\geq 3$. The latter hypothesis is absolutely necessary in this identification. For example, the automorphisms group of all bijections preserving orthogonality and the order on $\mathbf{L}$ identifies with the bijections on $\mathbf{P}$ preserving transition probabilities only if dim$(H)\geq 3$. 
Despite of the difficulties caused by $M_2(\mathbb{C})$, rank two algebras are used for quantum mechanics description of the spin state of spin-$\frac12$ particles, there is a counterexample for Uhlhorn's version of Wigner's theorem for such state space.

In this note we prove that in order that the description of the spin will be relativistic, it is not enough to preserve the projection lattice equipped with its natural partial order and orthogonality, but we also need to preserve the partial order set of all tripotents and orthogonality among them (a set which strictly enlarges the lattice of projections). Concretely, let $M$ and $N$ be two atomic JBW$^*$-triples not containing rank--one Cartan factors, and let $\mathcal{U} (M)$ and $\mathcal{U} (N)$ denote the set of all tripotents in $M$ and $N$, respectively. We show that each bijection $\Phi: \mathcal{U} (M)\to \mathcal{U} (N)$, preserving the partial ordering in both directions, orthogonality in one direction and satisfying some mild continuity hypothesis can be extended to a real linear triple automorphism. This, in particular, extends a result of Moln{\'a}r to the wider setting of atomic JBW$^*$-triples not containing rank--one Cartan factors, and provides new models to present quantum behavior.
\end{abstract}

\maketitle
\thispagestyle{empty}

\section{Introduction}

Together with H. Weyl, E.P. Wigner introduced group theory into physics and formalized the theory of symmetry. A symmetry is a transformation of a quantum structure that preserves a certain quantity or relation. A Wigner symmetry is mapping on the set $\mathbf{P}$ of rank--one projections of $B(H)$ preserving the transition probability between any two of these projections. The result known as Wigner's theorem, which has been labeled as a cornerstone of the mathematical formulation of quantum mechanics, plays a fundamental role in this theory. It was firstly enunciated in \cite{Wig31}, and it can be simply stated by saying that every bijective Wigner symmetry is induced by a unitary or an antiunitary operator. Given two projections $p= \xi\otimes \xi, q= \eta\otimes \eta\in\mathbf{P},$ the \emph{transition probability} between $p$ and $q$ is given by $tr(p q) = tr(pq^*) = tr(q p^*) = |\langle \xi, \eta\rangle|^2,$ where $tr(\cdot)$ stands for the trace on $B(H)$.
\smallskip

Given a complex Hilbert space $H$, the following mathematical objects are employed in the Hilbert space formulation of quantum mechanics:
\begin{enumerate}[$(1)$]\item The C$^*$-algebra $B(H)$ of bounded operators;
\item The Jordan algebra $B(H)_{sa}$ of bounded self-adjoint operators;
\item The set $\mathbf{P}$ of pure states on $H$, that is, the extreme points of the set of positive functionals of the unit ball of $B(H)_*=C_1(H)$, the space of trace class operators regarded as the predual of $B(H)$, which by trace duality corresponds to the rank--one projections $\xi\otimes\xi$ ($\xi\in H$, $\|\xi\|=1$) in $B(H)_*= C_1(H)$;
\item The convex set of normal states of $B(H)$, that is, the set of positive trace class operators of trace one;
\item The orthomodular lattice $\mathbf{L}$ of closed subspaces of $H$;
\item The partial algebra of positive operators bounded by the unit operator on $H$.
\end{enumerate}

The natural automorphisms of these mathematical models (i.e., the bijections $f$ on these sets preserving the corresponding relevant structure:  associative product and involution, Jordan product, transition probability, convex combinations, orthogonality and order between subspaces, and the partially defined sum $E + F \leq  I$ if and only if  $f(E) + f(F) \leq  I$ and in this case $f(E+F) = f(E) +f(F)$, respectively) represent the symmetry groups of quantum mechanics and are endowed with natural topologies induced by the probabilistic structure of quantum mechanics. It is shown, for example, in \cite{CasdeVilahtiLevrero97} that these six symmetry groups are all isomorphic when dim$(H)\geq 3$. The last restriction exclude rank two, where there are no more than two orthogonal projections.\smallskip

The reader is referred to \cite{Wig31,Wig59,LomontMendelson63,Bargmann64} for the original and pioneering results and to \cite{BarviHam2017, Chev2007,Gyo2004,Molnar85,Molnar96,Molnar98,Molnar99,Molnar2000,SharAlmeida90} and \cite{Molnar2002} for more recent proofs and variants of this influencing result. The references \cite{Geher2014,Geher2016} contain a detailed exposition on the origins and history of Wigner theorem. Special credit must be paid to the contributions due to L. Moln{\'a}r and his group.\smallskip

Let $P_1(H)$ denote the set of all rank--one projections on $H$. The set $P_1(H)$ can be regarded as the set of minimal projections in $B(H)$ as well as the set $\mathbf{P}$ of all pure states in $B(H)_*= C_1(H)$.\smallskip

Based on the \emph{fundamental theorem of projective geometry}, U. Uhlhorn established a version of Wigner's famous theorem on the structure of quantum mechanical symmetry transformations which reads as follows:

\begin{theorem}\label{thm Uhlhorn}{\rm(Uhlhorn's theorem \cite{Uhlhorn63})} Let $H$ be a complex Hilbert space with dim$(H)\geq 3$. Then every bijective map $\Phi: P_1(H)\to P_1(H)$ which preserves orthogonality between tripotents in both directions, that is,
$$p q = 0 \hbox{ in } P_1(H) \hbox{ if and only if } \Phi(p) \Phi(q)=0,$$
is induced by a unitary or antiunitary operator on the underlying Hilbert space.
\end{theorem}

There is another recent contribution by L. Moln{\'a}r which introduces a new point of view and another version of Wigner's theorem. Let us first introduce some terminology. Henceforth the set of all partial isometries or tripotents on $H$ will be denoted by $PI(H)= \mathcal{U}(B(H))$. The symbol $PI_1(H) = \mathcal{U}_{min} (B(H))$ will stand for the set of all rank--1 or minimal partial isometries on $H$. We shall consider the standard partial ordering on $PI(H)= \mathcal{U}(B(H))$ given by $ e\leq u$ if and only if $u-e$ is a partial isometry orthogonal to $e$ (i.e., $(ee^*) (u-e)(u-e)^* =0$ $=(e^*e) (u-e)^*(u-e)$).

\begin{theorem}\label{t Molnar 2002}\cite[Theorem 1]{Molnar2002} Let $H$ be a complex Hilbert space with dim$(H)\geq 3$. Suppose that $\Phi : \mathcal{U}(B(H))\to \mathcal{U}(B(H))$ is a bijective transformation which preserves the partial ordering and the orthogonality between partial isometries in both directions. If $\Phi$ is continuous {\rm(}in the operator norm{\rm)} at a single element of $\mathcal{U}(B(H))$ different from $0$, then $\Phi$ extends to a real linear triple isomorphism. Moreover $\Phi$ can be written in one of the following forms:\begin{enumerate}[$(1)$]
\item There exist unitaries $U, V$ on $H$ such that $\Phi(R) = URV$ {\rm(}$R \in \mathcal{U}(B(H))${\rm)};
\item There exist antiunitaries $U, V$ on $H$ such that $\Phi(R) = URV$ {\rm(}$R \in \mathcal{U}(B(H))${\rm)};
\item There exist unitaries $U, V$ on $H$ such that $\Phi(R) = U R^*V$ {\rm(}$R \in \mathcal{U}(B(H))${\rm)};
\item There exist antiunitaries $U, V$ on $H$ such that $\Phi(R) = UR^* V$ {\rm(}$R \in \mathcal{U}(B(H))${\rm)}.
\end{enumerate}
\end{theorem}

As observed in \cite{Molnar2002}, the analogue result of Theorem \ref{t Molnar 2002} when the mapping $\Phi$ is restricted to the lattice of projections of $B(H)$ is usually known as the fundamental theorem of projective geometry. Another related tool is the so-called Mackey--Gleason problem. Let $W$ be a von Neumann algebra without type $I_2$ direct summand, and let $\mathcal{P}(W)$ stand for the lattice of all projections in $W$. Suppose that $\mu$ is a bounded finitely additive measure on $\mathcal{P}(W)$. The Mackey--Gleason problem asks whether $\mu$ extends to a linear functional on the whole $W$. A positive answer for nonnegative measures on $B(H)$ with $H$ separable was given by A. M. Gleason in \cite{Gleason57}. It is known that the problem admits no solution for the von Neumann algebra $M_2(\mathbb{C})$ of all $2\times 2$ complex matrices. In \cite{BuWri92}, L.J. Bunce and J.D.M. Wright gave a complete positive solution to the question posed by Mackey. A Jordan version employed in our arguments was published by the just quoted authors in \cite{BuWri89}.\smallskip

Quantum mechanic's model stimulated the study of non-commutative geometry. The model which motivated the study of C$^*$-algebras is the space $B(H)$ of all bounded linear operators on a complex Hilbert space $H$. A natural question is what are the subspaces of $B(H)$ which are the range of a contractive projection? It was shown in \cite{ArazFri78} that such subspaces can be classified and they are closed under a certain triple product.  Left and right weak$^*$ closed ideals of $B(H)$ are precisely subspaces of the form $ B(H)p$ and $pB(H),$ respectively, where $p$ is a projection in $B(H)$. The latter are identified with subspaces of operators of the form $B(p(H), H)$ and $B(H, p(H))$. However, given two complex Hilbert spaces $H$ and $K$ (where we can always assume that $K$ is a closed subspace of $H$), the Banach space $B(H,K)$, of all bounded linear operators from $H$ to $K$ is not, in general, a C$^*$-subalgebra of some $B(\tilde{H})$. Despite of this handicap, $B(H,K)$ is stable under products of the form \begin{equation}\label{eq triple product JCstar triple} \{x,y,z\} = \frac12\left( x y^* z + z y^* x\right)\ \ (x,y,z\in B(H,K)).
\end{equation}  Closed complex-linear subspaces of $B(H,K)$ which are closed for the triple product defined in \eqref{eq triple product JCstar triple} were called \emph{J$^*$-algebras} by L. Harris in \cite{Harris74,Harris81}. J$^*$-algebras include, in particular, all C$^*$-algebras, all JC$^*$-algebras and all ternary algebras of operators. Harris also proved that the open unit ball of every J$^*$-algebra enjoys a interesting holomorphic property, namely, it is a bounded symmetric domain (see \cite[Corollary 2]{Harris74}). In \cite{BraKaUp78}, R. Braun, W. Kaup and H. Upmeier extended Harris' result by showing that the open unit ball of every (unital) JB$^*$-algebra satisfies the same property.\smallskip

When the holomorphic-property ``being a bounded symmetric domain'' is employed to classify complex Banach spaces, the definitive result is due to W. Kaup, who, in his own words, \emph{``introduced the concept of a JB$^*$-triple and showed that every bounded symmetric domain in a complex Banach space is biholomorphically equivalent to the open unit ball of a JB$^*$-triple and in this way, the category of all bounded symmetric domains with
base point is equivalent to the category of JB$^*$-triples}'' (see \cite{Ka83} and subsection \ref{subsec: background} for the detailed definitions).\smallskip

The first examples of JB$^*$-triples include C$^*$-algebras and $B(H,K)$ spaces with respect to the triple product given in \eqref{eq triple product JCstar triple}, the latter are known as \emph{Cartan factors of type 1}.\smallskip

There are six different types of Cartan factors whose open unit balls are associated to classic Cartan domains; the first one has been introduced in the previous paragraph. In order to define the next two types, let $j$ be a conjugation (i.e., a conjugate-linear isometry of period 2) on a complex Hilbert space $H$. We consider a linear involution on $B(H)$ defined by $x\mapsto x^t:=jx^*j$. \emph{Cartan factors of type 2 and 3} are the JB$^*$-subtriples of $B(H)$ of all $t$-skew-symmetric and $t$-symmetric operators.\smallskip

A \emph{Cartan factor of type 4}, also called a \emph{spin factor}, is a complex Hilbert space $M$ provided with a conjugation (i.e., a conjugate-linear isometry of period-2) $x\mapsto \overline{x},$ triple product and norm given by \begin{equation}\label{eq spin product}
\{x, y, z\} = \langle x, y\rangle z + \langle z, y\rangle  x -\langle x, \overline{z}\rangle \overline{y},
\end{equation} and \begin{equation}\label{eq spin norm} \|x\|^2 = \langle x, x\rangle  + \sqrt{\langle x, x\rangle ^2 -|
\langle x, \overline{x}\rangle  |^2},
 \end{equation} respectively (see also \cite[Chapter 3]{Fri2005}). The \emph{Cartan factors of types 5 and 6} (also called \emph{exceptional} Cartan factors) are spaces of matrices over the eight dimensional complex algebra of Cayley numbers; the type 6 consists of all $3\times 3$ self-adjoint matrices and has a natural Jordan algebra structure, and the type 5 is the subtriple consisting of all $1\times 2$ matrices (see \cite{Ka97} for more details). A JB$^*$-triple is called atomic if it coincides with an $\ell_{\infty}$-sum of Cartan factors.\smallskip

It should be recalled that an element $e$ in a JB$^*$-triple $E$ is called a \emph{tripotent} if it is a fixed point for the triple product, that is, $\{e,e,e\} =e$. The set of all tripotents in $E$ will be denoted by $\mathcal{U}(E)$. Two tripotents $e$ and $v$ in $\mathcal{U}(E)$ are called orthogonal if the triple product $\{e,e,v\}$ vanishes. A partial order is defined on $\mathcal{U}(E)$ by the relation $e\leq u$ if $u-e$ is a tripotent which is orthogonal to $e$ (see subsection \ref{subsec: background} for more details).\smallskip

Let  $\Phi : \mathcal{U}(E) \to \mathcal{U}(F)$ be a bijection between the sets of tripotents in two JB$^*$-triples $E$ and $F$. We shall say that $\Phi$ \emph{preserves the partial ordering} or \emph{preserves the partial ordering in one direction} (respectively, \emph{preserves orthogonality} or \emph{preserves orthogonality in one direction}) if $e\leq u$ (respectively, $e\perp u$) in $\mathcal{U}(E)$ implies $\Phi (e) \leq \Phi (u)$ (respectively, $\Phi(e) \perp \Phi (u)$) in $\mathcal{U}(F)$. The mapping $\Phi$ \emph{preserves the partial ordering} \emph{in both directions} (respectively, \emph{orthogonality} \emph{in both directions}) when the equivalence $e\leq u \Leftrightarrow \Phi (e) \leq \Phi (u)$ (respectively, $e\perp u \Leftrightarrow \Phi (e) \perp \Phi (u)$) holds for all $e,u\in \mathcal{U}(E)$.\smallskip

From a pure mathematical point of view a very natural question motivated by Theorem \ref{t Molnar 2002} can be posed in the following terms: Let $M$ be a Cartan factor or an atomic JBW$^*$-triple. Suppose that $\Phi : \mathcal{U}(M) \to \mathcal{U}(M)$ is a bijective transformation which preserves the partial ordering and orthogonality between tripotents in both directions. Assume, additionally, that $\Phi$ is continuous at a single element of $\mathcal{U}(M)$ whose components in the different factors are non-zero. Can we extend $\Phi$ to a real linear triple isomorphism? We shall see later that some of the hypotheses can and will be relaxed.\smallskip

However, our motivation is not merely mathematical. In Section \ref{sec: physical motivations} we consider the quantum state space of the spin of a spin-$\frac12$ particle. We show that the spin domain represents the geometry of this state space. By embedding the Minkowski space--time into this domain we find the physical meaning of the pure states and the partial ordering among them. We shall later show that if we use Lorentz's transformation to transform the spin of a moving particle from the co-moving frame to the lab frame, the pure states should be replaced with pure atoms and projections with tripotents. Thus, for a relativistic description the lattice of projections with its natural ordering is not enough to deduce a Wigner's type theorem for such domain (which was also known from several counter--examples, compare \cite[Proposition 4.9 and Example 4.1]{CasdeVilahtiLevrero97} or \cite[comments in page 449]{Chev2007}), but we need to consider bijections on the set of  tripotents preserving the partial order and orthogonality among them.\smallskip

By employing a literally device of analepsis or flashback, we begin with a look at our main conclusion, established in Theorem \ref{t main atomic JBWtriples}, where we prove that given two atomic JBW$^*$-triples $\displaystyle M= \bigoplus_{i\in I}^{\ell_{\infty}} C_i$ and  $\displaystyle N = \bigoplus_{j\in J}^{\ell_{\infty}} \tilde{C}_j$, where $C_i$ and $C_j$ are Cartan factors with rank $\geq 2$, each bijective transformation $\Phi : \mathcal{U}(M) \to \mathcal{U}(N),$ which is continuous at a tripotent $u = (u_i)_i$ in $M$ with $u_i\neq 0$ for all $i$, 
and preserves the partial ordering in both directions and orthogonality between tripotents in one direction, admits an extension to a real linear triple isomorphism $T: M\to N$. Furthermore, $M$ decomposes as the direct sum of two orthogonal weak$^*$-closed ideals $M_1$ and $M_2$ such that $T|_{M_1}$ is complex-linear and $T|_{M_2}$ is conjugate-linear. Since triple automorphisms on $B(H)$ preserve any triple transition probability defined in terms of the triple product, our result, in particular, implies that orthogonality and bi-order preserving bijections also preserve transition probabilities.
\smallskip

The main result is obtained after a series of technical results and studies on particular cases. In Section \ref{sec: order preservers between tripotents in general JBWtriples} we explore the general properties of those bijections on tripotents preserving the partial ordering in both directions and orthogonality between tripotents in one direction. Proposition \ref{p order preservation in both directions gives orthogonality preservation} proves that every bijection between the sets of tripotents of two JB$^*$-triples preserving the partial ordering in both directions and orthogonality between tripotents in one direction must also preserve orthogonality in both directions. We show next that every such bijection under study maps zero to zero, preserves order minimal and maximal tripotents, and is additive on finite sums of mutually orthogonal tripotents (see Lemma \ref{c first consequences}). Assuming that the bijection $\Phi$ acts between the set of tripotents of two JBW$^*$-triples, we prove that it must preserve infima and suprema of families of tripotents and the weak$^*$ limits of series given by families of mutually orthogonal tripotents (see Lemma \ref{l infima and suprema sums of orthogonal families}).\smallskip

After showing that a bijection preserving the partial ordering in both directions and orthogonality between the sets of tripotents of two atomic JBW$^*$-triples maps (bijectively) the tripotents in each factor in the domain JBW$^*$-triple to the tripotents in a single factor of the codomain (cf. Lemma \ref{l Peirce 2 for fully non-complete tripotents}), we can restrict our study to those bijections preserving the partial ordering in both directions and orthogonality between the sets of tripotents of two Cartan factors. One of the key advances is in Section \ref{sec: order preservers between spin factors}, where we show the following: each bijection from the set of tripotents of a spin factor onto the set of tripotents of any other Cartan factor which preserves
the partial ordering in both directions and orthogonality between tripotents, and satisfying a mild continuity property, can be extended to a complex-linear or conjugate-linear triple isomorphism (see Theorem \ref{t biorder preserving from a spin into a Cartan factor}). As a first consequence of this result, we extend Moln{\'a}r's Theorem \ref{t Molnar 2002} by showing that for any complex Hilbert space $H$ with dim$(H)\geq 2$ and any Cartan factor $\tilde{C}$, every bijection $\Phi : \mathcal{U}(B(H))\to \mathcal{U}(\tilde{C})$
preserving the partial ordering in both directions and orthogonality between tripotents, which is continuous at a single non-zero element in $\mathcal{U}(B(H))$, extends to a real linear triple isomorphism (see Theorem \ref{t Molnar 2002 for rank 2 and weaker hypotheses}).\smallskip

The result for the spin factor is extended to other Cartan factors in Section \ref{sec:order preservers between other Cartan factors}. Those Cartan factors of rank $\geq 2$ admitting a unitary tripotent --i.e., those Cartan factors of rank $\geq 2$ which are JBW$^*$-algebras-- are jointly treated in Theorem \ref{t Molnar thm rank 2 Cartan factors with a unitary}. The proof in the case of Cartan factors with rank $\geq 3$ uses the Jordan version of the Bunce--Wright--Mackey--Gleason theorem in \cite{BuWri89}. The conclusion of Theorem \ref{t Molnar thm rank 2 Cartan factors with a unitary} covers all type 1 Cartan factors of the form $B(H)$ with dim$(H)\geq 2$, Cartan factors of type 2 with dim$(H)\geq 6$ even, or infinite, all type 3 Cartan factors with dim$(H)\geq 3,$ and the exceptional Cartan factor of type 6. The remaining Cartan factors, namely, the rectangular type 1 Cartan factors with rank larger than or equal to two, the type 2 Cartan factors non-admitting a unitary element, and the exceptional type 5 Cartan factor are treated in Theorems \ref{t BHK rectangular}, \ref{t type 2 CF odd dimensional}, and \ref{t exceptional type 5 CF}, respectively.\smallskip

After the conclusions in this note, it seems very natural to ask whether every bijection between the sets of tripotents of two Cartan factors with rank $\geq 2$ preserving the partial ordering in both directions must preserve orthogonality (in both directions) automatically.

\subsection{Notation and background}\label{subsec: background}

A complex Banach space $E$ is called a \emph{JB$^*$-triple} if it admits a continuous triple product $\J \cdot\cdot\cdot :
E\times E\times E \to E,$ which is symmetric and bilinear in the first and third variables, conjugate-linear in the middle one,
and satisfies the following axioms:
\begin{enumerate}[{\rm (a)}] \item (Jordan identity)
$$L(a,b) L(x,y) = L(x,y) L(a,b) + L(L(a,b)x,y)
 - L(x,L(b,a)y)$$ for $a,b,x,y$ in $E$, where $L(a,b)$ is the operator on $E$ given by $x \mapsto \J abx;$
\item $L(a,a)$ is a hermitian operator with non-negative spectrum for all $a\in E$;
\item $\|\{a,a,a\}\| = \|a\|^3$ for each $a\in E$.\end{enumerate}

Let $E$ be a JB$^*$-triple. An element $e\in E$ is a \emph{tripotent} if $\{e,e,e\}= e$. If we fix a tripotent $e$ in $E$, we can find a decomposition of the space in terms of the eigenspaces of the operator $L(e,e)$ given in the following terms:
\begin{equation}\label{Peirce decomp} {E} = {E}_{0} (e) \oplus  {E}_{1} (e) \oplus {E}_{2} (e),\end{equation} where ${
E}_{k} (e) := \{ x\in {E} : L(e,e)x = {\frac k 2} x\}$ is a subtriple of ${E}$ called the \emph{Peirce-$k$ subspace} ($k=0,1,2$). \emph{Peirce-$k$ projection} is the name given to the natural projection of ${E}$ onto ${E}_{k} (e)$ and it is usually denoted by $P_{k} (e)$. 
Triple products among elements in different Peirce subspaces obey certain laws known as \emph{Peirce arithmetic}. Concretely,
the inclusion $\J {{E}_{k}(e)}{{E}_{l}(e)}{{E}_{m}(e)}\! \subseteq {E}_{k-l+m} (e),$ and the identity $\J {{E}_{0}(e)}{{E}_{2} (e)}{{E}}\! =\! \J {{E}_{2} (e)}{{E}_{0} (e)}{{E}}\! =\! \{0\},$ hold for all $k,l,m\in \{0,1,2\}$, where ${E}_{k-l+m} (e) = \{0\}$ whenever $k-l+m$ is not in $\{0,1,2\}$. The Peirce-$2$ subspace ${E}_{2} (e)$ is a unital JB$^*$-algebra with respect to the product and involution given by $x \circ_e y = \J xey$ and $x^{*_e} = \J exe,$ respectively. 
\smallskip

A tripotent $e$ in $E$ is called \emph{algebraically minimal} (respectively, \emph{complete} or \emph{algebraically maximal}) if  $E_2(e)=\CC e \neq \{0\}$ (respectively, $E_0 (e) =\{0\}$). We shall say that $e$ is a \emph{unitary tripotent} if $E_2(e) =E$. The symbols $\mathcal{U} (E)$, $\mathcal{U}_{min} (E)$, and $\mathcal{U}_{max} (E)$ will stand for the sets of all tripotents, all minimal tripotents, and all complete tripotents in $E$, respectively.\smallskip

A JB$^*$-triple might contain no non-trivial tripotents, that is the case of the JB$^*$-triple $C_0[0,1]$ of all complex-valued continuous functions on $[0,1]$ vanishing at $0$. In a JB$^*$-triple $E$ the extreme points of its closed unit ball are precisely the complete tripotents in $E$ (cf. \cite[Lemma 4.1]{BraKaUp78}, \cite[Proposition 3.5]{KaUp77} or \cite[Corollary 4.8]{EdRutt88}). Thus, every JB$^*$-triple which is also a dual Banach space contains an abundant set of tripotents. JB$^*$-triples which are additionally dual Banach spaces are called \emph{JBW$^*$-triples}. Each JBW$^*$-triple admits a unique (isometric) predual and its triple product is separately weak$^*$ continuous (cf. \cite{BarTi}).\smallskip

A JBW$^*$-triple is called \emph{atomic} if it coincides with the w$^*$-closure of the linear span of its minimal tripotents. A very natural example is given by $B(H),$ where each minimal tripotent is of the form $\xi\otimes \xi$ with $\xi$ in the unit sphere of $H$. Furthermore, every Cartan factor is an atomic JBW$^*$-triple. It is known that building upon these examples we can exhaust all possible cases since every atomic JBW$^*$-triple is an $\ell_{\infty}$-sum of Cartan factors (cf. \cite[Proposition 2 and Theorem E]{FriRu86}).\smallskip

The notion of orthogonality between tripotents is an important concept in the theory of JB$^*$-triples. Suppose $e$ and $v$ are two tripotents in a JB$^*$-triple $E$. According to the standard notation (see, for example \cite{loos1977bounded,Batt91}) we say that $e$ is \emph{orthogonal} to $u$ ($e\perp u$ in short) if $\{e,e,u\}=0$. It is known that $e\perp u$ if and only if $\{u,u,e\}=0$ (and the latter is equivalent to any of the next statements $ L(e,u) = 0,$ $L(u,e) = 0,$  $e \in E_0(u),$ $u\in E_0(e)$ cf. \cite[Lemma 3.9]{loos1977bounded}). It is worth to remark that two projections $p$ and $q$ in a C$^*$-algebra $A$ regarded as a JB$^*$-triple are orthogonal if and only if $p q =0$  (that is, they are orthogonal in the usual sense).\smallskip

We can also speak about orthogonality for pairs of general elements in a JB$^*$-triple $E$. We shall say that $x$ and $y$ in $E$ are \emph{orthogonal} ($x\perp y$ in short) if $L(x,y)=0$ (equivalently $L(y,x)=0$, compare \cite[Lemma 1.1]{BurFerGarMarPe} for several reformulations). Two orthogonal elements $a$ and $b$ in JB$^*$-triple $E$ are $M$-orthogonal, that is, $\|a+b\| = \max\{\|a\|, \|b\|\}$ (see \cite[Lemma 1.3$(a)$]{FriRu85}). \smallskip

Building upon the relation ``being orthogonal'' we can define a canonical order $\leq$ on tripotents in $E$ given by $e\leq u$ if and only if $u-e$ is a tripotent and $u-e \perp e$. This partial order is precisely the order considered by L. Moln{\'a}r in Theorem \ref{t Molnar 2002} and it provides an important tool in JB$^*$-triples (see, for example, the recent papers \cite{HamKalPePfi20,HamKalPePfi20Groth,HamKalPe20} where it plays an important role). This relations enjoys several interesting properties, for example, $e\leq u$ if and only if $e$ is a projection in the JB$^*$-algebras $E_2(u)$ (cf. \cite[Lemma 3.2]{Batt91} or \cite[Corollary 1.7]{FriRu85} or \cite[Proposition 2.4]{HamKalPe20}). In particular, if $e$ and $p$ are tripotents (i.e., partial isometries) in a C$^*$-algebra $A$ regarded as a JB$^*$-triple with the triple product in \eqref{eq triple product JCstar triple} and $p$ is a projection, the condition $e\leq p$ implies that $e$ is a projection in $A$ with $e\leq p$ in the usual order on projections (i.e., $p e = e$).\smallskip

A non-zero tripotent $e$ in $E$ is called (\emph{order}) \emph{minimal} (respectively, (\emph{order}) \emph{maximal}) if $0\leq u \leq e$ for a tripotent $u$ in $E$ implies that $u =e$ (respectively, $e \leq u$ for a tripotent $u$ in $E$ implies that $u =e$). Clearly, every algebraically minimal tripotent is (order) minimal but the reciprocal implication does not necessarily hold, for example, the unit element in $C[0,1]$ is order minimal but not algebraically minimal. In the C$^*$-algebra $C_0[0,1]$ of all continuous functions on $[0,1]$ vanishing at 0, the zero tripotent is order maximal but it is not algebraically maximal. In the setting of JBW$^*$-triples these pathologies do not happen, that is, in a JBW$^*$-triple order and algebraic maximal (respectively, minimal) tripotents coincide (cf. \cite[Corollary 4.8]{EdRutt88} and \cite[Lemma 4.7]{Batt91}).\smallskip

We shall say that a tripotent $e$ in a JB$^*$-triple $E$ has rank--$n$ (denoted by $r(e) =n$) if it can be written as the sum of $n$ orthogonal minimal tripotents in $E$. The rank of a JB$^*$-triple is a more technical notion. A subset $S$ of $E$ is called \emph{orthogonal} if $0 \notin S$ and $x \perp y$ for every $x\neq y$ in $S$. The minimal cardinal number $r$ satisfying $\hbox{card}(S) \leq r$ for every orthogonal subset $S \subseteq E$ is called the \emph{rank} of $E$ (cf. \cite{Ka97}, \cite{BuChu92} and \cite{BeLoPeRo} for basic results on the rank of a Cartan factor and a JBW$^*$-triple and its relation with reflexivity). It is known that for each tripotent $e$ in a Cartan factor $C$ we have $r(e) = r(C_2(e))$ (see, for example, \cite[page 200]{Ka97}).\smallskip

A triple homomorphism between JB$^*$-triples $E$ and $F$ is a linear map $T:E\to F$ such that $T\{a,b,c\} = \{T(a),T(b), T(c)\}$ for all $a,b,c\in E$. A triple isomorphism is a bijective triple homomorphism. Clearly, the inclusion $T(\mathcal{U} (E)) \subseteq \mathcal{U} (F)$ holds for each triple homomorphism $T$, while the equality  $T(\mathcal{U} (E)) =\mathcal{U} (F)$ is true for every triple isomorphism $T$. Every injective triple homomorphism is an isometry (see \cite[Lemma 1]{BarDanHor88}). Actually a deep result established by Kaup in \cite[Proposition 5.5]{Ka83} proves that a linear bijection between JB$^*$-triples is a triple isomorphism if and only if it is an isometry. Therefore, each triple isomorphism $T: E\to F$ induces a surjective isometry $T|_{\mathcal{U} (E)}: \mathcal{U} (E)\to \mathcal{U} (F)$ which preserves orthogonality and partial order in both directions.

\section{Physical motivation for bijections preserving the order between sets of tripotents}\label{sec: physical motivations}

We show that the four-dimensional complex spin domain can be used to describe the spin of an electron. However, if we want the model to be Lorentz invariant, the order structure of the projection lattice is not enough. The spin state in a co-moving frame of an electron is described by a projection.  Transformation of the spin state from the co-moving frame of a moving electron to the lab frame transforms projections into projections only under Galilean transformations, but if we use  Lorentz  transformations, the projections are mapped into tripotents. Thus, a relativistic theory of the spin of an electron should preserve partial ordering of tripotents in a spin factor.\smallskip

Let $M$ be a complex spin factor as defined in the introduction. Let $\langle \cdot, \cdot \rangle$ and $x\mapsto \overline{x}$ denote, respectively, the inner product and the conjugation on $M$ for which the triple product is given by $$\{a,b,c\} = \langle a, b\rangle c + \langle c,b\rangle a-\langle a,\bar{ c}\rangle \bar{b}. $$  The real subspace $M_{_\mathbb{R}}=\{a\in M: \, a=\bar{a}\},$ called the real part of $M$, is a real Hilbert space with respect to the inner product $( a, b) = \Re\hbox{e}\langle a, b\rangle$ ($a,b\in M_{_\mathbb{R}}$), and $M = M_{_\mathbb{R}}\oplus i M_{_\mathbb{R}}$. We shall denote by $S_{_\mathbb{R}}$ the unit sphere of $M_{_\mathbb{R}}$. It is known (see  \cite[Theorem in page 196]{HervIs92} or \cite[Section 3.1.3]{Fri2005}) that for any triple automorphism $T$ of $M$, there is a complex number $\lambda\in \mathbb{T}=\{\lambda\in\mathbb{C}:  |\lambda|=1\}$ and a unitary operator $U\in B(M_{_\mathbb{R}})$ such that $T(a +ib) =\lambda (U(a) + i U(b))$ for all $a, b\in M_{_\mathbb{R}}$.  Conversely, for any $\lambda\in \mathbb{T}$ and each unitary $U\in B(M_{_\mathbb{R}})$, the map $T(a +ib) =\lambda (U(a) + i U(b))$ ($a, b\in M_{_\mathbb{R}}$) is a triple automorphism on $M$.\smallskip

The set of tripotents $\mathcal{U}(M)$ consists of two types, minimal and maximal, the latter being of rank 2. By using the partial order on the tripotents, for any non-zero tripotent $u$, we can define an \textit{order interval} $(0,u]=\{v: v\le u\}\subset \mathcal{U}(M)$ as the set of tripotents which are less or equal to $u$. If $u$ is a minimal tripotent, $(0,u]=\{u\}$. An element $u$ in $M$ is a maximal tripotent if and only if
 \begin{equation}\label{maxTripGood}
  \hbox{there is a $\lambda\in \mathbb{T}$ and an element $a\in S_{_\mathbb{R}}$ such that }  u=\lambda a,
 \end{equation}  (see \cite[Section 3.1.4]{Fri2005} or \cite[Lemma 6.1]{HamKalPe20} or \cite[Section 3]{KalPe2019}). Thus, the set $\mathcal{U}_{max}(M)$ of maximal tripotents can be identified with $\mathbb{T}\times S_{_\mathbb{R}}.$\smallskip

Suppose $x,y\in S_{_\mathbb{R}}$, with $\langle x, y\rangle =0$. It is known (and easy to check) that the element $v=\frac12 (x+iy)$ is a minimal tripotent in $M$. Actually, every minimal tripotent in $M$ is of the form
\begin{equation}\label{minTripGood}
   v=\frac{\lambda}{2} (x+iy) = \frac{(\alpha x-\beta y) + (\beta x+\alpha y) }{2} = \frac{1}{2} (a+i b),  \end{equation}
where $\lambda = \alpha +i \beta\in \mathbb{T}$, $a= \alpha x-\beta y$ and $b=\beta x+\alpha y$ are in $S_{_\mathbb{R}}$, with $\langle x, y\rangle = \langle a, b\rangle =0$ (see, for example, \cite[Lemma 6.1]{HamKalPe20}).\smallskip

\begin{figure}[ht!]
\centering
 \scalebox{0.17}{\includegraphics{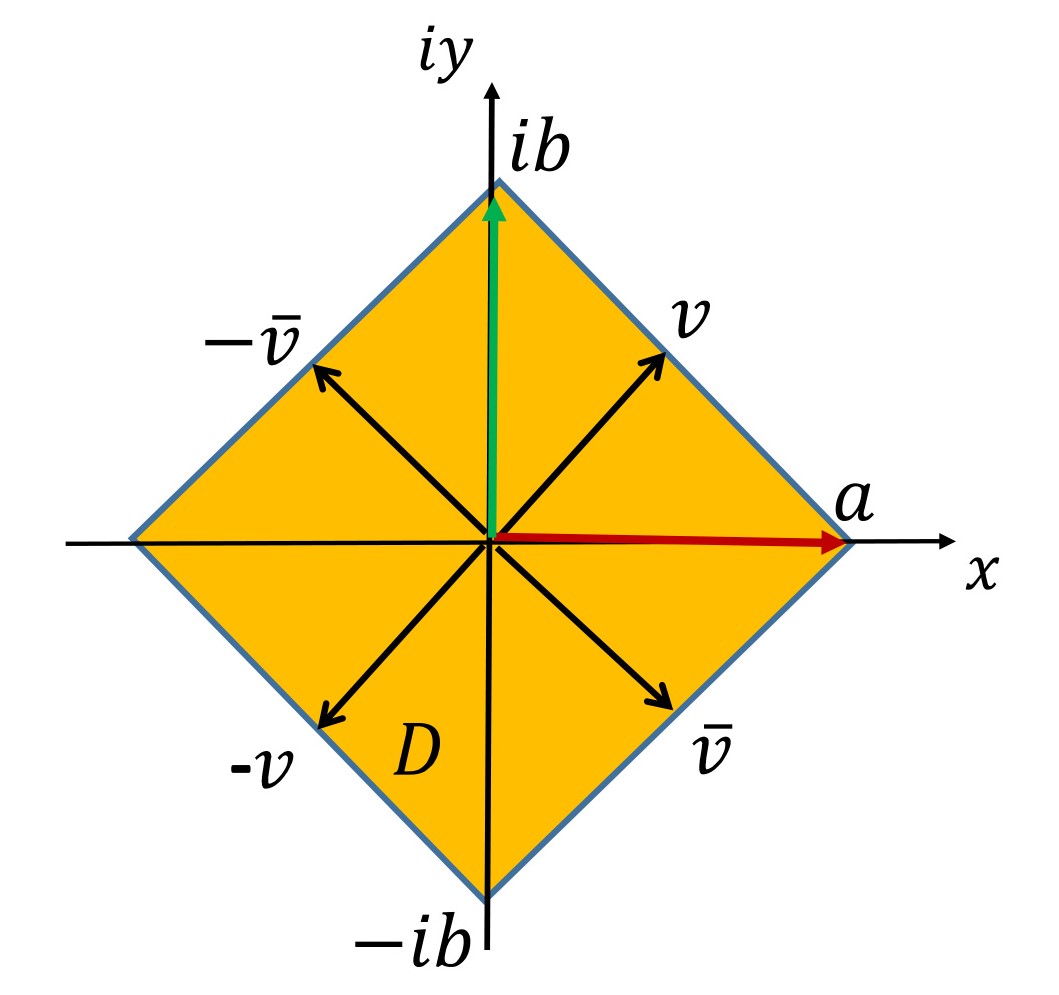}}
 \caption{\small Decomposition of a minimal tripotent $v=\frac{1}{2} ({a}+i{b})$ into its real and imaginary parts. The tripotent $\bar{v}=\frac{1}{2} ({a}-i{b})$ is orthogonal to $v$, with $v+\bar{v} =a$ and $v-\bar{v} =ib$. The tripotent $v$ is the intersection $(0,a]\cap (0,ib]$ of the intervals $(0,a]$ and $(0,ib]$.}\label{spinTr1}
\end{figure}

Any minimal tripotent decomposes into its real and imaginary parts in the form
  \begin{equation}\label{minTripabGood}
   v= \frac{1}{2} (a+i b),  \end{equation}
   where $a,b$ in $S_{_\mathbb{R}}$ and $\langle a, b\rangle =0$. The minimal tripotent $v$ is the average of a maximal tripotent from $S_{_\mathbb{R}}$ and a maximal tripotent from $iS_{_\mathbb{R}}$ (cf. \cite[Section 3.1.4]{Fri2005}, see Figure \ref{spinTr1}).

\begin{figure}[h!]
\centering
 \scalebox{0.33}{\includegraphics{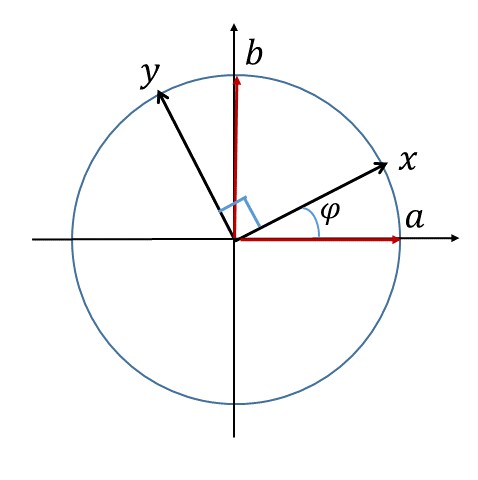}}
 \caption{\small Decomposition (\ref{minTripGood}) of  a  minimal tripotent $v=\frac{1}{2} (a+ib)=\frac{e^{-i\varphi}}{2} (x+iy)$. The maximal tripotents  $x,y$ are rotations by an angle $\varphi$ of ${a},{b}$ in the plane  $a,b$. }\label{spinTr2}
\end{figure}

 Furthermore, for each minimal tripotent $v=\frac{{\lambda}}{2} (x+iy)$, the element $\tilde{v}=\frac{{\lambda}}{2} (x-iy)$ is also a minimal tripotent which is orthogonal to $v$. Since $v+\tilde{v}=\lambda x,$ we have $v\leq \lambda x$ and $v\in (0,\lambda x]$. Also, $-\tilde{v}$  is a minimal tripotent, and  $v-\tilde{v}=i\lambda y$, implying that $v\leq \lambda iy$ and $v\in (0,i\lambda y]$. Thus, the tripotent $v$ is also the unique element in the intersection $(0,\lambda x]\cap (0,i\lambda y]$ of the intervals generated by the maximal tripotents $\lambda x,$ and $i\lambda y$. For the geometric connections between $x,y$ and $a,b$ for a tripotent $v=\frac{\lambda}{2} (x+iy)$, the reader is referred to Figure \ref{spinTr2}.


A seminal version of our next result is contained in \cite[Proposition 6.3$(1)$]{HamKalPe20}. Here we provide some additional details.

\begin{lemma}\label{l order spinGood} Let $u=\lambda a$ be a maximal tripotent in $M$, with $a\in S_{_\mathbb{R}}$ and $\lambda\in \mathbb{T}$. If $v$ is a minimal tripotent in $M$ such that $v\leq u$, then there is an element $b\in S_{_\mathbb{R}}$ with $ \langle a, b\rangle =0$ such that $v=\lambda \frac{1}{2} (a+i b)$ and $v$ is the average of a maximal tripotent from $\lambda S_{_\mathbb{R}}$ and a maximal tripotent from $i \lambda S_{_\mathbb{R}}$. Moreover, $\tilde{v}=\lambda \frac{1}{2} (a-i b)= \lambda \overline{\frac{1}{2} (a+i b)}$ is a minimal tripotent orthogonal to $v$ and $ u = v+\tilde{v} = \lambda a.$
\end{lemma}

\begin{proof} Since multiplication by $\bar{\lambda}$ is a triple automorphism, it is enough to prove the result for $\lambda=1$.  According to \eqref{minTripGood}, we can write $ v=\frac{1}{2} (x+ib)$ for $ x,b\in S_{_\mathbb{R}}$, where $ \langle x, b\rangle =0.$ The assumption $v\leq u$ implies that there is a minimal tripotent $\tilde{v}$ orthogonal to $v$ such that $u=v+\tilde{v}$. But any minimal tripotent $\tilde{v}$ orthogonal to $v$ is of the form $\tilde{v}=\mu \bar{v}$ for some $\mu\in \mathbb{T}$. We therefore have
$$\begin{aligned} u &= a = \frac{1}{2} (x+i b)  + \mu \frac{{1}}{2} (x-i b)  = \frac{1+\mu}{2} x + i \frac{1-\mu}{2} b \\
&= \left( \frac{1+\alpha}{2} x + \frac{\beta}{2} b \right) + i \left(\frac{\beta}{2} x + \frac{1-\alpha}{2} b\right),
 \end{aligned}$$ where $ \mu = \alpha+i \beta \in \mathbb{T}$. Having in mind that $a\in S_{_\mathbb{R}}$, we deduce that $\mu =1$ and $a=x$ as desired.
\end{proof}

 If $M$ is a spin domain, any non-zero and non-minimal tripotent $u$ is maximal and of the form given in \eqref{maxTripGood}, that is, $ u=\lambda a, \; \lambda\in \mathbb{T}, a\in S_{_\mathbb{R}},$ and our Lemma \ref{l order spinGood} implies that any tripotent in the interval $(0,u]$ is defined by a vector $b\in  S_{_\mathbb{R}}$ satisfying $ \langle a, b\rangle =0$. Thus $(0,u]$ is a sphere of dimension $\dim(M_{_\mathbb{R}}) -1$, with center at $\frac{1}{2}u$.\smallskip

For the description of the spin of an electron, we use the four-dimensional spin factor which we identify with the space $\mathbb{C}^4$. Denote the natural basis by $e_0,e_1,e_2,e_3$. Any element $x=\sum x_\mu e_\mu$ for some $x_\mu\in \mathbb{C}$. We assume that the conjugation is given by $\displaystyle \overline{x}=\sum\overline{x}_\mu e_\mu$, and the inner product is $\langle x, y \rangle=\sum_{\mu=0}^3 x_\mu \overline{y_\mu}$.\smallskip

We use a matrix representation (see \cite[\S 3.3.6]{Fri2005}) of this spin factor by associating with each basis element $e_\mu$ a $2\times 2$ complex matrix $\hat{e}_\mu$ defined by
\[\hat{e}_0=\left(
        \begin{array}{cc}
          1 & 0 \\
          0 & 1 \\
        \end{array}
      \right),\
     \hat{e}_1=\left(
        \begin{array}{cc}
          0 & -i \\
         -i & 0 \\
        \end{array}
      \right),\
      \hat{e}_2=\left(
        \begin{array}{cc}
          0 & -1 \\
         1 & 0 \\
        \end{array}
      \right),\
\hat{e}_3=\left(
        \begin{array}{cc}
          -i & 0 \\
          0 & i \\
        \end{array}
      \right).\]
  The matrices $\hat{e}_\mu$ can be identified with Pauli matrices  $\sigma_\mu$
  \[ \sigma_1=\left(
        \begin{array}{cc}
          0 & 1 \\
          1 & 0 \\
        \end{array}
      \right), \; \sigma_2=\left(
        \begin{array}{cc}
          0 & -i \\
          i & 0 \\
        \end{array}
      \right), \sigma_3=\left(
        \begin{array}{cc}
          1 & 0 \\
          0 & -1 \\
        \end{array}
      \right),\]
   as  $\hat{e}_j=-i\sigma_j$, for $j=1,2,3$, and $\hat{e}_0=I=\sigma_0$. The matrix representation of an arbitrary element $x$ is
  \begin{equation}\label{matembed}
   \hat{x}=\sum_{\mu=0}^{3} x_\mu \hat{e}_\mu=\left(
                               \begin{array}{cc}
                                 x_0-i x_3 &-x_2-ix_1 \\
                             x_2-i x_1 &  x_0+i x_3  \\
                               \end{array}
                             \right).
                             \end{equation}
    This leads to a simple formula for the determinant of any element $x$ as
    \begin{equation}\label{detDef}
      \det(x)=\det (\hat{x})=\sum_{\mu=0}^3 (x_\mu)^2=\langle x,\bar{x}\rangle\,.
    \end{equation}

The usual Quantum Mechanics description of the spin or angular momentum of an electron (see \cite[page 972]{Cohen-Tannoudji}) can be given as follows. We denote by $|z+> =|+>$ the \emph{state of the spin} of an electron moving up after passing a Stern--Gerlach apparatus in the $z$-direction and by $ |z-> =|->$ the \emph{state of the spin} of an electron moving down after passing the same apparatus. The \emph{general spin state} is described as a linear combination $|\chi>=c_+ |+>+c_-|->=(c_+,c_-) $, where $c_+$ and $c_-$ are complex numbers satisfying $|c_+|^2 +|c_-|^2=1.$\smallskip

It is customary (see \cite[page 171]{Ballentine}) to write the spin angular momentum operator  $\mathbf{S}=(S_1,S_2,S_3)$ as $S_j=\frac{1}{2}\hbar \sigma_j$, for $j=1,2,3,$, where
$\sigma_j$   are the Pauli matrices, mentioned above.  The states of  the spin of an electron  after passing a Stern--Gerlach apparatus in the $x$-direction are the eigenvectors of $S_1,$ $|x+>=\frac{1}{\sqrt{2}}(1,1) $ corresponding to the eigenvalue $\hbar/2$ and  $|x->=\frac{1}{\sqrt{2}}(1,-1)$ corresponding to the eigenvalue $-\hbar/2$. Similarly, the states of the spin passing a Stern--Gerlach apparatus in the  $y$-direction are the eigenvectors of $S_2,$  $|y+>=\frac{1}{\sqrt{2}}(1,i) $ corresponding to the eigenvalue $\hbar/2$ and $|y->=\frac{1}{\sqrt{2}}(1,-i) $ corresponding to the eigenvalue $-\hbar/2$.\smallskip

In the algebraic representation, these basic states $|j\pm>$ are identified with the projection  $P_{j\pm}$ on them as
$$P_{x+}=\frac{1}{2}\left(
           \begin{array}{cc}
             1 & 1 \\
             1 & 1 \\
           \end{array}
         \right),
         P_{x-}=\frac{1}{2}\left(
           \begin{array}{cc}
             1 & -1 \\
             -1 & 1 \\
           \end{array}
         \right),
         P_{y+}=\frac{1}{2}\left(
           \begin{array}{cc}
             1 & i \\
             -i & 1 \\
           \end{array}
         \right),$$
$$P_{y-}=\frac{1}{2}\left(
           \begin{array}{cc}
             1 & -i \\
             i & 1 \\
           \end{array}
         \right),
         P_{z+}=\left(
           \begin{array}{cc}
             1 & 0 \\
             0 & 0 \\
           \end{array}
         \right),
         P_{z-}=\left(
           \begin{array}{cc}
             0 & 0 \\
             0 & 1 \\
           \end{array}
         \right).$$

Note that all these $P_{j\pm}$ are projections of rank $1$, and thus are minimal in the projection lattice ordering.  For each $j$, we have $P_{j\pm}=\frac{1}{2}(\hat{e}_0\pm i\hat{e}_j)=\frac{1}{2}(\sigma_0\pm \sigma_j)$. They all belong to the interval $[0,I]=[0,\hat{e}_0]$, which consists of all projections in the space of $2\times 2$ complex matrices.\smallskip

 For any unit vector $\mathbf{b}$ in $\mathbb{R}^3$, the general \textit{spin state operator} $\varrho$ in the direction $\mathbf{b}$ is defined as
\begin{equation}\label{statOpa}
  \varrho=\frac{1}{2}\left(I+\sum_{j=1}^3 b_j\sigma_j\right)=\frac{1}{2}\left(\hat{e}_0+i\sum_{j=1}^3 b_j\hat{e}_j\right) \hbox{ (cf. \cite[page 172]{Ballentine})}.
\end{equation}

By using Lemma \ref{l order spinGood},   this formula reveals that a general spin state operator is a general minimal tripotent (in this case also a projection) in the interval $[0,I]=[0,\hat{e}_0]$ of the four-dimensional spin domain.\smallskip

To understand the physical meaning of this observation, we define a representation of Minkowski spacetime in the four-dimensional spin domain. We identify the Minkowski norm of a spacetime vector $x$ with the $\det(x)$ defined by \eqref{detDef}. The determinant of  the vector $\hat{e}_0=\sigma_0$ is 1, so we can identify it with the unit vector in the time direction.
For any $j=1,2,3$, the determinant of $i\hat{e}_j=\sigma_j$ is -1, the norm of a unit space-like  vector in Minkowski space. Thus, we can embed any spacetime vector $a$ with coordinates  $a_\mu\in \mathbb{R}$ into the spin domain as $\displaystyle \phi(a)=\sum_{\mu=0}^3 a_\mu\sigma_\mu$.  Under this embedding, it follows from \eqref{detDef} that $\det (\phi(a))$ is precisely the Minkowski norm of $a$.\smallskip

What is the meaning of formula  \eqref{statOpa} for a general spin operator $\varrho$?  The condition  $\varrho\in (0,\hat{e}_0]$ and the fact that  $\hat{e}_0=\sigma_0$ is the time direction show that the state $\varrho$ is stationary, and if it will not be disturbed, on a repeated measurement on time the result will not change. From Lemma \ref{l order spinGood}, we also see that $\varrho\in (0,i\sum b_j\hat{e}_j]=(0,\sum b_j\sigma_j]$. This condition expresses the fact $\varrho$  is an eigenvector of the spin measurement in the $\mathbf{b}$ direction. Since the spin is a pseudo-vector, it is reasonable to multiply $\mathbf{b}$ by the pseudo-scalar $i$, while decomposing it with respect to $\hat{e}_j$.   Note that the last ordering is not an ordering on the projection lattice.\smallskip

This description of the spin is valid for an electron at rest or moving with a low velocity with respect to the speed of light. On the other hand, if its velocity is large, we shall need to use spacetime transformations from a frame co-moving with the electron to our lab frame. It is known \cite[Chapter 1]{Fri2005}  that any spacetime transformation satisfying the Principle of Relativity must be either Galilean or Lorentz. Only for Galilean transformations the time remains invariant under the transformation, implying that a spin state $\varrho$ will remain a projection also in our lab frame. But if the transformations are Lorentz,  it is known, for example see \cite{FriRu01} and \cite{Fri2005}, that the Lorentz group acts on the four-dimensional spin domain by determinant-preserving transformations.  For such transformations, the time unit vector is not preserved if the relative velocity between the systems is not zero. In this case, the identity matrix $I$ will not be the identity after the transformation, but will be a maximal tripotent. Since a partial isometry $v$ is a projection only if $v<I$, the Lorentz transformation in general will not preserve projections but will preserve only the order among tripotents.\smallskip

It is known that a bijection preserving the order on the projection lattice of $M_2(\mathbb{C})$ does not necessarily preserve transition probabilities (see \cite[Example 4.1]{CasdeVilahtiLevrero97}), which should be observer-independent. As we shall show later, if a transformation preserves the partial ordering and orthogonality between tripotents, which is physical, as we have shown above, the transformations are triple automorphisms and thus preserve any physical phenomena which can be expressed by the triple product. 
\smallskip

The connection of the spin triple product to a new description of the relativistic spin of an electron and the state space of two-state systems, introduced here, has been further explored in \cite{Fri2021}. The transition probabilities coincide with the quantum mechanics predictions and agree with the experimental results testing quantum mechanics predictions based on Bell's inequality.\smallskip

 Why can we associate a triple product structure with the states of a quantum system in the first place? The measuring process defines some geometric properties on the state space. It was shown in \cite{FriRu89} and \cite{FriRu93} that a state space satisfying some physically significant properties implied by the measuring process is the predual of a JBW$^*$-triple.

\section{Order preserving bijections between sets of tripotents in JBW$^*$-triples}\label{sec: order preservers between tripotents in general JBWtriples}

This section is devoted to study the properties of those bijections between the set of tripotents of two general JB$^*$-triples preserving the partial ordering in both directions and orthogonality between elements. The first results will be focused on families of tripotents which are close to commute, that is, families of mutually orthogonal tripotents, or families of tripotents which are ordered by the natural partial order, or tripotents $e$ and $u$ satisfying that $u \in \mathbb{T} e$.\smallskip

\subsection{Order preserving bijections and orthogonality} \ \smallskip

In our first result we shall adapt and extend an argument taken from \cite[Lemma 2.3]{FriHak88}.

\begin{proposition}\label{p order preservation in both directions gives orthogonality preservation} Let $M, N$ be JB$^*$-triples. Suppose that $\Phi : \mathcal{U}(M) \to \mathcal{U}(N)$ is a bijective transformation which preserves the partial ordering between tripotents in both directions. Suppose additionally that $\Phi$ preserves orthogonality. Then $\Phi$ preserves orthogonality between tripotents in both directions.
\end{proposition}

\begin{proof}  The argument is valid for general JB$^*$-triples, even if in this general case we might have $\mathcal{U}(M),\mathcal{U}(N)=\{0\}$. To obtain the desired conclusion it suffices to prove that $\Phi^{-1}: \mathcal{U}(N) \to \mathcal{U}(M)$ preserves orthogonality.\smallskip 	
	
Let us take $\tilde{e}\perp \tilde{v}$ in $\mathcal{U} (N)$. Since $\tilde{e},\tilde{v}\leq \tilde{e}+\tilde{v}$, it follows from the assumptions that $\Phi^{-1} (\tilde{e}), \Phi^{-1} (\tilde{v})\leq \Phi^{-1} (\tilde{e}+\tilde{v})$ --observe that $\Phi^{-1}$ is a bijection preserving the partial ordering--. It follows from the definition of the partial order that there exists a tripotent ${w}\in \mathcal{U}(M)$ which is orthogonal to $\Phi^{-1}(\tilde{v})$ and $\Phi^{-1} (\tilde{e}+\tilde{v}) = \Phi^{-1} (\tilde{v}) + {w}$. By the surjectivity of $\Phi^{-1}$, there exists a tripotent  $\tilde{w}\in \mathcal{U}(N)$ such that ${w} = \Phi^{-1} (\tilde{w})$.\smallskip

Having in mind that $\Phi$ preserves orthogonality, we deduce that $\tilde{w}= \Phi(w)\perp \Phi \Phi^{-1}(\tilde{v}) = \tilde{v}$. In particular, $\tilde{v}+\tilde{w}$ is a tripotent in $N$.\smallskip

Since $\Phi^{-1}$ preserves the partial order in both directions and $$\Phi^{-1} (\tilde{v} + \tilde{w}) \geq \Phi^{-1} (\tilde{v}) +\Phi^{-1} (\tilde{w}) = \Phi^{-1} (\tilde{e}+\tilde{v}),$$ we deduce that $\tilde{v}+\tilde{w} \geq \tilde{e}+\tilde{v}$, or equivalently, $\tilde{w}-\tilde{e} = (\tilde{v}+\tilde{w}) - (\tilde{e}+\tilde{v})$ is a tripotent orthogonal to $\tilde{e}+\tilde{v}$. Now, by applying that $\tilde{e}+\tilde{v}\geq \tilde{e}$, we conclude that $\tilde{w}-\tilde{e}$ is a tripotent orthogonal to $\tilde{e}$, and hence $\tilde{w}\geq \tilde{e}$. A new application of the fact that $\Phi^{-1}$ preserves the partial order shows that ${w} =\Phi^{-1} (\tilde{w}) \geq \Phi^{-1} (\tilde{e})$. We also know that $\Phi^{-1} (\tilde{v}) \perp  {w}$. The last two statements together prove that $ \Phi^{-1} (\tilde{v}) \perp \Phi^{-1} (\tilde{e})$, yielding that $\Phi^{-1}$ preserves orthogonality between tripotents.
\end{proof}

Proposition \ref{p order preservation in both directions gives orthogonality preservation} implies that we can relax the hypothesis concerning the  preservation of orthogonality in both directions in Theorem \ref{t Molnar 2002}.\smallskip

We continue this section with some general results on bijections between sets of tripotents which preserve
the partial ordering in both directions and orthogonality between elements.

\begin{lemma}\label{c first consequences} Let $M, N$ be JB$^*$-triples. Suppose that $\Phi : \mathcal{U}(M) \to \mathcal{U}(N)$ is a bijective transformation which preserves the partial ordering in both directions and orthogonality between tripotents. Then the following statements hold:
\begin{enumerate}[$(a)$]\item $\Phi (0) =0$;
\item A tripotent $e\in M$ is order minimal {\rm(}respectively, order maximal{\rm)} if and only if $\Phi (e)$ is order minimal (respectively, order maximal) in $N$;
\item For each tripotent $e\in M$ the set $M_0(e) \cap \mathcal{U} (M)$ is mapped onto $N_0(\Phi(e)) \cap \mathcal{U} (N)$ by $\Phi$;
\item Suppose $e_1,\ldots, e_m$ are mutually orthogonal tripotents in $M$. Then \begin{equation}\label{eq Phi is additive on finite families of mo trip} \Phi(e_1)+\ldots + \Phi(e_m) = \Phi (e_1+\ldots + e_m);
\end{equation}
\item A tripotent $e\in M$ has rank--$n$ if and only if $\Phi (e)$ has rank--$n$ in $N$.
\end{enumerate}
\end{lemma}

\begin{proof}$(a)$ Since $0\perp 0$, the hypotheses imply that $\Phi (0) \perp \Phi (0),$ and thus $\Phi (0) =0$.\smallskip

$(b)$ The conclusion is clear from the comments preceding this corollary and the hypotheses, while $(c)$ is a consequence of fact that $\Phi$ preserves orthogonality between tripotents in both directions (cf. Proposition \ref{p order preservation in both directions gives orthogonality preservation}), because the tripotents in $M_0(e) \cap \mathcal{U} (M)$ are precisely the tripotents which are ortogonal to $e$.\smallskip

$(d)$ We shall first prove the case $m=2$. By applying that $\Phi$  preserves orthogonality between tripotents in one direction we get $\Phi(e_1)\perp \Phi(e_2)$ and $\Phi(e_1), \Phi(e_2)\leq \Phi (e_1+e_2).$ Therefore $\Phi(e_1)+ \Phi(e_2) \leq \Phi (e_1+e_2).$\smallskip

Similarly, having in mind that $\Phi^{-1}$ preserves orthogonality (cf. Proposition \ref{p order preservation in both directions gives orthogonality preservation}), we have $\Phi^{-1}(\Phi(e_1)+ \Phi(e_2))\geq \Phi^{-1}(\Phi(e_1))+\Phi^{-1}( \Phi(e_2))=e_1+e_2.$
Now, applying $\Phi$ to this inequality leads to $\Phi(e_1)+ \Phi(e_2)\geq\Phi(e_1+e_2),$ and thus to the equality of both expressions. An easy induction argument gives the desired statement. \smallskip

Finally $(e)$ follows from $(b)$ and $(d)$.
\end{proof}

Our next result reveals how a bijection preserving the partial ordering in both directions and orthogonality between the sets of tripotents in two JBW$^*$-triples enjoys certain continuity properties with respect to the weak$^*$ topologies.\smallskip

Let $\{e_i\}_{i\in \Lambda}$ be an arbitrary family of mutually orthogonal tripotents in a JBW$^*$-triple $M$. We know from \cite[Corollary 3.13]{Horn87} or \cite[Proposition 3.8]{Batt91} that the family $(e_i)_i$ is summable with respect to the weak$^*$ topology of $M$, that is, if $\mathcal{F}$ denotes the directed set of all non-empty finite subsets of $\Lambda$ ordered by inclusion, the net $\displaystyle \left(\sum_{i\in F} e_i \right)_{F\in \mathcal{F}}$ converges in the weak$^*$ topology of $M$ and its limit is denoted by $\displaystyle \sum_{i\in \Lambda} e_i = \hbox{w$^*$-}\sum_{i\in \Lambda} e_i$. It is further known that $\displaystyle \sum_{i\in \Lambda} e_i$ is precisely the supremum of the family $\{e_i\}_{i\in \Lambda}$. Since a bijection preserving the partial order in both directions between the sets of tripotents of two JBW$^*$-triples must preserve infima and suprema of families, the next lemma is a consequence of the above comments Proposition \ref{p order preservation in both directions gives orthogonality preservation} and Lemma \ref{c first consequences} (see also \cite[Theorem 3.6]{Batt91}).

\begin{lemma}\label{l infima and suprema sums of orthogonal families} Let $M, N$ be JBW$^*$-triples. Suppose that $\Phi : \mathcal{U}(M) \to \mathcal{U}(N)$ is a bijective transformation which preserves the partial ordering in both directions and orthogonality between tripotents. Then the following statements hold:\begin{enumerate}[$(a)$]\item Let $\{u_i : i\in\Lambda\}$ be a family of tripotents in $M$ with infimum $\displaystyle \wedge_{i\in \Lambda} u_i$. Then $\displaystyle \Phi (\wedge_{i\in \Lambda} u_i)$ is the infimum of the set $\{\Phi(u_i) : i\in\Lambda\}\subset \mathcal{U} (N);$
\item Let $\{u_i : i\in\Lambda\}$ be a family of tripotents in $M$. If the supremum $\displaystyle \vee_{i\in \Lambda} u_i$ exists in $\mathcal{U} (M)$, then the supremum of $\{\Phi(u_i) : i\in\Lambda\}\subset \mathcal{U} (N)$ also exists in $\mathcal{U} (N)$ and coincides with $\displaystyle\Phi\left( \vee_{i\in \Lambda} u_i \right)$;
\item Let $\{u_i : i\in\Lambda\}$ be a family of mutually orthogonal tripotents in $M$. Then $\{\Phi(u_i) : i\in\Lambda\}$ is a family of mutually tripotents in $N$ and $$\Phi\left(\sum_{i\in \Lambda} u_i \right) = \sum_{i\in \Lambda} \Phi(u_i);$$
\item Assuming that $M$ is an atomic JBW$^*$-triple, if $T: M\to N$ is a weak$^*$ continuous linear map such that $T(v)=\Phi(v)$ for all minimal tripotent $v\in M$, the maps $T$ and $\Phi$ coincide on $\mathcal{U}(M).$
\end{enumerate}
\end{lemma}

\begin{proof} The statements $(a)$, $(b)$ and $(c)$ have been already justified. For the last statement we simply observe that every tripotent $u$ in an atomic JBW$^*$-triple is the supremum of a family of mutually orthogonal minimal tripotents and any such family is weak$^*$ summable with limit $u$.
\end{proof}

For each subset $B$ of a JB$^*$-triple $E,$ we shall write $B^\perp$ for the \emph{(orthogonal) annihilator of $B$} given by $
B^\perp= B^{\perp}_{_E}:=\{ z \in E : z \perp x , \forall x \in B \}.$ It is known that $\{e\}^{\perp}= E_0 (e)$ and the inclusions $$ E_2(e) \oplus  E_1(e)\supseteq \{e\}_{_E}^{\perp \perp}= E_0(e)^{\perp} \supseteq E_2(e)$$ hold for every tripotent $e$ in $E$ (see \cite[Proposition 3.3]{BurGarPe11}). Contrary to what seems intuitive, the identity $\{e\}_{_E}^{\perp \perp}= E_2(e)$ need not be true (cf. \cite[Remark 3.4]{BurGarPe11}). However, if $e$ is a non-complete tripotent in a Cartan factor $C,$ the identity $\{e\}^{\perp \perp}=E_0(e)^\perp= E_2(e)$ always holds (see \cite[Lemma 5.6]{Ka97}). Furthermore, suppose that $\displaystyle M= \bigoplus_{i\in I} C_i$ is an atomic JBW$^*$-triple. A tripotent $e$ in $M$ will be called \emph{fully non-complete} if $\pi_i (e)$ is a non-complete tripotent in $C_i$ for each $i\in I$, where $\pi_i$ stands for the natural projection of $M$ onto $C_i$. If $M$ reduces to a single Cartan factor, a tripotent $e\in M$ is fully non-complete if and only if it is non-complete. The arguments in the proof of \cite[Lemma 2.3]{Pe2020} show that $\{e\}^{\perp\perp} = M_2(e)$ for each fully non-complete tripotent $e\in M$.\smallskip

A subspace $I$ of a JB$^*$-triple $E$ is said to be an \emph{ideal} if $\{ I,E,E\}\subset I$ and $\{ E,I,E\}\subset I$ (see \cite[\S 4]{Horn87}). Two subsets $A, B\subset E$ are called \emph{orthogonal} ($A\perp B$ in short) if $A\subset B^{\perp}$. Two ideals $I$ and $J$ are orthogonal if and only if $I \cap J = \{0\}$. If a JBW$^*$-triple $M$ decomposes in the form $M = I\oplus J$, where $I$ and $J$ are ideals, these two ideals must be weak$^*$closed \cite[Lemma 4.3]{Horn87}, moreover if a JB$^*$-triple $E$ admits two closed subtriples $I$ and $J$ satisfying $E = I \oplus J$, it is known that $E= I \oplus^{\ell_\infty} J$ if and only if $I$ and $J$ are ideals \cite[Lemma 4.4]{Horn87}. That is, the weak$^*$ closed $M$-ideals in a JBW$^*$-triple are precisely its weak$^*$ closed  ideals. A JBW$^*$-triple $M$ is a \emph{factor} if it is indecomposable, that is, $M$ cannot be decomposed as the direct ($\ell_{\infty}$-)sum of two orthogonal ideals (cf \cite[Definition 4.8]{Horn87}).\smallskip

In the next lemma we state a characterization needed in subsequent results.

\begin{lemma}\label{l characterization of non-factors} Let $M$ be a JBW$^*$-triple. Then  $M$ is not a factor if and only if there exist $A, B\subseteq \mathcal{U} (M)\backslash\{0\}$  with $A\perp B$ and $A\cup B = \mathcal{U} (M)\backslash\{0\}$.
\end{lemma}

\begin{proof} If $M$ is not a factor we can find two orthogonal non-zero weak$^*$ closed triple ideals $I$ and $J$ of $M$ such that $M = I \oplus J$. The ideals $I$ and $J$ are non-zero JBW$^*$-subtriples and thus the sets $A =  \mathcal{U} (I)\backslash\{0\}$ and $B=\mathcal{U} (J)\backslash\{0\}$ are contained in $\mathcal{U} (M)\backslash\{0\}$, $A\perp B$ and $\mathcal{U} (M)\backslash\{0\} = A\cup B,$ because each tripotent in $M$ writes uniquely as the sum of a tripotent in $I$ and tripotent in $J$. So, the ``only if'' implication holds.\smallskip

Suppose now that there exist $A, B\subseteq \mathcal{U} (M)\backslash\{0\}$  with $A\perp B$ and $A\cup B = \mathcal{U} (M)\backslash\{0\}$. We set $I = A^{\perp}$ and $J = B^{\perp}$. Clearly $B\subset I,$ $A\subset J,$ and thus $I$ and $J$ are non-zero.  It can be deduced from the separate weak$^*$ continuity of the triple product of $M$ that $I$ and $J$ are weak$^*$ closed. We shall next prove that $I$ and $J$ are ideals. We claim that $I$ (respectively, $J$) coincides with the norm closure of the linear span of all elements in $B$ (respectively, in $A$). Namely, fix $a\in I$. Since tripotents in $M$ are norm-total (cf. \cite[Lemma 3.11]{Horn87}) and $A\cup B = \mathcal{U} (M)\backslash\{0\}$, for each $\varepsilon >0$ there are $\lambda_1, \mu_1, \ldots, \lambda_n, \mu_n\in \mathbb{C}$ and mutually orthogonal tripotents $u_1, v_1 \ldots, u_n, v_n$ with $u_j\in A$ and $v_j\in B$ for all $j,$ such that $\displaystyle \left\| a - \left(\sum_{j=1}^n \lambda_j u_j + \mu_j v_j\right) \right\|<\varepsilon.$ By recalling that orthogonal elements are $M$-orthogonal, and applying that $\displaystyle a - \sum_{j=1}^n \mu_j v_j\perp  \sum_{j=1}^n \lambda_j u_j$ we arrive at $\displaystyle \left\| a - \left(\sum_{j=1}^n \mu_j v_j\right) \right\|<\varepsilon,$ which gives the desired conclusion for $I$. The statement for $J$ follows by similar arguments. The norm-totality of the set of tripotents in $M$ together with the fact that $A\cup B = \mathcal{U} (M)\backslash\{0\}$ prove that $M = I \oplus J$ and the rest is clear.
\end{proof}

We next consider bijections between sets of tripotents in two atomic JBW$^*$-triples (i.e., $\ell_{\infty}$-sums of Cartan factors). The next lemma shows that we can restrict ourself to the case of a single Cartan factor.

\begin{lemma}\label{l Peirce 2 for fully non-complete tripotents} Let $\displaystyle M= \bigoplus_{i\in I}^{\ell_{\infty}} C_i$ and  $\displaystyle N = \bigoplus_{j\in J}^{\ell_{\infty}} \tilde{C}_j$ be JBW$^*$-triples, where $C_i$ and $\tilde{C}_j$ are JBW$^*$-triple factors. Suppose that $\Phi : \mathcal{U}(M) \to \mathcal{U}(N)$ is a bijective transformation which preserves the partial ordering in both directions and orthogonality between tripotents. Then we have:
\begin{enumerate}[$(a)$]\item For each $i_0 \in I$ there exists a unique $\sigma(i_0)\in J$ such that $\Phi(\mathcal{U}(C_{i_0})) = \mathcal{U} (\tilde{C}_{\sigma({i_0})})$;
\end{enumerate}\smallskip
\noindent Assuming that $M$ and $N$ are atomic JBW$^*$-triples --i.e.,  $C_i$ and $C_j$ are Cartan factors for all $i,j$-- the following statements also hold:
\begin{enumerate}[$(b)$]
\item[$(b)$] $\Phi$ maps fully non-complete tripotents in $M$ to fully non-complete tripotents in $N$;
\item[$(c)$] For each fully non-complete tripotent $e$ in $M$, $\Phi$ maps the set $M_2(e) \cap \mathcal{U} (M)$ onto the set $N_2(\Phi(e)) \cap \mathcal{U} (N)$.
\end{enumerate}
\end{lemma}

\begin{proof}  We can assume, thanks to Proposition \ref{p order preservation in both directions gives orthogonality preservation}, that $\Phi$ preserves orthogonality in both directions. \smallskip
	
$(a)$ Every tripotent in $N$ writes in the form $(\tilde{e}_j)_{j\in J}$, where each $e_j$ is a tripotent in $\tilde{C}_j,$ and similarly for every tripotent in $M$. Fix $i_0\in I$. Suppose we can find $e, v\in \mathcal{U}(C_{i_0})$ such that $\Phi (e)$ has non-zero component in some $\tilde{C}_{j_1}$ and $\Phi(v)$ has non-zero component in some $\tilde{C}_{j_2}$ with $j_1\neq j_2$. Set $$A = \Phi^{-1} (\mathcal{U}(\tilde{C}_{j_1})\backslash\{0\}) \cap \mathcal{U}(C_{i_0}), \hbox{ and }$$ $$B = \Phi^{-1} (\{\tilde{e}\in N : \hbox{ the component of $\tilde{e}$ in }  \mathcal{U}(\tilde{C}_{j_1}) \hbox{ is zero}\})\cap \mathcal{U}(C_{i_0}).$$ Clearly, $A,B\subset \mathcal{U}(C_{i_0})\backslash\{0\}$ with $A\perp B$ and $A\cup B = \mathcal{U}(C_{i_0})\backslash\{0\}$ by the assumptions on $\Phi$. Lemma \ref{l characterization of non-factors} implies that $C_{i_0}$ is not a factor, which contradicts our hypotheses. Therefore there exists a unique $\sigma(i_0)\in J$ such that $\Phi(\mathcal{U}(C_{i_0})) \subseteq \mathcal{U} (\tilde{C}_{\sigma({i_0})})$, the equality follows from the same argument applied to $\Phi^{-1}$. \smallskip

$(b)$ Let $e= (e_i)_{i\in I}$ be a fully non-complete tripotent in $M$, that is, for each $i\in I$ there exists at least a minimal tripotent $v_{i}\in C_i$ such that $e_{i}\perp v_{i}$. By $(a)$, the fact that $\Phi$ preserves orthogonality, and Lemma \ref{c first consequences}$(b)$, we deduce that $\Phi(e_i) \perp \Phi(v_i)$ in $\tilde{C}_{j_{i}}$ and $\Phi(v_i)$ is a minimal tripotent in $\tilde{C}_{j_{i}}$. Lemma \ref{l infima and suprema sums of orthogonal families} assures that $\displaystyle \Phi(e) = \hbox{w$^*$-}\sum_{i} \Phi(e_i)$ with $\Phi(e_i) \perp \Phi(v_i)$, for all $i\in I$, which proves that $\Phi(e)$ is fully non-complete.\smallskip

$(c)$ Suppose $e$ is a fully non-complete tripotent in $M$. Let $u$ be a tripotent in $M_2(e)=\{e\}^{\perp\perp} = M_0(e)^{\perp}$ (cf. \cite[proof of Lemma 2.3]{Pe2020}). By Lemma \ref{c first consequences} we have $$\displaystyle \Phi\left( M_0(e) \cap \mathcal{U} (M)\right)  = N_0(\Phi(e)) \cap \mathcal{U} (N).$$ Thus, each tripotent $\widetilde{v} = \Phi (v) \in N_0(\Phi(e)) \cap \mathcal{U} (N)$ ($v\in M_0(e) \cap \mathcal{U} (M)$) must be orthogonal to $\Phi(e)$. Therefore  $\widetilde{v} = \Phi (v)$ must be orthogonal to $\Phi(u)$.\smallskip

The Peirce 0-subspace $N_0 (\Phi(e))$ is a JBW$^*$-triple, and hence every element there can be approximated in norm by a finite linear combination of mutually orthogonal tripotents in $N_0(\Phi(e))$ (cf. \cite[Lemma 3.11]{Horn87}). We have seen above that every tripotent $\widetilde{v}$ in $N_0(\Phi(e)) \cap \mathcal{U} (N)$ is orthogonal to $\Phi(u)$. We can therefore conclude that $\Phi(u)\in N_0(\Phi(e))^{\perp} = \{\Phi (e)\}^{\perp\perp} = N_2(\Phi(e)),$ where in the last equality we applied that $\Phi(e)$ is fully non-complete (cf. \cite[Lemma 2.3]{Pe2020}).
\end{proof}

Up to this point we did not need to distinguish between Cartan factors of rank--one from those of rank bigger than or equal to $2$. As we shall see in the next remark, the set of tripotents in a rank--one Cartan factor is too small to guarantee that a bijective transformation on this set preserving the partial order and orthogonality between tripotents in both directions admits an extension to a surjective real linear mapping.

\begin{remark}\label{r rank 1 for op bijections}{\rm Let $H$ be a complex Hilbert space regarded as a type 1 Cartan factor of the form $B(H, \mathbb{C})$, that is we consider its Hilbert norm and the triple product $$\{a,b,c\} = \frac12 (\langle a, b\rangle c + \langle c,b\rangle a), \ \ (a,b,c\in H).$$ It is easy to see that $\mathcal{U} (H) := S(H) \cup \{0\},$ where $S(H)$ stands for the unit sphere of $H$. Let $\Phi : \mathcal{U} (\mathbb{C}) = \mathbb{T}\cup \{0\}\to \mathcal{U} (\mathbb{C})= \mathbb{T}\cup \{0\}$ be the bijection defined by $$\Phi (e^{it}) =\left\{\begin{array}{cc}
                                                                                             e^{i(\pi-t)}, & \hbox{If } 0<t<\pi \\
                                                                                             -1, & \hbox{If } t=0 \\
                                                                                             1, & \hbox{If } t=\pi \\
                                                                                             e^{it}, & \hbox{If } -\pi <t <0
                                                                                           \end{array}
 \right., \hbox{ and } \Phi (0) =0.$$ Clearly, $\Phi$ preserves the partial order and orthogonality between tripotents in both directions because the unique possible relation is $p\leq e$, where $e$ is a non-zero (minimal and maximal) tripotent in $\mathbb{C}$ (i.e., and element in $\mathbb{T}$), and clearly $0\leq \Phi (e)$. If $T: \mathbb{C}\to \mathbb{C}$ is a real linear mapping satisfying $T|_{\mathbb{T}\cup \{0\}} = \Phi$. The conditions $T(1) = -1$, $T(i) = i$, and  $T\left(\frac{1}{\sqrt{2}} -i \frac{1}{\sqrt{2}}\right) = \frac{1}{\sqrt{2}} -i \frac{1}{\sqrt{2}}$ are incompatible with the real linearity of $T$.\smallskip

Let $M$ be a JBW$^*$-triple of rank--1, which can be identified with a complex Hilbert space. Let $\Phi : \mathcal{U}(M) \to \mathcal{U}({M})$ be a bijective transformation which preserves the partial ordering and orthogonality between tripotents in both directions. A non-zero element $u$ in $M$ is a tripotent if and only if it has norm one. Thus, the set of non-zero tripotents can be identified with the unit sphere $S$ in $M$. Since any tripotent in $M$ is minimal, the preservation of the partial ordering and orthogonality relation does not impose any restriction on the map $\Phi $. Thus, the collection of all bijective transformations preserving the partial ordering and orthogonality between tripotents in both directions coincide with the set of all bijections on $S\cup \{0\}$.
}\end{remark}

\subsection{Order preserving bijections acting on the circular orbit of a tripotent}\ \smallskip

Our next goal is to consider tripotents which are obtained as scalar multiples of a given tripotent by an element in the unit sphere of the complex field.\smallskip

We begin by recalling some terminology. Let $u,v$ be two tripotents in a JB$^*$-triple $E$. We shall say that $u$ and $v$ are \emph{collinear} (written $u\top v$) if $u\in E_1(v)$ and $v\in E_1(u)$. The tripotent $u$ \emph{governs} the tripotent $v$ ($u \vdash v$ in short) whenever $v\in E_{2} (u)$ and $u\in E_{1} (v)$. Following the standard sources --see, for example, \cite{DanFri87,Neher87}-- an ordered quadruple $(u_{1},u_{2},u_{3},u_{4})$ of minimal tripotents in a JB$^*$-triple $E$ is called a \emph{quadrangle} if $u_{1}\bot u_{3}$, $u_{2}\bot u_{4}$, $u_{1}\top u_{2}$ $\top u_{3}\top u_{4}$ $\top u_{1}$ and $u_{4}=2 \{{u_{1}},{u_{2}},{u_{3}}\}$ (the latter equality also holds if the indices are permutated cyclically, e.g. $u_{2} = 2 \{{u_{3}},{u_{4}},{u_{1}}\}$). A \emph{prequadrangle} is an ordered set $(u_1,u_2,u_3)$ of three tripotents such that $u_1\top u_2\top u_3$ and $u_1\perp u_3$. An ordered triplet $ (v,u,\tilde v)$ of minimal tripotents in $E$, is called a \emph{trangle} if $v\bot \tilde v$, $u\vdash v$, $u\vdash \tilde v$ and $ v = Q(u)\tilde v$.\smallskip

In the proof of \cite[Theorem 1]{Molnar2002} (see page 45), L. Moln{\'a}r stated the following property: for a finite dimensional complex Hilbert space $H_n$, a partial isometry $a\in B(H_n)$ is equal to the identity multiplied by a scalar from $\mathbb{T}$ if and only if for every rank--one projection $p\in B(H_n)$ we have that $\lambda p \leq a$ for some $\lambda\in \mathbb{T}$ depending on $p$. Our next result is an extension of this result to the setting of Cartan factors.

\begin{lemma}\label{l when a tripotent is a complex multiple of another tripotent} Let $u$ and $v$ be non-zero tripotents in a Cartan factor $C$. Then the following statements are equivalent:\begin{enumerate}[$(a)$]\item  There exists $\mu \in \mathbb{T}$ such that $\mu v\leq u$;
\item There exists $\gamma\in \mathbb{C}$ such that $u = \gamma v + P_0(v) (u)$
\item For each minimal tripotent $e\leq v$ there exists $\lambda\in \mathbb{T},$ depending on $e$, such that $\lambda e \leq u$.
\end{enumerate}
\end{lemma}

\begin{proof} The implication $(a)\Rightarrow (b)$ is clear from the definition of the partial order $\leq$; it is explicitly proved in \cite[Corollary 1.7]{FriRu85}. The implication $(b)\Rightarrow (c)$  is clear.\smallskip

We shall finally prove $(c)\Rightarrow (a).$ Let us distinguish several cases. If $v$ is minimal the conclusion trivially holds. We can therefore assume that $v$ has rank at least two. Let us find, via Zorn's lemma, a (possibly finite) maximal family $\{e_i: i\in I\}$ of mutually orthogonal minimal tripotents with $e_i\leq u$ (i.e., mutually orthogonal minimal projections in the JB$^*$-algebra $C_2(u)$)--just have in mind that $C$ equals the w$^*$-closure of the linear span of its minimal tripotents \cite[Theorem 2]{FriRu85} and \cite[\S 2]{FriRu86}. It follows from the maximality of the family $\{e_i: i\in I\}$ that $\displaystyle v = \sum_{i\in I} e_i$, where the sum converges with respect to the weak$^*$ topology.\smallskip

By hypothesis, for each $i\in I$, there exists $\lambda_i\in \mathbb{T}$ such that $\lambda_i e_i \leq u$.  Since $\{\lambda_i e_i: i\in I\}$ is a family of mutually orthogonal minimal tripotents the sum $\displaystyle \sum_i \lambda_i e_i$ converges in the weak$^*$-topology to a tripotent in $C$. We claim that \begin{equation}\label{eq u is an upper bound for the lambdai ei} \displaystyle \sum_{i\in I} \lambda_i e_i\leq u.
 \end{equation} By \cite[Proposition 3.8$(iii)$]{Batt91} it suffices to prove that $\lambda_i e_i\leq u$ for all $i\in I$, which holds by hypotheses. It follows from \eqref{eq u is an upper bound for the lambdai ei} that \begin{equation}\label{eq first form of u} u  = \sum_{i\in I} \lambda_i e_i  +P_0(v) (u).
 \end{equation}

We shall finally show that $\lambda_{i} = \lambda_j$ for all $i,j\in I$. Fix $i\neq j$ in $I$. Lemma 3.10 in \cite{FerPe18Adv} assures that one of the following statements holds:\begin{enumerate}[$(i)$]\item There exist minimal tripotents $v_2,v_4$ in $C$ and $\gamma\in \mathbb{T}$ such that $(e_i,v_2,\gamma e_j,v_4)$ is a quadrangle;
\item There exists a rank two tripotent $w\in C$ and $\gamma\in \mathbb{T}$ such that $(e_i, w, \gamma e_j)$ is a trangle.
\end{enumerate}

In case $(i),$ by considering $w = v_2 +v_4$, we get a trangle of the form $(e_i, w, \gamma e_j)$, so it suffices to consider the second case.  It is known that the element $\displaystyle  e = \frac{e_i+w+\gamma e_j}{2}$ is a minimal tripotent in $C$ (see, for example, \cite[Lemma 2.3 and Remark 2.6]{FerMarPe04} or more indirectly \cite[Proposition 5]{FriRu85}). It follows from the assumptions on $u$ that there exists $\lambda\in \mathbb{T}$ such that $\lambda e \leq u$, and hence $\{e,e,u \} = e$. By applying \eqref{eq first form of u}, the fact that the elements in the family $\{e_i: i \in I\}\cup \{P_0(v) (u)\}$ are mutually collinear, Peirce arithmetic and the properties of trangles we get $\{ e_i, w, e_i \}=\{ e_j, w, e_j\}=0,$ $\{ e_i, e_i,w \} = \frac12 w =\{ e_j, e_j,w \},$ $\{ e_i, w, e_j \}= \frac12 w,$ $\{w,w, e_j \} = e_j$, $\{w,w, e_i\} = e_i$, and
$$ \begin{aligned}
\lambda e &= \lambda \frac{e_i+w+\gamma e_j}{2} = \{e,e,u\} = \left\{ \frac{e_i+w+\gamma e_j}{2}, \frac{e_i+w+\gamma e_j}{2}, u\right\} \\
&= \left\{ \frac{e_i+w+\gamma e_j}{2}, \frac{e_i+w+\gamma e_j}{2}, \lambda_i e_i +\lambda_j e_j \right\}\\
&= \frac14 \left\{ e_i+w+\gamma e_j, e_i+w+\gamma e_j, \lambda_i e_i +\lambda_j e_j \right\}\\
& = \frac14 \left( \left\{ e_i, e_i, \lambda_i e_i +\lambda_j e_j \right\} + \left\{ e_i, w, \lambda_i e_i +\lambda_j e_j \right\} + \left\{ e_i, \gamma e_j, \lambda_i e_i +\lambda_j e_j \right\}\right. \\
&+\left. \left\{ w, e_i, \lambda_i e_i +\lambda_j e_j \right\} + \left\{ w, w, \lambda_i e_i +\lambda_j e_j \right\} + \left\{ w, \gamma e_j, \lambda_i e_i +\lambda_j e_j \right\}\right.\\
&\left.+ \left\{ \gamma e_j, e_i, \lambda_i e_i +\lambda_j e_j \right\} + \left\{ \gamma e_j, w, \lambda_i e_i +\lambda_j e_j \right\} + \left\{ \gamma e_j, \gamma e_j, \lambda_i e_i +\lambda_j e_j \right\} \right)\\
& = \frac14 \left( 2\lambda_i e_i + 2 \lambda_j  e_j +\lambda_i w + \frac{\gamma \lambda_i + \overline{\gamma} \lambda_j}{2} w \right) ,
\end{aligned}$$ which implies that $\lambda= \lambda_i = \lambda_j$ and $\gamma =1$. In such a case $u = \lambda v + P_0(v)(u)$ (with $\lambda =\lambda_i$ for $i$ arbitrary in $I$). This finishes the proof.
\end{proof}

Given a subset $\mathcal{A}$ in a JB$^*$-triple $E$, the symbol $\mathcal{A}_{E}^{\perp}$ will stand for the set of all elements in $E$ which are orthogonal to every element in $\mathcal{A}$.

\begin{lemma}\label{l scalar multiples of minimal and non-minimal tripotents} Let $\Phi : \mathcal{U}(C) \to \mathcal{U}(\tilde{C})$ be a bijective transformation which preserves the partial ordering in both directions and orthogonality between tripotents, where $C$ and $\tilde{C}$ are Cartan factors with rank bigger than or equal to $2$. Then if $u$ is a tripotent in $C$ and $\lambda\in \mathbb{T}$, the element $\Phi(\lambda u)$ lies in $\mathbb{T} \Phi(u)$. Actually, $\Phi (\mathbb{T} u ) = \mathbb{T} \Phi(u)$.
\end{lemma}

\begin{proof} We can clearly assume that $u$ is  non-zero (cf. Lemma \ref{c first consequences}$(a)$).\smallskip
	
Let us first assume that $u$ is a minimal tripotent. Since a minimal tripotent in a Cartan factor of rank $\geq 2$ is never complete, it follows from Lemma \ref{l Peirce 2 for fully non-complete tripotents}$(c)$ that $\Phi (\mathbb{T} u) \cup \{0\} = \Phi (C_2(u)\cap \mathcal{U} (C)) = \tilde{C}_2(\Phi(u))\cap \mathcal{U} (\tilde{C}) = \mathbb{T} \Phi(u)\cup \{0\}$ (see also Lemma \ref{c first consequences}$(b)$).\smallskip

Suppose now that $u$ is a non-zero tripotent in $\tilde{C}$. Fix $\lambda\in \mathbb{T}$. Let $\tilde{e}$ be any minimal tripotent in $\tilde{C}$ with $\tilde{e}\leq \Phi(u)$. Applying that $\Phi$ is surjective and Lemma \ref{c first consequences}$(b)$ we can deduce the existence of a minimal tripotent $e\in C$ such that $\Phi(e) = \tilde{e}$ with $e\leq u$. Clearly, $\lambda e\leq \lambda u$, and by the hypotheses on $\Phi$ and what we proved in the first paragraph, we have $$\mu \tilde{e}= \mu \Phi(e) = \Phi (\lambda e) \leq \Phi(\lambda u),$$ for some $\mu\in \mathbb{T}$. It follows from Lemma \ref{l when a tripotent is a complex multiple of another tripotent} that there exists $\gamma\in \mathbb{T}$ such that $$\Phi(\lambda u) = \gamma \Phi(u) + P_0(\Phi(u)) (\Phi (\lambda u)).$$
If $u$ is maximal or complete, $\Phi(u)$ satisfies the same property (cf. Lemma \ref{c first consequences}), and thus $\Phi(\lambda u) = \gamma \Phi(u)$. \smallskip

Suppose next that $u$ (equivalently, $\Phi(u)$) is non-complete. For each minimal tripotent $\tilde{v}\in \tilde{C}$ with $\Phi(v)= \tilde{v}\perp \Phi(u)$, the hypotheses on $\Phi$ and Proposition \ref{p order preservation in both directions gives orthogonality preservation} assure that $v\perp u \Leftrightarrow v\perp \lambda u\Leftrightarrow \tilde{v} =\Phi(v) \perp \Phi (\lambda u)$. We have therefore shown that $\{\Phi(u) \}_{_{\tilde{C}}}^{\perp}  = \{\Phi(\lambda u) \}_{_{\tilde{C}}}^{\perp}$ (just apply that every element in $\{\Phi(u) \}_{_{\tilde{C}}}^{\perp}$ or in $\{\Phi(\lambda u) \}_{_{\tilde{C}}}^{\perp}$ can be approximated in norm by finite linear combinations of mutually orthogonal minimal tripotents in the corresponding set), and hence $\Phi(\lambda u) = \gamma \Phi(u)$.
\end{proof}

We can now define a key notion for our purposes. Let $\Phi : \mathcal{U}(C) \to \mathcal{U}(\tilde{C})$ be a bijective transformation which preserves the partial ordering in both directions and orthogonality between tripotents, where $C$ and $\tilde{C}$ are Cartan factors with rank bigger than or equal to $2$. Suppose that $u$ is a non-zero tripotent in $C$. By Lemma \ref{l scalar multiples of minimal and non-minimal tripotents}, for each $\lambda\in \mathbb{T},$ there exists a unique $f_u(\lambda)\in \mathbb{T}$ such that $\Phi (\lambda u ) = f_u(\lambda) \Phi(u)$. Clearly, the mapping $f_u : \mathbb{T}\to \mathbb{T}$ is a bijection and $f_u (1) = 1$. We have therefore proved that \begin{equation}\left\{\begin{array}{l}
                                                                                              \label{eq first definition of fu} \hbox{for each non-zero tripotent $u\in C$ there exists a bijection } \\
                                                                                              \hbox{$f_u : \mathbb{T}\to \mathbb{T}$ satisfying } \Phi (\lambda u ) = f_u(\lambda) \Phi(u), \hbox{ for all } \lambda\in \mathbb{T}.
                                                                                            \end{array}\right.
\end{equation}

The next proposition is devoted to analyze the first properties of the maps $f_u$.

\begin{proposition}\label{p all maps fu coincide} Let $\Phi : \mathcal{U}(C) \to \mathcal{U}(\tilde{C})$ be a bijective transformation which preserves the partial ordering in both directions and orthogonality between tripotents, where $C$ and $\tilde{C}$ are Cartan factors with rank bigger than or equal to $2$.
Then the following statements hold:\begin{enumerate}[$(a)$]\item If $e_1$ and $e_2$ are two orthogonal non-zero tripotents in $C$, the maps $f_{e_1}$ and $f_{e_2}$ coincide;
\item If $e$ and $u$ are two non-zero tripotents in $C$ with $e\leq u$, the maps $f_{e}$ and $f_{u}$ coincide;
\item If $e$ is a non-zero and non-complete tripotent in $C$ and $\lambda\in \mathbb{T}$, the mappings $f_e$ and $f_{\lambda e}$ coincide;
\item If $u$ is a non-zero tripotent in $C$ and $\lambda\in \mathbb{T}$, the mappings $f_u$ and $f_{\lambda u}$ coincide;
\item If $u$ is a non-zero tripotent in $C$, the mapping $f_{u}: \mathbb{T}\to \mathbb{T}$ is multiplicative. In particular $f_u(-1) = -1,$ and $\Phi (-v) = -\Phi (v)$ for every tripotent $v\in C$. Furthermore, $f_u \left(\overline{\lambda}\right) = \overline{f_u(\lambda)}$, for all $\lambda \in \mathbb{T},$ $f_u (i) \in\{\pm i\}$ and $\Phi (i v) \in \{\pm i \Phi (v)\}$ for every tripotent $v\in C$.
\end{enumerate}
\end{proposition}

\begin{proof}$(a)$ and $(b)$ The tripotent $u = e_1 +e_2$ has rank--2. We consider the maps $f_{e_1}, f_{e_2}, f_{u}: \mathbb{T}\to \mathbb{T}$. By Lemma \ref{c first consequences}$(d)$ and the definition of these maps (cf. Lemma \ref{l scalar multiples of minimal and non-minimal tripotents} and \eqref{eq first definition of fu}) we have $$\begin{aligned}f_u(\lambda) (\Phi(e_1) +\Phi(e_2)) & =f_u(\lambda ) \Phi(e_1 +e_2) = f_u(\lambda ) \Phi(u) \\
&= \Phi (\lambda u) = \Phi(\lambda e_1) + \Phi(\lambda e_2) = f_{e_1} (\lambda) \Phi( e_1) + f_{e_2} (\lambda) \Phi( e_2),
\end{aligned}$$ with $\Phi(e_1) \perp \Phi(e_2)$, yielding that $f_{e_1} (\lambda)=  f_{e_2} (\lambda) = f_{u} (\lambda)$ for all $\lambda\in \mathbb{T}$.\smallskip

$(c)$ Suppose $e$ is a non-zero and non-complete tripotent in $C$ and $\lambda\in \mathbb{T}$. Since $C$ has rank $\geq 2$, we can find a non-zero tripotent $u$ such that $e\perp u$. Since $\lambda e \perp u$, it follows from $(a)$ that $f_{e} = f_{u} = f_{\lambda e}$.\smallskip

$(d)$ If $u$ is non-complete the desired conclusion follows from $(c)$. We assume that $u$ is complete. In this case there exists a non-zero tripotent $e\in C$ with $e\leq u$.  By applying $(b)$ and $(c)$ we get $f_{u} = f_{e}= f_{\lambda e} = f_{\lambda u}$.\smallskip

$(e)$ Pick a non-zero tripotent $u$, $\lambda,\mu \in \mathbb{T}$. We deduce from $(d)$ and \eqref{eq first definition of fu} that $$f_{u} (\lambda \mu) \Phi (u) = \Phi (\lambda \mu u) = f_{\mu u} (\lambda) \Phi(\mu u) = f_{u} (\lambda) f_u (\mu) \Phi( u),$$ therefore $f_{u} (\lambda \mu) = f_{u} (\lambda) f_u (\mu)$ for all $\lambda, \mu$ as above. Since $1= f_{u} (1)= f_{u} (-1) f_u (-1)$ and $f_u : \mathbb{T}\to \mathbb{T}$ is a bijection with $f_u(1) = 1$, we get $f_u (-1) =-1$ as desired.\smallskip

Moreover, the equality $1= f_u(1) = f_u(\lambda \overline{\lambda}) = f_u(\lambda) f_u(\overline{\lambda})$ gives $f_u \left(\overline{\lambda}\right) = \overline{f_u(\lambda)}$. Finally $-1 = f_u(-1) = f_u (i i ) =  f_u (i)  f_u (i ),$ which implies that $f_u(i) = \pm i$.
\end{proof}

Let $M$ be a complex spin factor as defined in the introduction. Let $\langle \cdot, \cdot \rangle$ and $x\mapsto \overline{x}$ denote the inner product and the conjugation on $M$ for which the triple product is given by $$\{a,b,c\} = \langle a, b\rangle c + \langle c,b\rangle a-\langle a,\bar{ c}\rangle \bar{b}. $$  The real subspace $M_{_\mathbb{R}}=\{a\in M: \, a=\bar{a}\},$ called the real part of $M$, is a real Hilbert space with respect to the inner product $( a, b) = \Re\hbox{e}\langle a, b\rangle$ ($a,b\in M_{_\mathbb{R}}$), and $M = M_{_\mathbb{R}}\oplus i M_{_\mathbb{R}}$. We shall denote by $S_{_\mathbb{R}}$ the unit sphere of $M_{_\mathbb{R}}$. It is known (see \cite[Theorem in page 196]{HervIs92} or \cite[Section 3.1.3]{Fri2005}) that for any triple automorphism $T$ of $M$ there is a complex number $\lambda\in \mathbb{T}$ and a unitary operator $U\in B(M_{_\mathbb{R}})$ such that $T(a +ib) =\lambda (U(a) + i U(b))$ for all $a+i b\in M$.  Conversely, for any $\lambda\in \mathbb{T}$ and each unitary $U\in B(M_{_\mathbb{R}})$ the map $T(a +ib) =\lambda (U(a) + i U(b))$ ($a+i b\in M$) is a triple automorphism on $M$.\smallskip

The one-dimensional spin factor coincides with $\mathbb{C}$. The term ``factor'' is not appropriate in the case in which dim$(M)=2$, because in this case $M$ is precisely $\mathbb{C}\oplus^{\infty} \mathbb{C},$ which is not a factor. For this reason, spin factors are always assumed to have dimension at least three. Particular examples of spin factors, worth to be recalled here, include the 3-dimensional spin factor $S_2(\mathbb{C})$ of all symmetric 2 by 2 complex matrices, and the von Neumann factor $M_2(\mathbb{C})$ of all 2 by 2 complex matrices which is a 4-dimensional spin factor (cf.  \cite[\S 7]{AlfShulStor78}, \cite[\S 6]{HOS}, \cite[page 82]{FriRu85}, \cite[\S 6]{HamKalPe20} or \cite[\S 3]{KalPe2019}).\smallskip

We can improve now the conclusion in the previous Proposition \ref{p all maps fu coincide} by using the basic structure of spin factors.

\begin{proposition}\label{p all maps fu coincide new} Let $\Phi : \mathcal{U}(C) \to \mathcal{U}(\tilde{C})$ be a bijective transformation which preserves the partial ordering in both directions and orthogonality between tripotents, where $C$ and $\tilde{C}$ are Cartan factors with rank bigger than or equal to $2$. Then the maps $f_{e}$ and $f_{u}$, defined in \eqref{eq first definition of fu}, coincide whenever $e$ and $u$ are two non-zero tripotents in $C$. Therefore
\begin{equation}\label{eq the mapping f is uniform for all nonzero tripotents}\left\{\begin{array}{l}
                                                                                \hbox{there exists a bijective group homomorphism $f:\mathbb{T}\to \mathbb{T}$} \\
                                                                                \hbox{satisfying $f(-1) = -1,$ $f \left(\overline{\lambda}\right) = \overline{f(\lambda)}$, for all $\lambda \in \mathbb{T},$}\\
                                                                                \hbox{$f (i) \in\{\pm i\}$, and $\Phi (\lambda u ) = f(\lambda) \Phi (u)$ for all $u\in \mathcal{U} (C)$ and $\lambda\in \mathbb{T}$.}
                                                                              \end{array}\right.
\end{equation}
\end{proposition}

\begin{proof} Let us first assume that $e$ and $u$ are minimal tripotents. In this case, applying Lemma 3.10 in \cite{FerPe18Adv} we conclude that one of the following statements holds:\begin{enumerate}[$(i)$]\item There exist minimal tripotents $e_2,e_3,e_4$ in $C$ such that $(e,e_2,e_3,e_4)$ is a quadrangle and $u$ is a linear combination of $e,e_2,e_3,$ and $e_4$;
\item There exist a minimal tripotent $\tilde e\in C$ and a rank two tripotent $w\in C$ such that $(e, w,\tilde e)$ is a trangle and $u$ is a linear combination of $e, w,$ and $\tilde e$.
\end{enumerate}

We shall deal with both cases in parallel. The tripotent $e+e_3$ (respectively, $e+\tilde{e}$) has rank--2. Lemma 3.8 in \cite{KalPe2019} affirms that the JB$^*$-subtriple $M= C_2(e+e_3)$ (respectively, $C_2(e+\tilde{e})$) is a spin factor. Clearly, $M= C_2(e+e_3)$ contains $e$ and $u$.  We therefore focus on the spin factor $M$ (which clearly has rank--2) with the structure recalled at the beginning of this section.\smallskip

The element $e+e_3$ (respectively, $e+\tilde{e}$) has rank--2 in $M$, and thus it is a complete and unitary tripotent in $M$. We can assume that $e+e_3 = \lambda x$ (respectively, $e+\tilde{e}= \lambda x$) for some $\lambda\in \mathbb{T}$ and $x\in S_{_\mathbb{R}}$. Any other complete tripotent in $M$ is of the form $\mu y$ for some $\mu\in \mathbb{T}$ and $y\in S_{_\mathbb{R}}$. Let us find $z\in S_{_\mathbb{R}}$ such that $\langle x, z\rangle =0$ and $y = t x + s z$ for unique real numbers $s,t$ with $s^2 + t^2 =1$.\smallskip

We shall first show that \begin{equation}\label{eq fz and fx coincide}\hbox{ the mappings } f_x, f_z, f_y, f_{\lambda x}, f_{\mu y} \hbox{ and } f_e \hbox{ coincide.}
\end{equation}

The minimal tripotents $v = \frac{x+ i z}{2}$ and $\pm\overline{v} = \pm \frac{x- i z}{2}$ are orthogonal in $M$ with $v+\overline{v} = x$ and $v-\overline{v} = i z$. Since $v\leq x$ and $v\leq i z$, it follows from Proposition \ref{p all maps fu coincide}$(b)$ that $f_{v} = f_x$ and $f_{v} = f_{i z}$. Since Proposition \ref{p all maps fu coincide}$(c)$ assures that $f_{i z} = f_{z}$, we conclude that $f_x = f_z$.\smallskip

Take the element $\gamma = t - i s\in \mathbb{T}$. The minimal tripotents $\gamma v = \gamma \frac{x+ i z}{2}$ and $\pm \overline{\gamma}\ \overline{v} = \pm \overline{\gamma} \frac{x- i z}{2}$ are orthogonal in $M$ with $\gamma v+ \overline{\gamma}\ \overline{v} = t x + s z =y,$ and thus $\gamma v\leq y$. Therefore Proposition \ref{p all maps fu coincide}($(b)$ and $(c)$) proves that $f_y = f_{\gamma v} = f_{v} =f_x$.\smallskip

A new application of Proposition \ref{p all maps fu coincide}($(c)$ and $(b)$) implies that $f_{x} = f_{\lambda x} = f_{e+e_3} = f_{e}$ (respectively, $f_{x} = f_{\lambda x} = f_{e+\tilde{e}} = f_e$) because $e\leq e+e_3$ (respectively, $e\leq e+ \tilde{e}$). The just quoted result also implies that $f_{y} = f_{\mu y}$.\smallskip

Since $u$ is a minimal tripotent in the spin factor $M$, we can always find a unitary (i.e., rank--2 tripotent), say $\mu y$ (with $\mu\in \mathbb{T}$ and $y\in S_{_\mathbb{R}}$) such that $u\le \mu y$, and hence $f_{u}=f_{\mu y}$ (cf. Proposition \ref{p all maps fu coincide}$(b)$). An application of our conclusion in \eqref{eq fz and fx coincide} implies that $f_e= f_{u}$.\smallskip

We have therefore shown that $f_e= f_{u}$ whenever $u$ and $e$ are minimal tripotents in $C$.\smallskip

Finally if $u$ and $e$ are two non-zero tripotents in $C$, we can find two minimal tripotents $v_1,v_2\in C$ with $v_1\leq e$ and $v_2\leq u$. By combining Proposition \ref{p all maps fu coincide}$(b)$ with our conclusion for minimal tripotents we conclude that $f_e = f_{v_1} = f_{v_2} = f_u$.
\end{proof}

\begin{remark}\label{r existence of the mapping f for atomic without rank-one Cartan factors} If we combine Lemma \ref{l Peirce 2 for fully non-complete tripotents} with the previous Proposition \ref{p all maps fu coincide new} it follows that the conclusion of the latter result also holds when $C$ and $\tilde{C}$ are replaced with two atomic JBW$^*$-triples not containing Cartan factors of rank--one.
\end{remark}

\begin{remark}\label{r on continuity assumptions} Let $\Phi : \mathcal{U}(C) \to \mathcal{U}(\tilde{C})$ be a bijective transformation which preserves the partial ordering in both directions and orthogonality between tripotents, where $C$ and $\tilde{C}$ are Cartan factors with rank bigger than or equal to $2$. Let $f:\mathbb{T} \to \mathbb{T}$ be the bijection given by Proposition \ref{p all maps fu coincide new}. Let us consider the following statements:
\begin{enumerate}[$(a)$]\item $\Phi$ is continuous at every non-zero tripotent;
\item $\Phi$ is continuous at a non-zero tripotent;
\item For each non-zero tripotent $u\in C$ the mapping $\Phi|_{\mathbb{T} u}: \mathbb{T} u\to \mathcal{U}(\tilde{C})$ is continuous at $u$;
\item There exists a non-zero tripotent $u\in C$ such that the mapping $\Phi|_{\mathbb{T} u}: \mathbb{T} u\to \mathcal{U}(\tilde{C})$ is continuous at $u$;
\item The mapping $f$ is continuous at $1$;
\item The mapping $f$ is continuous.
\end{enumerate} Then the implications $(a)\Rightarrow (b) \Rightarrow (c)\Leftrightarrow (d)\Leftrightarrow (e)\Leftrightarrow (f)$ hold. Furthermore, if any of the previous statements holds then the mapping $f$ is the identity or the conjugation on $\mathbb{T}$. \smallskip

The implications $(a)\Rightarrow (b) \Rightarrow (d)\Leftarrow (c)$ and $(f)\Rightarrow (e)$ are clear. For each non-zero tripotent $u$ in $C$, the identity $\Phi (\lambda u) = f(\lambda) \Phi(u)$ holds for all $\lambda\in \mathbb{T}$ and each non-zero tripotent $u\in C$ {\rm(}cf. \eqref{eq the mapping f is uniform for all nonzero tripotents}{\rm)}, and thus the implications $(d)\Rightarrow (e) \Rightarrow (c)$ hold. Finally if $f:\mathbb{T}\to \mathbb{T}$ is continuous at $1$, it must be continuous at every point because it is a group homomorphism, hence $(e)\Rightarrow (f)$.\smallskip

It is well known that every continuous group homomorphism on $\mathbb{T}$ must be of form $f(\lambda) = \lambda^n$ for some $n\in \mathbb{Z}$ {\rm(}cf. \cite[\S 4.2.1]{DymMckean72}{\rm)}, but being $f$ a bijection the possibilities reduce to the identity or the conjugation on $\mathbb{T}$.
\end{remark}

We state next an identity principle which will be applied in different subsequent arguments.

\begin{proposition}\label{p identity principle} Let $\Phi : \mathcal{U}(M) \to \mathcal{U}({N})$ be a bijective transformation which preserves the partial ordering in both directions and orthogonality between tripotents, where $M$ and $N$ are atomic JBW$^*$-triples not containing Cartan factors of rank--one. Suppose $T:M\to N$ is a weak$^*$ continuous complex-linear operator such that $T(e) = \Phi(e)$ for all minimal tripotent $e\in M$.  Then $\Phi (\lambda u ) = \lambda \Phi(u)$  and $T(u) = \Phi(u),$ for all $\lambda\in \mathbb{T}$, $u\in \mathcal{U}(M)$. Consequently, $T$ is an isometric triple isomorphism.
\end{proposition}

\begin{proof} Let $f: \mathbb{T}\to \mathbb{T}$ be the bijection given by Proposition \ref{p all maps fu coincide new} and Remark \ref{r existence of the mapping f for atomic without rank-one Cartan factors}. Fix a minimal tripotent $e\in M$ and $\lambda\in \mathbb{T}$. Applying the linearity of $T,$ the hypotheses, and the fact that $e$ and $\lambda e$ are minimal tripotents we get $$f(\lambda) \Phi(e) = \Phi (\lambda e) = T(\lambda e) = \lambda T(e) = \lambda T(e)= \lambda \Phi(e),$$ yielding that $f$ is the identity on $\mathbb{T},$ and hence $\Phi (\lambda u ) = \lambda \Phi(u)$ for all $\lambda\in \mathbb{T}$, $u\in \mathcal{U}(M)$ by the just quoted proposition.\smallskip

In an atomic JBW$^*$-triple $M$, every non-zero tripotent $u$ can be written as the supremum of a family $\{e_i : i \in I\}$ of mutually orthogonal minimal tripotents in $M$, which is precisely the limit of the series $\displaystyle \sum_{i\in I} e_i$ in the weak$^*$ topology. By Lemma \ref{l infima and suprema sums of orthogonal families}$(c)$ and the hypotheses we have $$\Phi (u) = \Phi\left( \displaystyle \sum_{i\in I} e_i \right) =  \sum_{i\in I} \Phi(e_i)  = \sum_{i\in I} T(e_i)  = T\left( \sum_{i\in I} e_i \right)  = T(u).$$

It is part of the theory of JB$^*$-triples that the set of all tripotents in a JBW$^*$-triple $M$ is norm-total, that is, every element in $M$ can be approximated in norm by a finite linear combination of mutually orthogonal tripotents in $M$ (cf. \cite[Lemma 3.11]{Horn87}), and hence our mapping $T$ must preserve the cube of every element in $M$, that is, $T\{x,x,x\} = \{T(x), T(x), T(x)\}$ for all $x\in M$. Therefore, a standard polarization identity implies that $T$ is a triple homomorphism.\smallskip

A triple version of the celebrated Kaplansky--Cleveland theorem, established in \cite{FerGarPe2012}, asserts that a non-necessarily continuous triple homomorphism from a JB$^*$-triple to a normed Jordan triple has closed range whenever it is continuous \cite[Corollary 18]{FerGarPe2012}. As a consequence of this result, the triple homomorphism has closed range. Furthermore, since $T$ coincide with $\Phi$ at every tripotent of $M$ and $\Phi$ is bijective, it follows from the norm-totality of the set of tripotents in $M$ that $T$ is surjective.\smallskip

Finally, since ker$(T)$ is a weak$^*$ closed triple ideal of $M$, if it is non-zero we can find a non-trivial tripotent $u\in \hbox{ker}(T)\cap \mathcal{U} (M)$, what would imply that $0= T(u) = \Phi(u)$ contradicting the injectivity of $\Phi$.
\end{proof}

\section{Order preserving bijections on complex spin factors}\label{sec: order preservers between spin factors}

We continue with our study on bijections between the sets of tripotents of two spin factors preserving the partial ordering in both directions and orthogonality between tripotents.

\begin{lemma}\label{min_trip spin} Let $M$ and $\tilde{M}$ be complex spin factors with dimension at least three.  Let $\Phi : \mathcal{U}(M) \to \mathcal{U}(\tilde{M})$ be a bijective transformation which preserves the partial ordering in both directions and orthogonality between tripotents. Suppose $a,b\in S_{_\mathbb{R}}$ with  $\langle a, b\rangle =0$ satisfy $\Phi (a) \in S_{_\mathbb{R}}$. Let  $v= \frac{1}{2} (a+i b)$ be a minimal tripotent in $M$. Then $\Phi (i b) \in i S_{_\mathbb{R}}$, $ \langle \Phi({a}), \Phi (i b)\rangle =0$  and
\begin{equation}\label{phi_v}
  \Phi(v)= \frac{1}{2} (\Phi({a})+\Phi({ib})),\;\; \Phi(\bar{v})= \frac{1}{2} (\Phi({a})-\Phi({ib})).
\end{equation}
\end{lemma}

\begin{proof} Since  $v$ is a minimal tripotent and $v\leq a$ in $M$, by the hypothesis on $\Phi$, $\Phi (v)$ is a minimal tripotent in $\tilde{M}$ with $\Phi (v) \leq \Phi (a) \in \widetilde{S}_{_\mathbb{R}}$.  By Lemma \ref{l order spinGood} there exists $\tilde{b}\in \widetilde{S}_{_\mathbb{R}}$ with $ \langle \Phi({a}), \tilde{b}\rangle =0$ such that $\Phi(v)= \frac{1}{2} (\Phi({a})+i \tilde{b}),$ $\tilde{v}= \frac{1}{2} (\Phi({a})-i \tilde{b})$ is another minimal tripotent orthogonal to $\Phi(v)$ and $ \Phi(a) = \Phi(v) + \tilde{v}.$\smallskip

Now, Lemma \ref{c first consequences}$(d)$ affirms that $$\Phi(v) + \Phi (\overline{v}) = \Phi (a) = \Phi(v) + \tilde{v} ,$$ yielding that $\Phi (\overline{v}) = \tilde{v} = \frac{1}{2} (\Phi({a})-i \tilde{b})$. By combining Proposition \ref{p all maps fu coincide}$(e)$ and the just quoted corollary we get $$\begin{aligned}\Phi(i b)&= \Phi (v - \overline{v})= \Phi(v) - \Phi (\overline{v}) =  i \tilde{b} \in i S_{_\mathbb{R}},
\end{aligned}$$ and clearly $ \langle \Phi({a}), \Phi (i b)\rangle =0$,  as claimed. By substituting the latter identity for $\Phi (i b)$ in the formulae for  $\Phi(v)$ and $\Phi(\bar{v})$ we get (\ref{phi_v}).
\end{proof}

Similar techniques to those employed in the proof of Proposition \ref{p all maps fu coincide new} can be applied to get the following improved version of Lemma \ref{l Peirce 2 for fully non-complete tripotents}.

\begin{lemma}\label{l Peirce 2 for fully non-complete tripotents improved} Let $\Phi : \mathcal{U}(C) \to \mathcal{U}(\tilde{C})$ be a bijective transformation which preserves the partial ordering in both directions and orthogonality between tripotents, where $C$ and $\tilde{C}$ are Cartan factors with rank bigger than or equal to $2$. Let $e$ be a rank--2 tripotent in $C$.
Then $\Phi$ maps the set $C_2(e) \cap \mathcal{U} (C)$ onto the set $\tilde{C}_2(\Phi(e)) \cap \mathcal{U} (\tilde{C})$.
\end{lemma}

\begin{proof} We observe that if $C$ has rank--3 the conclusion is a consequence of Lemma \ref{l Peirce 2 for fully non-complete tripotents}$(c)$.\smallskip

In any case, the Peirce-2 subspace $M=C_2(e)$ is a spin factor (cf. \cite[Lemma 3.8]{KalPe2019}), and we keep the notation employed in this section. Let us fix a complete (equivalently, rank--2 or unitary) tripotent $w$ in $M$. We can assume that $e = \lambda x$ and $ w = \mu y$ for some $\lambda, \mu \in \mathbb{T}$, $x,y\in S_{_\mathbb{R}}$. The minimal tripotents $v = \frac{x+iy}{2}$ and $\overline{v} = \frac{x-iy}{2}$ are orthogonal with $v+\overline{v} = x$ and $v-\overline{v} = iy$. By Lemma \ref{c first consequences}$(d)$ and Proposition \ref{p all maps fu coincide}$(e)$ we get $$\Phi (x) = \Phi (v) +\Phi (\overline{v}), \hbox{ and } \Phi (i y ) = \Phi (v) -\Phi (\overline{v}).$$ The first identity guarantees that $\Phi (v)$ and $\Phi (\overline{v})$ lie in $\tilde{C}_2(\Phi(e))$, and thus the second one proves that $\Phi (i y )$ lies in $\tilde{C}_2(\Phi(e))$. The previous Proposition \ref{p all maps fu coincide new} affirms that $\Phi (i y )= f(i) \Phi (y )$, and consequently $\Phi(w) = \Phi (i y )\in \tilde{C}_2(\Phi(e))$. \smallskip

We have proved that for each complete or unitary tripotent $w\in M = C_2(e)$ we have $\Phi(w) \in \tilde{C}_2(\Phi(e))$. The set $\mathcal{U} (M)$ reduces to rank--2 or complete tripotents, minimal tripotents and zero. Given a minimal tripotent $v\in M$, the tripotents $v \pm \overline{v}$ have rank--2, and hence, by the previous conclusion, $\Phi (v \pm \overline{v}) \in \tilde{C}_2(\Phi(e))$. A new combination of Lemma \ref{c first consequences}$(d)$ and Proposition \ref{p all maps fu coincide}$(e)$ proves that $\Phi (v), \Phi(\overline{v}) \in \tilde{C}_2(\Phi(e))$, which concludes the proof.
\end{proof}

The key result of this section is the following theorem.

\begin{theorem}\label{t spin order 2Good} Let $M$ and $\tilde{M}$ be complex spin factors of dimension $n\geq 3$. Let $\Phi : \mathcal{U}(M) \to \mathcal{U}(\tilde{M})$ be a bijective transformation which preserves the partial ordering in both directions and orthogonality between tripotents. Suppose there exists a non-zero tripotent $u\in M$ such that $\Phi|_{\mathbb{T} u}$ is continuous at $u$. Then there exists a complex-linear or conjugate-linear triple isomorphism $T: M\to \tilde{M}$ such that $\Phi (e) = T(e)$ for all $e\in \mathcal{U} (M)$. More concretely, there exist $\lambda_0\in \mathbb{T}$ and a surjective real linear isometry $U\in B(M_{_\mathbb{R}},\tilde{M}_{_\mathbb{R}})$ such that $\Phi =\lambda_0 \tilde{U}$ or $\bar{\Phi} =\lambda_0 \tilde{U},$ where $\tilde{U}$ is the {\rm(}isometric{\rm)} triple isomorphism from $M$ onto $\tilde{M}$ defined by $\tilde{U} (a+i b) = U(a) + i U(b)$ for all $a+i b \in M$.
\end{theorem}

\begin{proof}Let us pick $x_1\in S_{_\mathbb{R}}$. Since $x_1$ is a maximal tripotent, also $\Phi (x_1)$ must be a maximal tripotent (cf. Lemma \ref{c first consequences}$(b)$), and by \eqref{maxTripGood} there is $\lambda_0\in \mathbb{T}$ and $\tilde{e}_1\in \tilde{S}_{_\mathbb{R}}$ such that $\Phi (x_1)= \lambda_0 \tilde{e}_1$. The mapping ${\Phi}_{1} =\overline{\lambda_0} \Phi : \mathcal{U}(M) \to \mathcal{U}(\tilde{M})$ is a bijection which preserves the partial ordering and orthogonality between tripotents in both directions (cf. Proposition \ref{p order preservation in both directions gives orthogonality preservation}) and ${\Phi}_1 (x_1)= \tilde{e}_1\in S_{_\mathbb{R}}$. To prove the desired conclusion it suffices to show that ${\Phi}_1|_{S_{_\mathbb{R}}},$ or its conjugate, admits an extension to a linear surjective isometry in $B(M_{_\mathbb{R}},\tilde{M}_{_\mathbb{R}})$.\smallskip

By combining  Lemma \ref{min_trip spin} (i.e., $\Phi_1 (i b) \in i \tilde{S}_{_\mathbb{R}}$ for all $b\in {S}_{_\mathbb{R}}$) with the conclusion in Proposition \ref{p all maps fu coincide}$(e)$ (i.e., $\Phi_1 ( b) = \Phi_1 (i (-i) b) \in \{\pm i \Phi_1 (i b)\}$ for all $b\in {S}_{_\mathbb{R}}$) we obtain:
\begin{equation}\label{eq first property thm spin b}\hbox{Suppose $b\in S_{_\mathbb{R}}$ satisfies $\langle x_1, b\rangle =0$. Then $\Phi_1 (b) \in S_{_\mathbb{R}}$.}
\end{equation}

We claim next that the following property also holds:
\begin{equation}\label{eq first property thm spin c} \Phi_1 (S_{_\mathbb{R}}) = \tilde{S}_{_\mathbb{R}}.
\end{equation} In order to prove the claim we pick an arbitrary $c\in S_{_\mathbb{R}}$. Since $M_{_\mathbb{R}}$ (equivalently $M$) has dimension $\geq 3$, there exists $b\in S_{_\mathbb{R}}$ with $\langle a,b\rangle = \langle c,b\rangle =0$. We deduce from \eqref{eq first property thm spin b} that $\Phi_1 (b)\in \tilde{S}_{_\mathbb{R}}$, and a new application of \eqref{eq first property thm spin b} to the pair $b,c$ gives $\Phi_1 (c)\in \tilde{S}_{_\mathbb{R}}$, which concludes the proof of the claim--the equality follows from the same argument applied to $\Phi_1^{-1}$.\smallskip

We also have $\Phi_1(0) =0$ (cf. Lemma \ref{c first consequences}$(a)$). We shall next show that
\begin{equation}\label{eq first property thm spin d} \langle a, b\rangle =0 \hbox{ in } S_{_\mathbb{R}} \Longleftrightarrow \langle \Phi_1 (a), \Phi_1 (b) \rangle =0 \hbox{ in } \tilde{S}_{_\mathbb{R}}.
\end{equation} ($\Rightarrow$) As in the first paragraph of this proof, the elements $a$ and $v = \frac{a + i b}{2}$ are tripotents with $a$ complete, $v$ minimal, and $v\leq a$. Lemma \ref{min_trip spin} implies that $\Phi_1 \left( \frac{a \pm i b}{2} \right) = \frac{\Phi_1(a) \pm \Phi_1(i b)}{2}$, with $\langle \Phi_1(a), \Phi_1(ib)\rangle =0$. An application of Proposition \ref{p all maps fu coincide}$(e)$ implies that $$\Phi_1(b) \in \{\pm i \Phi(ib)\} = \{\mp \tilde{b}\},$$ and hence $\langle \Phi_1 (a), \Phi_1 (b) \rangle =0,$ as we wanted.\smallskip

($\Leftarrow$) This implication can be obtained by simply applying the previous argument to $\Phi_1^{-1}$.\smallskip

Only two steps separate us from the final conclusion. The next one is a straight consequence of Proposition \ref{p all maps fu coincide new}. We recall that, by the just quoted result, there exists a bijective mapping $f: \mathbb{T}\to \mathbb{T}$ which is a group homomorphism satisfying the properties stated in \eqref{eq the mapping f is uniform for all nonzero tripotents}. Therefore the next statements hold:
\begin{enumerate}[$(1)$]\item If $\Phi_1 (i a) = i \Phi_1 (a)$ for some $a\in S_{_\mathbb{R}}$ (this proves that $f(\lambda) =\lambda$ for all $\lambda$), it follows that $\Phi_1 (i b)= f(i) \Phi_1 (b) = i \Phi_1 (b)$ for all $b\in S_{_\mathbb{R}}$;
\item If $\Phi_1 (i a) = - i \Phi_1 (a)$ for some $a\in S_{_\mathbb{R}}$  (this proves that $f(\lambda) =\overline{\lambda}$ for all $\lambda$), it follows that $\Phi_1 (i b)=  f(i) \Phi_1 (b) =- i \Phi_1 (b)$ for all $b\in S_{_\mathbb{R}}$.
\end{enumerate}

Assume that we can find $a\in S_{_{\mathbb{R}}}$ with $\Phi_1(i a) = i \Phi_1(a)$, and hence $\Phi_1(i b) = i \Phi_1(b)$ for all $b\in S_{_{\mathbb{R}}}$. Let $\{e_i : i\in I\}$ be an orthonormal basis of $M_{_{\mathbb{R}}}$. We deduce from \eqref{eq first property thm spin d} and \eqref{eq first property thm spin c} that $\{\Phi_1(e_i) : i\in I\}$ is an orthonormal basis of $\tilde{M}_{_{\mathbb{R}}}$. The linear mapping $U: M_{_{\mathbb{R}}}\to \tilde{M}_{_{\mathbb{R}}},$ defined by $\displaystyle U\left(\sum_{i\in I} \alpha_i e_i \right) = \sum_{i\in I} \alpha_i \Phi_1\left(e_i \right),$ is a surjective isometry, and hence a surjective isometry in $B(M_{_{\mathbb{R}}},\tilde{M}_{_{\mathbb{R}}})$. Set $\tilde{U}: M \to M$, $\tilde{U} (a +i b) = U(a) + i U(b)$. Clearly $\tilde{U}$ is a linear bijection from $M$ onto $\tilde{M}$.\smallskip

We shall next prove that \begin{equation}\label{eq U and Phi1 coincide on SR}\hbox{$U(x) = \Phi_1(x)$ and $U(\frac{x+i y}{2}) = \Phi_1\left(\frac{x+i y}{2}\right)$ for all $x,y\in S_{_{\mathbb{R}}}$ with $\langle x, y\rangle =0$.}
\end{equation}

In order to get the first statement we shall first establish the following property:
\begin{equation}\label{eq linearity on 2-components} \begin{aligned} & \hbox{Given $a,b\in S_{_\mathbb{R}}$ with $\langle a, b\rangle =0$ and $s,t\in \mathbb{R}$ with $s^2 +t^2 =1$}\\
& \hbox{we have $\Phi_1 ( s a + t b ) = s \Phi_1 (a) + t \Phi_1 (b)$.}
\end{aligned}
\end{equation}

By hypothesis there exists a non-zero tripotent $u\in C$ such that $\Phi$ is continuous at $u$ (or simply that $\Phi|_{\mathbb{T} u}$ is continuous at $u$). The mapping $\Phi_1 = \overline{\lambda_0} \Phi$ enjoys the same continuity properties. As we have seen in Remark \ref{r on continuity assumptions}, in this case the mapping $f$ associated with $\Phi_1$ must be the identity or the conjugation, and according to what we have assumed $f(\lambda) = \lambda,$ for all $\lambda\in \mathbb{T}$.\smallskip

Let $v$ denote the minimal tripotent in $M$ given by $v= \frac{ a + i b}{2}$ and set $\gamma = s - i t\in \mathbb{T}$. By applying Lemma \ref{min_trip spin}\eqref{phi_v} we arrive at $$\Phi_1(v) = \frac{ \Phi_1(a) + \Phi_1(i b)}{2} =\frac{ \Phi_1(a) + f(i) \Phi_1( b)}{2} =\frac{ \Phi_1(a) + i \Phi_1( b)}{2}$$ and $$\Phi_1(\overline{v}) = \frac{ \Phi_1(a) - \Phi_1(i b)}{2} =\frac{ \Phi_1(a) - i \Phi_1( b)}{2}.$$ Since $\gamma v$ and $\overline{\gamma}\ \overline{v}$ are orthogonal minimal tripotents with $\gamma v + \overline{\gamma}\ \overline{v} = (s a +t b)$ and, as in many cases before, by Lemma \ref{c first consequences}$(d)$ and Proposition \ref{p all maps fu coincide new} we have $$\begin{aligned}
\Phi_1 (s a +t b) &= \Phi_1 (\gamma v + \overline{\gamma}\  \overline{v}) =   \Phi_1 (\gamma v) + \Phi_1 (\overline{\gamma}\ \overline{v}) =  f(\gamma) \Phi_1 ( v) +  f(\overline{\gamma}) \Phi_1 ( \overline{v})\\
& = \gamma \frac{ \Phi_1(a) + i \Phi_1( b)}{2} + \overline{\gamma} \frac{ \Phi_1(a) + i \Phi_1( b)}{2} = s \Phi_1 (a )+ t \Phi_1 (b).
\end{aligned}$$ This finishes the proof of \eqref{eq linearity on 2-components}.\smallskip

Back to \eqref{eq U and Phi1 coincide on SR}, let $\displaystyle x = \sum_{i\in I} t_i e_i$ be an arbitrary element in $S_{_\mathbb{R}}$. Fix $i_0\in I.$ The elements $a=e_{i_0}$ and $\displaystyle b =  \sum_{i\in I\backslash\{i_0\}} {t_i} \displaystyle\left(\sum_{j\neq i_0} t_j^2\right)^{-\frac12} e_i$ belong to $S_{_\mathbb{R}}$ with $\langle a , b\rangle =0$ and $x = t_{i_{0}} a + \displaystyle\left(\sum_{j\neq i_0} t_j^2\right)^{\frac12} b$. The conclusion in \eqref{eq linearity on 2-components} shows that $$\begin{aligned} \Phi_1(x) &= \Phi_1\left(  t_{i_{0}} a + \displaystyle\left(\sum_{j\neq i_0} t_j^2\right)^{\frac12} b \right) =  t_{i_{0}} \Phi_1(a) + \displaystyle\left(\sum_{j\neq i_0} t_j^2\right)^{\frac12} \Phi_1(b)\\
&=  t_{i_{0}} \Phi_1(e_{i_0}) + \displaystyle\left(\sum_{j\neq i_0} t_j^2\right)^{\frac12} \Phi_1(b).
\end{aligned}$$ Since $\langle \Phi_1(e_{i_0}), \Phi_1 (b)\rangle =0$ (cf. \eqref{eq first property thm spin d}), it follows that $\langle \Phi_1(x), \Phi_1 (e_{i_0})\rangle = t_{i_0}$, and the arbitrariness of $i_0$ implies that \begin{equation}\label{eq Hilbert sums in the sphere spin} \Phi_1(x) = \sum_{i\in I} t_i \Phi_1(e_i) = \sum_{i\in I} t_i U(e_i)  = U(x),
 \end{equation} yielding that the first part of \eqref{eq U and Phi1 coincide on SR} is true. For the second statement in \eqref{eq U and Phi1 coincide on SR}, fix $x,y\in S_{_{\mathbb{R}}}$ with $\langle x, y\rangle =0$ and apply, once again, Lemma \ref{min_trip spin}\eqref{phi_v} and the previous conclusion to get $$\Phi_1\left(\frac{x+i y}{2}\right) = \frac{\Phi_1(x)+i \Phi_1(y)}{2}= \frac{U(x)+i U(y)}{2} = U\left(\frac{x+i y}{2}\right).$$

The remaining tripotents in $M$ not covered in \eqref{eq U and Phi1 coincide on SR} are of the form $\gamma a$ with $a\in S_{_{\mathbb{R}}}$ and $\gamma \in \mathbb{T}$ (cf. \eqref{minTripGood}). However, in this case $\Phi_1( \gamma a ) = f(\gamma) \Phi_1(a) = \gamma U(a) = U(\gamma a)$.\smallskip

To conclude the proof we note that assuming $\Phi_1(i b) = - i \Phi_1(b)$ for all $b\in S_{_{\mathbb{R}}}$, the mapping $\overline{\Phi_1}$ satisfies that $\overline{\Phi_1}(i b) = i \overline{\Phi_1}(b)$ for all $b\in S_{_{\mathbb{R}}}$. By applying the conclusion in the previous paragraphs to $\overline{\Phi_1}$, we deduce the existence of a triple isomorphism $U$ from $M$ onto $\tilde{M}$ such that $\overline{\Phi_1} = U|_{\mathcal{U}(M)}.$
 \end{proof}

\begin{remark}\label{r role of the continuity hypothesis} Let us observe that the hypothesis concerning the continuity of the bijection $\Phi$ in Theorem \ref{t spin order 2Good} is only employed to guarantee that the mapping $f$ given by Proposition \ref{p all maps fu coincide new} satisfies $f(\lambda)= \lambda$ or $f(\lambda) = \overline{\lambda}$ for all $\lambda\in \mathbb{T}$.
\end{remark}

We shall see next that Theorem \ref{t spin order 2Good} admits a stronger, almost equivalent, statement.

\begin{theorem}\label{t biorder preserving from a spin into a Cartan factor} Let $\Phi : \mathcal{U}(M) \to \mathcal{U}(\tilde{C})$ be a bijective transformation which preserves the partial ordering in both directions and orthogonality between tripotents, where $M$ is complex spin factors of dimension $n\geq 3$ and $\tilde{C}$ is a Cartan factor. Then $\tilde{C}$ also is a complex spin factor.\smallskip

Suppose additionally that there exists a non-zero tripotent $u\in M$ such that $\Phi|_{\mathbb{T} u}$ is continuous at $u$. Then there exists a complex-linear or conjugate-linear triple isomorphism $T: M\to \tilde{C}$ such that $\Phi (e) = T(e)$ for all $e\in \mathcal{U} (M)$. More concretely, there exist $\lambda_0\in \mathbb{T}$ and a surjective real linear isometry $U\in B(M_{_\mathbb{R}},\tilde{C}_{_\mathbb{R}})$ such that $\Phi =\lambda_0 \tilde{U}$ or $\bar{\Phi} =\lambda_0 \tilde{U},$ where $\tilde{U}$ is the {\rm(}isometric{\rm)} triple isomorphism from $M$ onto $\tilde{C}$ defined by $\tilde{U} (a+i b) = U(a) + i U(b),$ for all $a+i b \in M$.
\end{theorem}

\begin{proof} It is well known that $M$ admits a unitary (rank--2) tripotent $u$.  Clearly $\mathcal{U} (M) = M_2(u) \cap \mathcal{U} (M)$. Lemma \ref{l Peirce 2 for fully non-complete tripotents improved} and the surjectivity of $\Phi$ prove that $$\mathcal{U} (\tilde{C}) = \Phi \left( \mathcal{U} (M)  \right) = \Phi \left( M_2(u) \cap \mathcal{U} (M) \right) = \tilde{C}_2(\Phi(u)) \cap \mathcal{U} (\tilde{C}).$$ It then follows that $\Phi(u)$ is a unitary tripotent in $\tilde{C}$. Lemma \ref{c first consequences}\eqref{eq Phi is additive on finite families of mo trip} affirms that $\Phi (u)$ has rank--2, and thus Lemma 3.8 in \cite{KalPe2019} implies that $\tilde{C} = \tilde{C}_2(\Phi(e))$ is a spin factor. The rest is clear from Theorem \ref{t spin order 2Good}.
\end{proof}

We can now rediscover and extend Moln{\'a}r's Theorem \ref{t Molnar 2002} even to the case of 2-dimensional complex Hilbert spaces and mappings satisfying a weaker hypothesis.

\begin{theorem}\label{t Molnar 2002 for rank 2 and weaker hypotheses} Let $H$ be a complex Hilbert space with dim$(H)\geq 2$. Suppose that $\Phi : \mathcal{U}(B(H))\to \mathcal{U}(\tilde{C})$ is a bijective transformation which preserves the partial ordering in both directions and orthogonality between tripotents, where $\tilde{C}$ is a Cartan factor. If $\Phi$ is continuous {\rm(}in the operator norm{\rm)} at a single element of $\mathcal{U}(B(H))$ different from $0$, then $\Phi$ extends to a real linear triple isomorphism.
\end{theorem}

\begin{proof} If dim$(H)\geq 3$ and $\tilde{C} = B(H)$ the conclusion follows from Theorem \ref{t Molnar 2002} and Proposition \ref{p order preservation in both directions gives orthogonality preservation}. If dim$(H)=2$ the Cartan factor $B(H)$ is a 4-dimensional spin factor and then the desired result is a consequence of our Theorem \ref{t biorder preserving from a spin into a Cartan factor}.
\end{proof}

\section{Order preserving bijections between arbitrary Cartan factors}\label{sec:order preservers between other Cartan factors}

In this section we shall study how to produce a complex-linear or conjugate-linear extension of every bijection preserving the partial ordering in both directions and orthogonality between the posets of tripotents in two arbitrary Cartan factors.\smallskip

Our first goal in this section is to prove that every bijective transformation between the sets of tripotents in two Cartan factors with rank bigger than or equal to $2$ which preserves the partial ordering in both directions and orthogonality between tripotents, must preserve quadrangles and trangles.

\begin{proposition}\label{p Peirce 2 for fully non-complete tripotents trangles and quadrangles} Let $\Phi : \mathcal{U}(C) \to \mathcal{U}(\tilde{C})$ be a bijective transformation which preserves the partial ordering in both directions and orthogonality between tripotents, where $C$ and $\tilde{C}$ are Cartan factors with rank bigger than or equal to $2$. Suppose that there exists a non-zero tripotent $u\in C$ such that $\Phi|_{\mathbb{T} u}$ is continuous at $u$. Then the following statements hold:
\begin{enumerate}[$(a)$]\item $\Phi$ maps quadrangles in $C$ to quadrangles in $\tilde{C}$;
\item $\Phi$ maps trangles in $C$ to trangles in $\tilde{C}$.
\end{enumerate}
\end{proposition}

\begin{proof} Let $(u_{1},u_{2},u_{3},u_{4})$ (respectively, $(v,u,\tilde v)$) be a quadrangle (respectively, a trangle) in $C$. The tripotent $e=u_1 +u_3$ (respectively, $e=v+\tilde{v}$) is a rank--2 tripotent in $M$. Lemma \ref{c first consequences}\eqref{eq Phi is additive on finite families of mo trip} implies that $\Phi (e)$ is a rank--2 tripotent. By \cite[Lemma 3.8]{KalPe2019} the JB$^*$-subtriples $C_2(e)$ and $\tilde{C}_2 (\Phi(e))$ are spin factors. Lemma \ref{l Peirce 2 for fully non-complete tripotents improved} and the hypotheses prove that $\Phi|_{\mathcal{U} (C_2(e))} : \mathcal{U} (C_2(e))\to \mathcal{U} (\tilde{C}_2 (\Phi(e)))$ is a bijection preserving the local order in both directions. The hypothesis concerning the continuity of $\Phi$ implies that the mapping $f$ given by Proposition \ref{p all maps fu coincide new} is the identity or the conjugation on $\mathbb{T}$. An application of Theorem \ref{t biorder preserving from a spin into a Cartan factor} shows the existence of a complex-linear or conjugate-linear isometric triple isomorphism $\tilde{U}: C_2(e) \to  \tilde{C}_2 (\Phi(e))$ whose restriction to $\mathcal{U} (C_2(e))$ coincides with $\Phi|_{\mathcal{U} (C_2(e))}$. Finally, the desired conclusion follows from the fact that the quadrangle $(u_{1},u_{2},u_{3},u_{4})$ (respectively, the trangle $(v,u,\tilde v)$) is contained in $C_2(e)$.
\end{proof}

\subsection{Cartan factors of rank bigger than or equal to three} \ \smallskip

The result established by L. Moln{\'a}r in Theorem \ref{t Molnar 2002} deals with the subset of type 1 Cartan factors of the form $B(H)$ with dim$(H)\geq 3$. Let us observe that this condition on the dimension of $H$ is equivalent to assume that $B(H)$ has rank greater than or equal to three. We shall see next that this is not an exclusive advantage of this kind of Cartan factors.\smallskip

We recall that a JB$^*$-triple $E$ is called \emph{abelian} if $L(a,b) L(c,d) = L(c,d) L(a,b)$ for all $a,b,c,d\in E$ (see, for example, \cite[$(1.4)$]{Horn87}, \cite[page 131]{Batt91}). Each single-generated JB$^*$-subtriple of $E$ is always abelian. A tripotent $e$ in $E$ is called \emph{abelian} if the Peirce-2 subspace $E_2(e)$ is an abelian JB$^*$-triple, equivalently, $E_2(e)$ is a commutative unital C$^*$-algebra. Clearly every minimal tripotent is abelian.\smallskip

In this section we shall employ some new results on finite tripotents established in \cite{HamKalPe20}. Let $e$ be a tripotent in a JBW$^*$-triple $M$. According to \cite[\S 3]{HamKalPe20}, we shall say that $e$ is {\em finite} if any tripotent $u\in M_2(e)$ which is complete in $M_2(e)$ is already unitary in $M_2(e)$. For example, every abelian (and hence every minimal) tripotent in $M$ is finite (see \cite[Lemma 3.2$(e)$]{HamKalPe20}).

\begin{proposition}\label{p tripotents of rank 3} Let $\Phi : \mathcal{U}(C) \to \mathcal{U}(\tilde{C})$ be a bijective transformation which preserves the partial ordering in both directions and orthogonality between tripotents, where $C$ and $\tilde{C}$ are Cartan factors. Let us additionally assume that there exists a non-zero tripotent $u\in C$ such that $\Phi|_{\mathbb{T} u}$ is continuous at $u$. Suppose $e$ is a tripotent in $C$ with rank $\geq 3$. Then there exists a complex-linear or conjugate-linear {\rm(}isometric{\rm)} triple isomorphism $T: C_2(e) \to \tilde{C}_2 (\Phi(e))$ satisfying $\Phi (v) = T(v)$ for every tripotent $v \in C_2(e)$. In particular, $$\Phi \left( C_2(u) \cap \mathcal{U} (C) \right) = \tilde{C}_2(\Phi(u)) \cap \mathcal{U} (\tilde{C}).$$
We can actually assume that $T$ is a complex-linear or a conjugate-linear Jordan $^*$-isomorphism between the JBW$^*$-algebras $C_2(e)$ and $\tilde{C}_2 (\Phi(e))$.
\end{proposition}

\begin{proof} Let $e$ be a tripotent satisfying the assumptions of the proposition. We know that $\Phi(e)$ is a tripotent in $\tilde{C}$ with rank $\geq 3$ (cf. Lemma \ref{c first consequences}\eqref{eq Phi is additive on finite families of mo trip}). The Peirce-2 subspace $C_2(e)$ is an atomic and factor JBW$^*$-algebra with unit $e$ and has rank at least three, in particular $C_2(e)$ does not contain a type $I_2$ part. The projections in $C_2(e)$ are precisely the tripotents $v$ in $C$ with $v\leq e$ (cf. \cite[Lemma 3.5$(i)$]{Batt91}). Let $\mathcal{P} (C_2(e))$ denote the lattice of projections in $C_2(e)$. It follows from the hypotheses and the previous comments that $\Phi|_{\mathcal{P} (C_2(e))} : \mathcal{P} (C_2(e))\to  \mathcal{P} (\tilde{C}_2(\Phi(e)))$ is a bijection preserving the partial ordering (and orthogonality) in both directions.\smallskip

Let $e_1,\ldots, e_m$ be a finite collection of mutually orthogonal projections in $C_2(e)$. It follows from Lemma \ref{c first consequences}$(d)$ that $\Phi (e_1+\ldots+e_m) = \Phi (e_1) +\ldots+\Phi(e_m)$. Since clearly $\|\Phi (p)\|\leq 1$ for every projection $p\in C_2(e)$, we can conclude that $\Phi|_{\mathcal{P} (C_2(e))} : \mathcal{P} (C_2(e))\to  \mathcal{P} (\tilde{C}_2(\Phi(e)))$ is a vector-valued finitely additive quantum measure on $C_2(e)$ in the terminology of \cite{BuWri89,BuWri94}. The arguments to obtain a vector-valued version of the Bunce--Wright--Mackey--Gleason theorem for von Neumann algebras from the scalar-valued version (see \cite[Theorem B implies Theorem A]{BuWri94}) are also valid to deduce a vector-valued Jordan version from the scalar-valued Jordan version of this Bunce--Wright--Mackey--Gleason theorem for JBW-algebras established in \cite[Theorem 2.1]{BuWri89}. Therefore, by the just commented vector-valued Bunce--Wright--Mackey--Gleason theorem, there exists a complex-linear bijection (and hence a Jordan $^*$-isomorphism between these two JBW$^*$-algebras) $T: C_2(e) \to \tilde{C}_2(\Phi(e))$ satisfying \begin{equation}\label{eq T and Phi conincide on projections}\hbox{$T(p ) = \Phi (p)$ for every projection $p$ in $C_2(e)$.}
\end{equation}

Let $f$ be the mapping given by Proposition \ref{p all maps fu coincide new} (see also \eqref{eq first definition of fu}). Now, by applying the hypothesis concerning the continuity of $\Phi$, we can deduce, as in the proof of Theorem \ref{t spin order 2Good} (see also Remark \ref{r role of the continuity hypothesis}), that $f(\lambda) = \lambda$ or $f(\lambda) = \overline{\lambda}$ for all $\lambda\in \mathbb{T}$. \smallskip

Unitaries in the JBW$^*$-algebra $C_2(e)$ are precisely the unitary tripotents in $C_2(e)$ (cf. \cite[Proposition 4.3]{BraKaUp78}). Let $u$ be a unitary tripotent in $C_2(e)$. Since the JBW$^*$-subalgebra of $C_2(e)$ generated by $u$ is a commutative von Neumann algebra and $C_2(e)$ is atomic, we can always find a spectral resolution of $u$ in terms of a family of mutually orthogonal minimal projections in $C_2(e)$, that is, there exists mutually orthogonal minimal projections $\{p_j: j\in J\}$ in $C_2(e)$ (which are minimal tripotents in $C$) and $\{\lambda_j: j\in J\}\subset \mathbb{T}$ such that $\displaystyle u = \hbox{w$^*$-}\sum_{j\in J} \lambda_j p_j$. Clearly $u$ is the supremum of the family $\{\lambda_j p_j: j\in J\}$ and hence $\Phi (u)$ must coincide with the supremum of the family $\{\Phi(\lambda_j p_j) : j \in J\}$ (cf. Lemma \ref{l infima and suprema sums of orthogonal families}$(b)$) and similarly, $T (u)$ must coincide with the supremum of the family $\{T(\lambda_j p_j) :j \in J\}.$ For each $j\in J$ we have \begin{equation}\label{eq T and f Phi on multiples of minimal projections}
T(\lambda_j p_j)= \lambda_j T( p_j), \hbox{ and } \Phi(\lambda_j p_j) = f(\lambda_j) \Phi(p_j).
\end{equation}

We shall distinguish two cases: \begin{enumerate}[$(1)$]
\item $f(\lambda) = \lambda,$ for all $\lambda\in \mathbb{T}$,
\item $f(\lambda) = \overline{\lambda},$ for all $\lambda\in \mathbb{T}$.
\end{enumerate}

We deal first with case $(1)$. Under this assumption \eqref{eq T and f Phi on multiples of minimal projections} particularizes in the form
$$T(\lambda_j p_j)= \lambda_j T( p_j), \hbox{ and } \Phi(\lambda_j p_j) = \lambda_j \Phi(p_j).$$ It then follows from \eqref{eq T and Phi conincide on projections} that $T(\lambda_j p_j)=  \Phi(\lambda_j p_j)$ for all $j$. Since $\Phi (u)$ is the supremum of $\{\Phi(\lambda_j p_j) : j \in J\},$ and $T (u)$ is the supremum of $\{T(\lambda_j p_j) :j \in J\},$ it follows that $\Phi (u) = T(u)$. We have therefore proved that \begin{equation}\label{eq T and Phi coincide on unitaries in C2(e)} T(u) = \Phi (u),\hbox{ for all unitary } u\in C_2(e).
\end{equation}

Let $e_i$ be a minimal tripotent in $C_2(e)$. Since $e_i$ is a finite tripotent in the latter JBW$^*$-algebra (cf. \cite[Lemma 3.2$(e)$]{HamKalPe20}), by applying Proposition 7.5 in \cite{HamKalPe20} we deduce the existence of a unitary $u\in C_2(e)$ such that $e_i \leq u$. Clearly, the elements $\pm e_i + (u-e_i)$ are unitaries in $C_2(e)$, and thus \eqref{eq T and Phi coincide on unitaries in C2(e)} guarantees that $$T(\pm e_i + (u-e_i)) = \Phi (\pm e_i + (u-e_i)).$$ Having in mind that $\pm e_i \perp (u-e_i)$, Lemma \ref{c first consequences}$(d)$ and Proposition \ref{p all maps fu coincide}$(e)$ give $$ \pm T( e_i)  + T(u-e_i) = T(\pm e_i + (u-e_i)) = \Phi (\pm e_i + (u-e_i))= \pm \Phi ( e_i ) + \Phi (u-e_i),$$ which proves that $T(e_i) = \Phi (e_i).$ The arbitrariness of $e_i$ implies that \begin{equation}\label{eq T and Phi coincide on minimal tripotents in C2(e)} T(w) = \Phi (w),\hbox{ for all minimal tripotent } w\in C_2(e).
\end{equation}

Having in mind that every tripotent in $C_2(e)$ is the supremum of a family of mutually orthogonal minimal tripotents in $C_2(e)$, we deduce from \eqref{eq T and Phi coincide on minimal tripotents in C2(e)} and Lemma \ref{l infima and suprema sums of orthogonal families}$(b)$ that $T(v) = \Phi (v)$ for all tripotent $v\in C_2(e)$ (this can be also deduced from Proposition \ref{p identity principle}).\smallskip

Suppose finally that $(2)$ holds, that is, $f(\lambda) = \overline{\lambda},$ for all $\lambda\in \mathbb{T}$. Let $^{*_{e}}$ and $^{*_{\Phi(e)}}$ denote the involutions on the JBW$^*$-algebras $C_2(e)$ and $\tilde{C}_2 (\Phi(e))$, respectively. We know by construction that $T(x^{*_{e}}) = T(x)^{*_{\Phi(e)}}$ for all $x\in C_2(e)$. The mapping $R: C_2(e)\to \tilde{C}_2 (\Phi(e))$, $R(x) = T(x^{*_{e}})$ is a conjugate-linear Jordan $^*$-isomorphism between these two JBW$^*$-algebras. By repeating the arguments given in case $(1)$ we arrive to the conclusion that $R(v) = \Phi (v)$ for all tripotent $v\in C_2(e)$.
\end{proof}

The next results are straightforward consequence of the result we have just proved. The first one is direct from Lemmata \ref{l Peirce 2 for fully non-complete tripotents improved}, \ref{l scalar multiples of minimal and non-minimal tripotents} and the previous proposition.

\begin{corollary}\label{c Peirce 2 for fully non-complete tripotents final} Let $\Phi : \mathcal{U}(C) \to \mathcal{U}(\tilde{C})$ be a bijective transformation which preserves the partial ordering in both directions and orthogonality between tripotents, where $C$ and $\tilde{C}$ are Cartan factors with rank bigger than or equal to $2$. Suppose that there exists a non-zero tripotent $u\in C$ such that $\Phi|_{\mathbb{T} u}$ is continuous at $u$. Let $e$ be a tripotent in $C$. Then $\Phi$ maps the set $C_2(e) \cap \mathcal{U} (C)$ onto the set $\tilde{C}_2(\Phi(e)) \cap \mathcal{U} (\tilde{C})$.
\end{corollary}

\begin{proof} If $e$ has rank one or two the result follows from Lemmata \ref{l scalar multiples of minimal and non-minimal tripotents} and \ref{l Peirce 2 for fully non-complete tripotents improved}, respectively. The remaining cases are given by Proposition \ref{p tripotents of rank 3}.\end{proof}

We shall next prove that the hypothesis concerning the rank of the tripotent $e$ in Proposition \ref{p tripotents of rank 3} can be somehow relaxed.

\begin{corollary}\label{c tripotents of rank 3} Let $\Phi : \mathcal{U}(C) \to \mathcal{U}(\tilde{C})$ be a bijective transformation which preserves the partial ordering in both directions and orthogonality between tripotents, where $C$ and $\tilde{C}$ are Cartan factors with rank $\geq 3$. Let us additionally assume that there exists a non-zero tripotent $u\in C$ such that $\Phi|_{\mathbb{T} u}$ is continuous at $u$. Suppose $e$ is a tripotent in $C$. Then there exists a complex-linear or a conjugate-linear Jordan $^*$-isomorphism $T: C_2(e) \to \tilde{C}_2 (\Phi(e))$ satisfying $\Phi (v) = T(v)$ for every tripotent $v \in C_2(e)$.
\end{corollary}

\begin{proof} If $e$ has rank $\geq 3$ the conclusion follows from Proposition \ref{p tripotents of rank 3}. Otherwise there exists a tripotent $\tilde{e}$ with rank $\geq 3$ such that $e\leq \tilde{e}$. Applying the previous Proposition \ref{p tripotents of rank 3} to $\tilde{e}$ we deduce the existence of a complex-linear or a conjugate-linear Jordan $^*$-isomorphism $T: C_2(\tilde{e}) \to \tilde{C}_2 (\Phi(\tilde{e}))$ satisfying $\Phi (v) = T(v)$ for every tripotent $v \in C_2(\tilde{e})$. Since $C_2({e})$ is a JBW$^*$-subalgebra of $C_2(\tilde{e})$ with $T(C_2(e))= \tilde{C}_2 (\Phi(e))$, the mapping $T|_{C_2(e)}: C_2(e) \to \tilde{C}_2 (\Phi(e))$ satisfies the desired property.
\end{proof}

Next we have another generalization, in a new direction, of the result established by Moln{\'a}r in Theorem \ref{t Molnar 2002}.

\begin{theorem}\label{t Molnar thm rank 2 Cartan factors with a unitary} Let $\Phi : \mathcal{U}(C) \to \mathcal{U}(\tilde{C})$ be a bijective transformation which preserves the partial ordering in both directions and orthogonality between tripotents, where $C$ and $\tilde{C}$ are Cartan factors with rank $\geq 2$. Suppose $C$ admits a unitary tripotent. Let us additionally assume that there exists a non-zero tripotent $w\in C$ such that $\Phi|_{\mathbb{T} w}$ is continuous at $w$. Then there exists a complex-linear or a conjugate-linear triple isomorphism $T: C \to \tilde{C}$ satisfying $\Phi (v) = T(v)$ for every tripotent $v \in C$.
\end{theorem}

\begin{proof} Since $C$ admits a unitary tripotent $u$, we conclude that $C = C_2(u)$ and $\mathcal{U} (C) = \mathcal{U} (C_2(u))$. If $u$ has rank--2, it follows from \cite[Lemma 3.8]{KalPe2019} that $\mathcal{U} (C) = \mathcal{U} (C_2(u))$ is a spin factor, and thus the desired statement follows from Theorem \ref{t biorder preserving from a spin into a Cartan factor}.\smallskip

Assume next that $u$ has rank $\geq 3$. By Corollary \ref{c tripotents of rank 3} there exists a complex-linear or a conjugate-linear Jordan $^*$-isomorphism $T: C= C_2(u) \to \tilde{C}_2 (\Phi(u))$ satisfying $\Phi (v) = T(v)$ for every tripotent $v \in C_2(e)$. The surjectivity of $\Phi$ proves that $$ \mathcal{U} (\tilde{C})  = \Phi ( \mathcal{U} (C)) = T ( \mathcal{U} (C)) = \mathcal{U} (\tilde{C}_2 (\Phi(u))),$$ yielding that $\Phi (u)$ is a unitary tripotent in $\tilde{C}$ which concludes the proof.
\end{proof}

Type 1 Cartan factors of the form $B(H)$ with dim$(H)\geq 2$, Cartan factors of type 2 with dim$(H)\geq 6$ even, or infinite, all type 3 Cartan factors with dim$(H)\geq 3,$ and the exceptional Cartan factor of type 6 admit a unitary tripotent and have rank $\geq 3$ (cf. \cite[Proposition 2]{HoMarPeRu} and \cite[Table 1 in page 210]{Ka97}). The type 3 Cartan factor of rank--2 is precisely the 3-dimensional spin factor $S_2(\mathbb{C})$.

\subsection{Rectangular type 1 Cartan factors} \ \smallskip

In this section we shall study the case of rectangular type 1 Cartan factors with rank bigger than or equal to two. Along this section we shall focus on rectangular Cartan factors of the form $B(H,K),$ where $H$ and $K$ are two complex Hilbert spaces with dim$(H)\neq \hbox{dim} (K)$ and both of them are $ \geq 2$. We can assume that $K$ is a proper closed subspace of $H$.\smallskip

Following the notation from the influencing paper \cite{BuChu92}, given a Cartan factor $C$ of type $j\in \{1,\ldots, 6\},$ the \emph{elementary} JB$^*$-triple $K_j$ of type $j$ (also called \emph{nuclear} JB$^*$-triple of type $j$ in \cite{DanFri87}) is defined in the following terms: ${K}_1 = {K} (H_1, H_2)$; ${K}_i = C \cap {K}(H)$ when $C$ is of type $i = 2 , 3$, and ${K}_i = C$ if the latter is of type $ 4, 5,$ or $6$. Here ${K} (H_1, H_2)$ stands for the space of compact linear operators from $H_1$ to $H_2$ and ${K} (H)= {K} (H, H)$. Obviously, if ${K}$ is an elementary JB$^*$-triple of type $j$, its bidual is precisely a Cartan factor of type $j$.\smallskip

At this stage we need to recall some coordinatization theorems for Jordan triple systems ``covered'' by a ``grid'' developed in papers by K. McCrimmon, and K. Meyberg \cite{McCrimMey81}, T. Dang and Y. Friedman \cite{DanFri87}, E. Neher \cite{Neher87} and G. Horn \cite{Horn87c, Horn87b} (see also the monograph \cite[\S 6]{Fri2005}). A \emph{grid} in a JB$^*$-triple $E$ is a family formed by minimal and rank two tripotents in $E$ built up of quadrangles of minimal tripotents or trangles of the form $(v,u, \tilde v)$ with $v$ and $\tilde v$ minimal, where  all the non-vanishing triple products among the elements of the grid are those associated to these types of trangles and quadrangles.\smallskip

The results in \cite{McCrimMey81,DanFri87,Neher87} and \cite{Horn87c} show that every Cartan factor $C$ admits a (rectangular, symplectic, hermitian, spin, or exceptional) grid $\mathcal{G}$ such that the elementary JB$^*$-triple $K$ associated with $C$ is precisely the norm closed linear span of the grid $\mathcal{G}$, and $C$ being the weak$^*$-closure of $K$ is nothing but the weak$^*$-closure of the linear span of $\mathcal{G}$ (compare, for example, \cite[Structure Theorem IV.3.14]{Neher87} or \cite[\S 2]{DanFri87}).\smallskip

Let us describe one of these famous grids with more detail. Let $\Lambda$ and $\Gamma$ be two index sets. A family of minimal tripotents $\{ u_{ij} : i\in \Lambda, j\in \Gamma\}$ is called a \emph{rectangular grid} if the following properties hold:\begin{enumerate}[$(i)$]\item $u_{jk}$, $u_{il}$ are collinear if they share a common row index ($j=i$) or a column index ($k=l$), and are orthogonal otherwise;
\item $(u_{jk}, u_{jl}, u_{il}, u_{ik})$ is a quadrangle for all $j\neq i$, $k\neq l$;
\item all other types of triple products (i.e., those which are not of the form $L(x,x)(y)$ or $\{x,y,z\}$, where $(x,y,z)$ is a prequadrangle) vanish.
\end{enumerate}

Given two complex Hilbert spaces $H$ and $K$ and two norm-one elements $\xi\in K$, $\eta\in H$ the symbol $\xi\otimes \eta$ will stand for the tripotent in $B(H,K)$ defined by $\xi\otimes \eta (\zeta) = \langle \zeta, \eta\rangle \xi.$ Let $\{\xi_i: i \in \Lambda\}$ and $\{\eta_j : j\in \Gamma\}$ be orthonormal basis of $K$ and $H$, respectively. It is not hard to check that the set $\{ u_{ij} = \xi_i\otimes \eta_j : i\in \Lambda, j\in \Gamma\}$ is a rectangular grid in $B(H,K)$ whose linear span is weak$^*$ dense (see for example \cite[pages 313-317]{DanFri87}). Furthermore, every Cartan factor admitting a rectangular grid whose linear span in weak$^*$-dense is isometrically isomorphic to some $B(H,K)$ \cite[Proposition in page 314]{DanFri87}.

\begin{theorem}\label{t BHK rectangular} Let $K$ be a proper subspace of a complex Hilbert space $H$ with dim$(K)\geq 2,$ let $C= B(H,K),$ and let $\tilde{C}$ be another Cartan factor. Suppose $\Phi : \mathcal{U}(C) \to \mathcal{U}(\tilde{C})$ is a bijective transformation which preserves the partial ordering in both directions and orthogonality between tripotents. Assume that there exists a non-zero tripotent $u\in C$ such that $\Phi|_{\mathbb{T} u}$ is continuous at $u$. Then there exists a complex-linear or conjugate-linear triple isomorphism $T: C\to \tilde{C}$ such that $\Phi (e) = T(e)$ for all $e\in \mathcal{U} (C)$.
\end{theorem}

\begin{proof} To simplify the arguments we shall distinguish two cases:  dim$(H)=\infty$ and dim$(H)<\infty.$ Although both results are deeply technical, the infinite-dimensional case admits a simpler solution.\smallskip

\emph{Case $I$.} We shall first assume that \emph{dim$(H)=\infty$}.  Let $\{\eta_i: i \in \Lambda\}$ and $\{\xi_j : j\in \Gamma\}$ be orthonormal basis of $K$ and $H$, respectively. Since dim$(K)\leq$dim$(H)$ and the latter is infinite, we can find a family $\{\Gamma_k: k\in J\}$ of mutually disjoint subsets of $\Gamma$ such that $|\Gamma_k | = |\Lambda|\geq 2$ for all $k\in J$ (for the last inequality we applied that dim$(K)\geq 2$), and $\Gamma = \bigcup_{k\in J} \Gamma_k$. The set $\{ u_{ij} = \eta_i\otimes \xi_j : i\in \Lambda, j\in \Gamma\}$ is a rectangular grid in $B(H,K)$, and Proposition \ref{p Peirce 2 for fully non-complete tripotents trangles and quadrangles} assures that $\{ \Phi(u_{ij}) = \Phi (\eta_i\otimes \xi_j) : i\in \Lambda, j\in \Gamma\}$ is a rectangular grid contained in $\tilde{C}$. Let ${K}(C)$ and ${K}(\tilde{C})$ denote the elementary or nuclear JB$^*$-subtriples of $C$ and $\tilde{C}$ generated by the grids $\{ u_{ij} = \eta_i\otimes \xi_j : i\in \Lambda, j\in \Gamma\}$ and $\{ \Phi(u_{ij}) = \Phi(\eta_i\otimes \xi_j) : i\in \Lambda, j\in \Gamma\},$ respectively, and let $T: {K}(C)\to {K}(\tilde{C})$ be a triple isomorphism defined by $T(u_{ij}) = \Phi (u_{ij})$ for all $i\in \Lambda, j\in \Gamma$. Let us observe that weak$^*$-closure of ${K}(C)$ is the whole $C$. It is known that mapping $T$ admits an extension to a weak$^*$-continuous triple isomorphism, denoted by the same symbol $T$, from $B(H,K)$ onto the weak$^*$ closure, $\overline{{K}(\tilde{C})}^{w^*},$ of ${K}(\tilde{C})$ in $\tilde{C}$ (cf. \cite[Lemma 1.14]{DanFri87}, \cite[\S 3]{Horn87c} or \cite[Extension Theorem 3.17 and the arguments in the proof of the Isomorphism Theorem 3.18]{Neher87}). By construction \begin{equation}\label{eq T and Phi coincide on the first grid} T(u_{ij}) = \Phi (u_{ij}) \hbox{ for all } i\in \Lambda, j\in \Gamma.
\end{equation}

In order to simplify the arguments, for each $k\in J$, let $H_k$ stand for the closed subspace of $H$ generated by the orthonormal system $\{\xi_j : k\in \Gamma_k\}$. Since dim$(H_k)= \hbox{dim}(K)$ for all $j$, the subtriple $B(H_k, K)$ is isometrically isomorphic to $B(K),$ and hence it admits a unitary element denoted by $u_k$. We observe that $C_2 (u_k) = B(H_k,K)$.\smallskip

If dim$(K)=2$ we apply Theorem \ref{t biorder preserving from a spin into a Cartan factor} and Lemma \ref{l Peirce 2 for fully non-complete tripotents improved}, while if dim$(K)\geq 3$ we employ Corollary \ref{c tripotents of rank 3}, to deduce that $\Phi : \mathcal{U} (C_2(u_k)) \to \mathcal{U} (\tilde{C}_2(\Phi(u_k)))$ is a bijection preserving the partial order in both directions and orthogonality in one direction, and hence there exists a complex-linear or a conjugate-linear triple isomorphism $T_k : C_2(u_k) \to \tilde{C}_2(\Phi(u_k))$ such that
\begin{equation}\label{eq Phi and Tk coincide on the commond domain} T_k(w) = \Phi(w), \hbox{ for all } w\in \mathcal{U} (C_2(u_k)) \hbox{ ($k\in J$).}
\end{equation}
Furthermore, the set $\{ u_{ij} = \xi_i\otimes \eta_j : i\in \Lambda, j\in \Gamma_k\}$ is a rectangular grid in $C_2 (u_k)$, and by \eqref{eq T and Phi coincide on the first grid}, $T_k (u_{ij}) =\Phi (u_{ij}) = T (u_{ij}),$ for all $i\in \Lambda, j\in \Gamma_j$. Clearly, the mappings $f$ given by Proposition \ref{p all maps fu coincide new} for $\Phi$ and for $\Phi|_{\mathcal{U} (C_2(u_k))}$ coincide for all $k\in J.$ Consequently, the maps $T_k$ are all complex-linear or conjugate-linear at the same time.\smallskip

\emph{Case $I.a$.} Let us assume that the mapping $f$ is the identity on $\mathbb{T}$, and thus all $T_k$ are complex-linear and \begin{equation}\label{eq Tk and the restriction on C2ek coindice} \hbox{$T_k = T|_{C_2(u_k)}$ for all $k\in J$.}
 \end{equation} In order to finish the proof in this case we shall prove that $T$ is a triple isomorphism  whose restriction to $\mathcal{U} (C)$ coincides with $\Phi$. It suffices to show that
$T(e) = \Phi (e)$ for all minimal tripotent $e\in C$ (cf. Proposition \ref{p identity principle} and have in mind that the surjectivity of $T$ on the whole $\tilde{C}$ is automatic if we recall that after this conclusion the image of $T$ contains all minimal tripotents in $\tilde{C}$). We shall split the arguments in several steps.\smallskip

Let $e_1$ and $e_2$ be two collinear minimal tripotents in $C$ and suppose that $\alpha_1, \alpha_2$ are two complex numbers such that $|\alpha_1|^2 + |\alpha_2|^2 =1$. The elements $e_1$ and $e_2$ (respectively, $\Phi (e_1)$ and $\Phi (e_2)$, which are mutually collinear minimal tripotents by Theorem \ref{t biorder preserving from a spin into a Cartan factor} or Proposition \ref{p Peirce 2 for fully non-complete tripotents trangles and quadrangles}) generate a JB$^*$-subtriple of $\tilde{C}$ which is isometrically isomorphic to a 2-dimensional complex Hilbert space (see \cite[Lemma in page 306]{DanFri87}). Actually, by combining \cite[Lemma 3.10]{FerPe18Adv} and Theorem \ref{t biorder preserving from a spin into a Cartan factor} or Proposition \ref{p Peirce 2 for fully non-complete tripotents trangles and quadrangles}, and the fact that $f$ is the identity mapping we have \begin{equation}\label{eq Phi is linear on 2 elements of mc min trip} \Phi \left( \alpha_1 e_1 + \alpha_2 e_2 \right) = \alpha_1 \Phi (e_1) + \alpha_2 \Phi (e_2).
\end{equation}

Now, let $\{ e_j : j \in I\}$ be a family of mutually collinear minimal tripotents in $C$. We claim that for each family $\{ \alpha_j : j \in I\}\subset \mathbb{C}$ with $\displaystyle \sum_{j\in I} |\alpha_j|^2=1$ we have \begin{equation}\label{eq Phi is linear on families of mc min trip} \Phi \left( \sum_{j\in I} \alpha_j e_j \right) =  \sum_{j\in I} \alpha_j \Phi (e_j).
\end{equation} Namely, by the arguments above $\{ \Phi(e_j) : j \in I\}$ is a family of mutually collinear minimal tripotents in $\tilde{C}$ and it generates a JB$^*$-subtriple which is isometrically isomorphic to a complex Hilbert space (see \cite[Lemma in page 306]{DanFri87}). Since $\displaystyle \sum_{j\in I} |\alpha_j|^2< \infty$, and hence $\{ j \in I : |\alpha_j|\neq 0\}$ is countable, an induction argument based on \eqref{eq Phi is linear on 2 elements of mc min trip}, like the one employed in the proof of Theorem \ref{t spin order 2Good}, gives the statement in \eqref{eq Phi is linear on families of mc min trip}. It should be remarked her that the formula established in \eqref{eq Phi is linear on families of mc min trip} does not really depend on the dimension of the Hilbert space $H$. It is worth to notice that in the particular setting of type 1 Cartan factors, for each family of mutually orthogonal minimal tripotents $\{ e_j : j \in I\}$ and each family $\{ \alpha_j : j \in I\}\subset \mathbb{C}$ with $\displaystyle \sum_{j\in I} |\alpha_j|^2=1$, the element $\displaystyle \sum_{j\in I} \alpha_j e_j $ is a minimal tripotent too. \smallskip

A minimal tripotent in $C$ is of the form $e= \eta\otimes \xi$ with $\eta$ and $\xi$ in the unit spheres of $K$ and $H$, respectively. Since $\{\xi_j: j \in \Gamma\}$ is an orthonormal basis of $H$ we can write $e$ in the form $$ e= \sum_{j \in \Gamma} \alpha_j \eta\otimes \xi_j, \hbox{ where } \sum_{j\in \Gamma} |\alpha_j|^2=1.$$ The minimal tripotents in the family $\{ \eta\otimes \xi_j : j\in \Gamma\}$ are mutually collinear, and  for each $j \in \Gamma_k$ ($k\in J$), the minimal tripotent $\eta\otimes \xi_j$ belongs to $C_2 (u_k) = B(H_k,K).$ Thus, by applying \eqref{eq Phi is linear on families of mc min trip}, \eqref{eq Phi and Tk coincide on the commond domain} and \eqref{eq Tk and the restriction on C2ek coindice} we get $$\begin{aligned}\Phi(e)&= \sum_{j \in \Gamma} \alpha_j \Phi(\eta\otimes \xi_j) = \sum_{k\in J} \sum_{j \in \Gamma_k} \alpha_j \Phi(\eta\otimes \xi_j) \\
&= \sum_{k\in J} \sum_{j \in \Gamma_k} \alpha_j T_k(\eta\otimes \xi_j) = \sum_{k\in J} \sum_{j \in \Gamma_k} \alpha_j T(\eta\otimes \xi_j) = T(e),
\end{aligned} $$ which concludes the proof in this Case $I$. \smallskip

\emph{Case $I.b$.} Assume next that $f$ is the conjugation on $\mathbb{T}$. Let us take two conjugations on $H$ and $K$, both denoted by the same symbol $\overline{\cdot}$. We define a conjugation (conjugate linear isometry) on $B(H,K)$ given by $$B(H,K) \to B(H,K), \ \ a\mapsto a^c \ \hbox{where } a^c (\xi):=\overline{a(\overline{\xi})},\ \forall\xi \in H.$$ It is easy to check that $(ab)^c = a^c b^c$ and $(a^c)^* = (a^*)^c$ for all $a,b\in B(H,K),$ and thus the mapping $a\mapsto a^c$ is a conjugate-linear triple automorphism on $B(H,K)$. It is easy to see that the mapping $\Phi_1: \mathcal{U}(B(H,K))\to \mathcal{U}(\tilde{C}),$ $\Phi_1(u) = \Phi(u^c)$ is a bijection preserving the partial order  in both directions and orthogonality between tripotents. Furthermore, the corresponding map $f$ given by Proposition \ref{p all maps fu coincide new} for $\Phi_1$ must be the identity on $\mathbb{T}$. By applying the Case $I$ to $\Phi_1$ we deduce the existence of a complex-linear triple isomorphism $T: B(H,K)\to \tilde{C}$ whose restriction to $\mathcal{U} (B(H,K))$ coincides with $\Phi_1$. The mapping $R: B(H,K)\to \tilde{C}$, $R(x) = T(x^c)$ is a conjugate-linear triple isomorphism whose  restriction to $\mathcal{U} (B(H,K))$ coincides with $\Phi$. This finishes the proof of Case $II$.\smallskip

\emph{Case $II$} The second part of this proof will be devoted to study the case in which dim$(H)<\infty.$ The argument is very similar to the one given in the infinite dimensional case. Let $\{\eta_i: i \in \Lambda=\{1,\ldots,d_1\}\}$ and $\{\xi_j : j\in \Gamma=\{1, \ldots, d_2\}\}$ be orthonormal basis of $K$ and $H$, respectively, where $2\leq d_1<d_2\in \mathbb{N}$. Let us write $d_2 = c_1 d_1 + r_1$ with $c_1,r_1\in \mathbb{N}\cup \{0\}$ and $r_1< d_1$. Let us find mutually disjoint subsets $\Gamma_1, \ldots \Gamma_{c_1}$ of $\Gamma$ and another set $\Gamma_{c_1+1}$ not disjoint with some of the previous ones such that $|\Gamma_k | = |\Lambda|= d_1 \geq2$ for all $k\in \{1,\ldots, c_1+1\}$ and $\Gamma = \bigcup_{k=1}^{c_1+1} \Gamma_k$. If $r_1=0$ we simply take $\Gamma_{c_1+1} = \emptyset$ and  the counter $k$ moves in $\{1,\ldots, c_1\}$ --this case is even easier or identical to Case $I$, so we shall focus only in the case in which $r_1\neq 0$, where the set $\Gamma_{c_1+1}$ is precisely the set formed by the last $d_1$ elements in $\Gamma$. \smallskip

The set $\{ u_{ij} = \eta_i\otimes \xi_j : i\in \Lambda, j\in \Gamma\}$ is a rectangular grid in $B(H,K)$, and Proposition \ref{p Peirce 2 for fully non-complete tripotents trangles and quadrangles} assures that $\{ \Phi(u_{ij}) = \Phi(\eta_i\otimes \xi_j) : i\in \Lambda, j\in \Gamma\}$ is a rectangular grid contained in $\tilde{C}$. As in the previous case, or even easier because we do not need to consider weak$^*$-closures, there exists a triple isomorphism $T: C\to K(\tilde{C})$ defined by $T(u_{ij}) = \Phi (u_{ij})$ for all $i\in \Lambda, j\in \Gamma$, where ${K}(\tilde{C})$ stands for the (finite dimensional) JBW$^*$-subtriple of $\tilde{C}$ generated by the rectangular grid $\{ \Phi(u_{ij}) = \Phi(\eta_i\otimes \xi_j) : i\in \Lambda, j\in \Gamma\}.$ That is \begin{equation}\label{eq T and Phi coincide on the first grid finite dimensional} T(u_{ij}) = \Phi (u_{ij}) \hbox{ for all } i\in \Lambda, j\in \Gamma.
\end{equation}

As in the case in which dim$(H) = \infty$, for each $k\in \{1,\ldots, c_1+1\}$, $H_k$ will denote the closed subspace of $H$ generated by the orthonormal system $\{\xi_j : j \in \Gamma_k\}$. Since dim$(H_k)= \hbox{dim}(K)$ for all $j$, the subtriple $B(H_k, K)$ is isometrically isomorphic to $B(K),$ and thus we can pick a unitary element denoted by $u_k$. We observe that $C_2 (u_k) = B(H_k,K)$.\smallskip

Here we find the dichotomy dim$(K)=2$ and dim$(K)\geq 3$. In the first case we apply Theorem \ref{t biorder preserving from a spin into a Cartan factor} and Lemma \ref{l Peirce 2 for fully non-complete tripotents improved}, while in the second one we use Corollary \ref{c tripotents of rank 3}, to deduce that $\Phi : \mathcal{U} (C_2(u_k)) \to \mathcal{U} (\tilde{C}_2(\Phi(u_k)))$ is a bijection preserving the partial order  in both directions and orthogonality between tripotents, and hence there exists a complex-linear or a conjugate-linear triple isomorphism $T_k : C_2(u_k) \to \tilde{C}_2(\Phi(u_k))$ such that
\begin{equation}\label{eq Phi and Tk coincide on the commond domain finite dimensional} T_k(w) = \Phi(w), \hbox{ for all } w\in \mathcal{U} (C_2(u_k)) \hbox{ ($k\in  \{1,\ldots, c_1+1\}$).}
\end{equation}

The arguments given in the case dim$(H)=\infty$ can be literally followed to conclude that the maps $T_k$ are all complex-linear or conjugate-linear at the same time, and this property is determined by the mapping $f$ given by Proposition \ref{p all maps fu coincide new} which is the same for all $T_k$ and $T$.\smallskip

\emph{Case $II.a$.} By assuming that $f$ is the identity on $\mathbb{T}$, we actually deduce, as in the case in which dim$(H)= \infty$, that
\begin{equation}\label{eq Tk and the restriction on C2ek coindice finite dimensional} \hbox{$T_k = T|_{C_2(u_k)}$ for all $k\in \{1,\ldots, c_1+1\}$,}
 \end{equation} and for each (finite) family of mutually collinear minimal tripotents  $\{ e_j : j \in I\}$ in $C$ and each finite collection $\{ \alpha_j : j \in I\}\subset \mathbb{C}$ with $\displaystyle \sum_{j\in I} |\alpha_j|^2=1$ we have \begin{equation}\label{eq Phi is linear on families of mc min trip finite dimensional} \Phi \left( \sum_{j\in I} \alpha_j e_j \right) =  \sum_{j\in I} \alpha_j \Phi (e_j)
\end{equation} (the validity of the latter formula in the finite dimensional case has been already justified). In order to finish, we observe that any minimal tripotent $e= \eta\otimes \xi$ in $C$ (with $\eta$ and $\xi$ in the unit spheres of $K$ and $H$, respectively) writes as a finite sum of the form $$ e= \sum_{j=1}^{d_2} \alpha_j \eta\otimes \xi_j, \hbox{ where } \sum_{j=1}^{d_2} |\alpha_j|^2=1.$$ The minimal tripotents in the set $\{ \eta\otimes \xi_j : j\in \Gamma\}$ are mutually collinear, and  for each $j \in \Gamma_k$ ($k\in \{1,\ldots, c_1+1\}$), the minimal tripotent $\eta\otimes \xi_j$ belongs to $C_2 (u_k) = B(H_k,K).$ Thus, by applying \eqref{eq Phi is linear on families of mc min trip finite dimensional}, \eqref{eq Tk and the restriction on C2ek coindice finite dimensional} and \eqref{eq Phi and Tk coincide on the commond domain finite dimensional} we get $$\begin{aligned}\Phi(e)&= \sum_{j \in \Gamma} \alpha_j \Phi(\eta\otimes \xi_j) = \sum_{k=1}^{c_1} \sum_{j \in \Gamma_k} \alpha_j \Phi(\eta\otimes \xi_j) + \sum_{j=c_1+1}^{d_2} \alpha_j \Phi(\eta\otimes \xi_j)\\
&= \sum_{k=1}^{c_1} \sum_{j \in \Gamma_k} \alpha_j T_k(\eta\otimes \xi_j) + \sum_{j=c_1+1}^{d_2} \alpha_j T_{c_1+1}(\eta\otimes \xi_j) \\
&= \sum_{k=1}^{c_1} \sum_{j \in \Gamma_k} \alpha_j T(\eta\otimes \xi_j) + \sum_{j=c_1+1}^{d_2} \alpha_j T(\eta\otimes \xi_j) = T(e),
\end{aligned} $$ which concludes the proof in this Case $I.a$. \smallskip

\emph{Case $II.b$.} Finally, if the mapping $f$ is the conjugation on $\mathbb{T}$, an identical argument to that given in the Case $I.b$ works here too.
\end{proof}

\subsection{Type 2 Cartan factors not admitting a unitary element} \ \smallskip

After the results in the previous section, the Cartan factors which are not covered by our conclusions reduce to the exceptional type 5 Cartan factor and those Cartan factor of type 2 which do not admit a unitary element. This section is devoted to present the results in the latter type. We recall first the basic notions, associated with each conjugation $j$ on a complex Hilbert space $H$ we can consider the linear involution on $B(H)$ defined by $x\mapsto x^t:=jx^*j$. A Cartan factor of type 2 is the JB$^*$-subtriple $B(H)_{a}$ of $B(H)$ of all $t$-skew-symmetric operators. As we recalled before, $B(H)_{a}$ admits a unitary element when dim$(H)$ is even or infinite (cf. \cite[Proposition 2]{HoMarPeRu}), so the result is covered by our conclusion in Theorem \ref{t Molnar thm rank 2 Cartan factors with a unitary}. We therefore reduce our study to type 2 Cartan factors $B(H)_{a}$ with dim$(H)$ odd.\smallskip

The next lemma has been borrowed from the proof of \cite[Theorem 4.5]{FerPe18Adv}.

\begin{lemma}\label{l Peirce-1 subspace of a minimal trip in a type 2 Cf is a rectangular type 1 Cf}\cite[proof of Theorem 4.5]{FerPe18Adv} Let $C=B(H)_{a}$ be a type 2 Cartan factor. Suppose $u$ is a minimal tripotent in $B(H)_{sa}$. Then the Peirce-1 subspace $C_1(u)$ is isometrically triple isomorphic to a rectangular type 1 Cartan factor of rank--2, it is actually triple isomorphic to $B(K_1,K_2)$, where $K_1$ and $K_2$ are complex Hilbert spaces with dim$(K_1)=2$ and dim$(K_2)=$dim$(H)-2$.
\end{lemma}

\begin{proof} For the explicit argument see the third paragraph in the proof of \cite[Theorem 4.5]{FerPe18Adv}.
\end{proof}

We have avoided, up to this point, to study the behavior of a bijection preserving the partial ordering in both directions and orthogonality on tripotents belonging to a Peirce-1 subspace associated with a fixed tripotent, like we did for those in the Peirce-0 and Peirce-2 subspaces in Lemma \ref{c first consequences}, Lemma \ref{l Peirce 2 for fully non-complete tripotents improved} and Proposition \ref{p tripotents of rank 3}, respectively.

\begin{lemma}\label{l Peirce 1 for minimal tripotents} Let $\Phi : \mathcal{U}(C) \to \mathcal{U}(\tilde{C})$ be a bijective transformation which preserves the partial ordering in both directions and orthogonality between tripotents, where $C$ and $\tilde{C}$ are Cartan factors with rank bigger than or equal to $2$. Assume that there exists a non-zero tripotent $u\in C$ such that $\Phi|_{\mathbb{T} u}$ is continuous at $u$. Let $e$ be a minimal tripotent in $C$. Then $\Phi$ maps the set $C_1(e) \cap \mathcal{U} (C)$ onto the set $\tilde{C}_1(\Phi(e)) \cap \mathcal{U} (\tilde{C})$.
\end{lemma}

\begin{proof} We observe that $C_1(e)$ need not be a Hilbert space--i.e., a rank--1 factor-- (compare the previous Lemma \ref{l Peirce-1 subspace of a minimal trip in a type 2 Cf is a rectangular type 1 Cf}). Let $v$ be any minimal tripotent in $C_1(e)$. By a new application of \cite[Lemma 3.10]{FerPe18Adv} we deduce that one of the following statements holds:\begin{enumerate}[$(i)$]\item There exist minimal tripotents $e_2,e_3,e_4$ in $C$ such that $(e,e_2,e_3,e_4)$ is a quadrangle and $v$ is a linear combination of $e,e_2,e_3,$ and $e_4$;
\item There exist a minimal tripotent $\tilde e\in C$ and a rank two tripotent $u\in C$ such that $(e, u,\tilde e)$ is a trangle and $v$ is a linear combination of $e, u,$ and $\tilde e$.
\end{enumerate}

In the case $(i)$ (respectively, $(ii)$) we set $w=e+e_3$ (respectively, $w=e+\tilde{e}$). In any of the cases, $w$ is a ran-2 tripotent and $e,v\in C_2(w).$ Lemma \ref{l Peirce 2 for fully non-complete tripotents improved} asserts that $\Phi|_{\mathcal{U} (C_2(w))} : \mathcal{U} (C_2(w))\to \mathcal{U} (\tilde{C}_2 (\Phi(w)))$ is a bijection preserving the local order in both directions and orthogonality between tripotents. The continuity of $\Phi$ forces the mapping $f$ given by Proposition \ref{p all maps fu coincide new} for $\Phi$ to be the identity or the conjugation on $\mathbb{T}$. Theorem \ref{t biorder preserving from a spin into a Cartan factor} proves the existence of a complex-linear or conjugate-linear triple isomorphism $\tilde{U}: C_2(w) \to  \tilde{C}_2 (\Phi(w))$ whose restriction to $\mathcal{U} (C_2(w))$ coincides with $\Phi|_{\mathcal{U} (C_2(w))}$, the rest is clear because $\tilde{U}$ is a triple isomorphism and $e,v\in C_2(w)$.
\end{proof}

We are now in a position to describe the bijections preserving the partial ordering in both directions and orthogonality between tripotents from the set of tripotents of a type 2 Cartan factor not admitting a unitary element onto the set of tripotents of any other Cartan factors. As a novelty in the proof of this result, we do not employ the linear extension provided by a grid and its image. Instead of that we shall define a complex-linear or a conjugate-linear extension based on our new understanding of the bijection on the the tripotents belonging to the Peirce subspaces associated with a minimal tripotent.

\begin{theorem}\label{t type 2 CF odd dimensional} Let $C= B(H)_{a}$ be a type 2 Cartan factor, where $H$ is a finite dimensional complex Hilbert space with odd dimension, and let $\tilde{C}$ be another Cartan factor. Suppose $\Phi : \mathcal{U}(C) \to \mathcal{U}(\tilde{C})$ is a bijective transformation which preserves the partial ordering in both directions and orthogonality between tripotents. Assume additionally that there exists a non-zero tripotent $u\in C$ such that $\Phi|_{\mathbb{T} u}$ is continuous at $u$. Then there exists a complex-linear or conjugate-linear {\rm(}isometric{\rm)} triple isomorphism $T: C\to \tilde{C}$ such that $\Phi (e) = T(e)$ for all $e\in \mathcal{U} (C)$.
\end{theorem}

\begin{proof} We can clearly assume that dim$(H)\geq 5$ as it is well known that $ B(H)_{a}$ is isometrically isomorphic to a complex Hilbert space (it has rank--1). Let us begin with a simple observation, the hypothesis concerning the continuity of $\Phi$ assures that the mapping $f$ given by Proposition \ref{p all maps fu coincide new} for $\Phi$ is the identity or the conjugation on $\mathbb{T}$. In the second case, we can argue as in the final step of the proof Theorem \ref{t BHK rectangular} and composing $\Phi$ is a conjugate-linear triple automorphism on $C$ to lead ourself to te first case. So, we assume that the mapping $f$ is the identity on $\mathbb{T}$ and we shall find a complex-linear extension.\smallskip

Let us fix a minimal tripotent $u$ in $C$ and the Peirce decomposition \begin{equation}\label{eq Peirce decomposition in the type 2 theorem} C = C_2(u) \oplus C_1(u)\oplus C_0(u) = \mathbb{C} u \oplus C_1(u)\oplus C_0(u).
 \end{equation} There is no loos of generality in assuming that $u = \xi_0 \otimes \eta_0 - \eta_0\otimes \xi_0$, where $\xi_0,\eta_0$ are two orthonormal vectors in $H$ with $j(\eta_0) = \eta_0$, $j(\xi_0)= \xi_0$, where $j$ is the involution on $H$ employed for the definition of $B(H)_a=\{ a\in B(H) : a^t = j a^* j =-a\}$. Let $p_0$ denote the minimal projection $\xi_0\otimes\xi_0$ in $B(H)$, and let $H_0:=(1-p_0) (H)$. Clearly dim$(H_0) =$dim$(H)-1$ is even. Since $p_0^t = j p_0^* j = j \xi_0\otimes\xi_0 j = j(\xi_0)\otimes j(\xi_0)= \xi_0\otimes\xi_0 = p_0$ and $(1-p_0)^t = 1-p_0$. In particular, $H_0$ is a fixed subspace for $j$ and hence $\hat{\jmath} = j|_{H_0}$ is a conjugation on $H_0$. The subspace $D = \{a \in C : a = (1-p_0) a (1-p_0)\}$ is a JB$^*$-subtriple of $C$ which is isometrically triple isomorphic to the type 2 Cartan factor $$B(H_0)_{a}^{\hat{\jmath}} = \{ a\in B(H_0) : a^{\hat{t}} =\hat{\jmath} a^* \hat{\jmath} = -a\}.$$ Furthermore, the Peirce-0 subspace $C_0(u)$ is a JB$^*$-subtriple of $D\cong B(H_0)_{a}^{\hat{\jmath}}.$\smallskip

Now, we apply Lemma \ref{l Peirce 1 for minimal tripotents} to deduce that $\Phi|_{\mathcal{U}(C_1(u))} : \mathcal{U}(C_1(u)) \to \mathcal{U}(\tilde{C}_1(\Phi(u)))$ is a bijection preserving the partial ordering in both directions and orthogonality between tripotents. We observe that the mapping $f$ given by Proposition \ref{p all maps fu coincide new} for $\Phi|_{\mathcal{U}(C_1(u))}$ coincides with the one given for $\Phi,$ and hence it is the identity on $\mathbb{T}$. Since, by Lemma \ref{l Peirce-1 subspace of a minimal trip in a type 2 Cf is a rectangular type 1 Cf}, $C_1(u)$ is a rank--2 rectangular type 1 Cartan factor, it follows from Theorem \ref{t BHK rectangular} that there exists a complex-linear triple isomorphism \begin{equation}\label{eq def T1} \hbox{$T_1: C_1(u) \to \tilde{C}_1(\Phi(u))$, such that $T_1|_{\mathcal{U}(C_1(u))}$ coincides with $\Phi|_{\mathcal{U}(C_1(u))}$.}
\end{equation}

On the other hand, since dim$(H_0)$ is even and $D\cong B(H_0)_{a}^{\hat{\jmath}}$, we can find a unitary tripotent $u_2\in D$ (cf. \cite[Proposition 2]{HoMarPeRu}). In this case $D= C_2(u_2)$, and hence a combination of Lemma \ref{l Peirce 2 for fully non-complete tripotents improved} and Proposition \ref{p tripotents of rank 3} guarantees that $\Phi|_{\mathcal{U}(C_2(u_2))} : \mathcal{U}(C_2(u_2)) \to \mathcal{U}(\tilde{C}_2(\Phi(u_2)))$ is a bijection preserving the partial ordering in both directions and orthogonality between tripotents (and the mapping $f$ associated with it by Proposition \ref{p all maps fu coincide new} is the identity on $\mathbb{T}$). Since $C_2(u_2)$ admits a unitary element and has rank $\geq 2,$ Theorem \ref{t Molnar thm rank 2 Cartan factors with a unitary} implies the existence of a complex-linear (isometric) triple isomorphism \begin{equation}\label{eq def T2} \hbox{$T_2:  C_2(u_2)\to \tilde{C}_2(\Phi(u_2))$ satisfying $\Phi (w) = T_2(w)$ for all $w\in \mathcal{U}(C_2(u_2))$.}
\end{equation}

Since, by \eqref{eq Peirce decomposition in the type 2 theorem}, every element $x$ in $C$ decomposes uniquely in the form $x = \lambda u + P_1 (u) (x) + P_0(u) (x)$ and $P_0(u) (x) \in C_0(u)\subseteq D= C_2(u_2) \cong B(H_0)_{a}^{\hat{\jmath}},$ we can define a linear mapping $T: C\to \tilde{C}$ given by \begin{equation}\label{eq def of the linear mapping T in the type 2} T (x) = T(\lambda u + P_1 (u) (x) + P_0(u) (x)) = \lambda \Phi(u) + T_1(P_1 (u) (x)) + T_2 (P_0(u) (x)).
\end{equation} Clearly, $T$ is weak$^*$ continuous, and in order to show that $T(w) = \Phi(w)$ for all $w\in \mathcal{U} (C)$ it suffices to prove that $T(e) = \Phi(e)$ for every minimal tripotent $e$ in $C$ (cf. Proposition \ref{p identity principle}).\smallskip

Let us take a minimal tripotent $e\in C$. Lemma 3.10 in \cite{FerPe18Adv} assures that one of the following properties  holds: \begin{enumerate}[$(i)$]\item There exist minimal tripotents $u_2,u_3,u_4$ in $C$ such that $(u,u_2,u_3,u_4)$ is a quadrangle and $e$ is a linear combination of $u,u_2,u_3,$ and $u_4$, that is, $e = \alpha u + \beta u_2 + \gamma u_3 + \delta u_4$;
\item There exist a minimal tripotent $\tilde u\in C$ and a rank two tripotent $v\in C$ such that $(u, v,\tilde u)$ is a trangle and $e$ is a linear combination of $u, v,$ and $\tilde u$, that is $e = \alpha u + \beta v + \gamma \tilde{u}$.
\end{enumerate}

Each case must be treated independently. In $(i)$ we consider the rank--2 tripotent $u+u_3$ and the Peirce-2 subspace $C_2(u+u_3)$, which contains $u$ and $e$. We deduce from Theorem \ref{t Molnar thm rank 2 Cartan factors with a unitary} the existence of a complex-linear isometric triple isomorphism $R: C_2(u+u_3) \to \tilde{C}_2(\Phi(u+u_3))$ whose restriction to $\mathcal{U} (C_2(u+u_3))$ coincides with $\Phi|_{\mathcal{U} (C_2(u+u_3))}$ (recall that the mapping $f$ given by Proposition \ref{p all maps fu coincide new} for the mapping $\Phi|_{\mathcal{U} (C_2(u+u_3))}$ must coincide with the corresponding mapping for $\Phi$, and thus it is the identity). Therefore, having in mind that $e\in C_2(u+u_3)$, $u_2,u_4\in C_1(u)$ and $u_3\in C_0(u) \subseteq D= C_2(u_2)$, we deduce from the linearity of $R$ that
$$\begin{aligned}
\Phi (e) &=  R(e) = \alpha R(u)  + \beta R(u_2) + \gamma R(u_3) + \delta R(u_4)\\
&= \alpha \Phi(u)  + \beta \Phi(u_2) + \gamma \Phi(u_3) + \delta \Phi(u_4)\\
&= \alpha \Phi(u)  + \beta T_1 (u_2) + \gamma T_2 (u_3) + \delta T_1 (u_4) \\
&= \alpha T(u)  + \beta T(u_2) + \gamma T(u_3) + \delta T(u_4) = T(e),
\end{aligned}$$ where at the antepenultimate equality we applied \eqref{eq def T1} and \eqref{eq def T2}, and at the penultimate and last equalities the definition of $T$.\smallskip

In $(ii)$ we consider the rank--2 tripotent $u+\tilde{u}$ and the Peirce-2 subspace $C_2(u+\tilde{u})$, which contains $u$ and $e$. Theorem \ref{t Molnar thm rank 2 Cartan factors with a unitary} proves the existence of a complex-linear isometric triple isomorphism $R: C_2(u+\tilde{u}) \to \tilde{C}_2(\Phi(u+\tilde{u}))$ whose restriction to $\mathcal{U} (C_2(u+\tilde{u}))$ coincides with $\Phi|_{\mathcal{U} (C_2(u+\tilde{u}))}$ (recall that the mapping $f$ given by Proposition \ref{p all maps fu coincide new} for the mapping $\Phi|_{\mathcal{U} (C_2(u+\tilde{u}))}$ coincides with the corresponding mapping for $\Phi$, and thus it is the identity). Since $e\in C_2(u+\tilde{u})$, $v\in C_1(u)$ and $\tilde{u}\in C_0(u) \subseteq D= C_2(u_2)$, it follows from the linearity of $R$ that
$$\begin{aligned}
\Phi (e) &=  R(e) = \alpha R(u) + \beta R(v) + \gamma R(\tilde{u}) = \alpha \Phi(u) + \beta \Phi(v) + \gamma \Phi(\tilde{u}) \\
&= \alpha \Phi(u) + \beta T_1(v) + \gamma T_2(\tilde{u}) = \alpha T(u)  + \beta T(v) + \gamma T(\tilde{u}) = T(e),
\end{aligned}$$ where, as before, we applied \eqref{eq def T1}, \eqref{eq def T2} and \eqref{eq def of the linear mapping T in the type 2}, and the definition of $T$. This concludes the proof.
\end{proof}

\subsection{Exceptional type 5 Cartan factors}\ \smallskip

The unique Cartan factor not covered by the results in previous sections and subsections is the exceptional Cartan factor of type 5.  We shall see how the proof given in Theorem \ref{t type 2 CF odd dimensional} can be adapted after an appropriate modification and a more sophisticated structure results of the Peirce-1 subspace associated with a minimal tripotent in a type 5 Cartan factor taken from \cite{Fri2005}.

\begin{theorem}\label{t exceptional type 5 CF} Let $C$ be a type 5 Cartan factor, and let $\tilde{C}$ be another Cartan factor. Suppose $\Phi : \mathcal{U}(C) \to \mathcal{U}(\tilde{C})$ is a bijective transformation which preserves the partial ordering in both directions and orthogonality between tripotents. Assume additionally that there exists a non-zero tripotent $u\in C$ such that $\Phi|_{\mathbb{T} u}$ is continuous at $u$. Then there exists a complex-linear or conjugate-linear {\rm(}isometric{\rm)} triple isomorphism $T: C\to \tilde{C}$ such that $\Phi (e) = T(e)$ for all $e\in \mathcal{U} (C)$.
\end{theorem}

\begin{proof} As in previous cases (see Theorems \ref{t BHK rectangular} and \ref{t type 2 CF odd dimensional}), the hypothesis concerning the continuity of $\Phi$ assures that the mapping $f$ given by Proposition \ref{p all maps fu coincide new} for $\Phi$ is the identity or the conjugation on $\mathbb{T}$. We can always reduce to the first case by just composing $\Phi$ with a conjugate-linear triple automorphism on $C$ whose existence is well known, for example, from \cite{Ka97} and \cite{loos1977bounded}. We can therefore assume that the mapping $f$ is the identity on $\mathbb{T},$ and it suffices to find a complex-linear extension.\smallskip

Let us fix a minimal tripotent $u$ in $C$. By the structure results in \cite[Section 6.3.7, page 264]{Fri2005}, the Peirce-1 subspace $C_1(u)$ is isometrically triple isomorphic to the type 2 Cartan factor of all $5\times 5$ antisymmetric matrices (which is the only type 2 Cartan factor of rank 2 which is not a spin factor). We deduce from Lemma \ref{l Peirce 1 for minimal tripotents} that  $\Phi$ maps $C_1(u) \cap \mathcal{U} (C)$ onto the set $\tilde{C}_1(\Phi(u)) \cap \mathcal{U} (\tilde{C})$, and hence Theorem \ref{t type 2 CF odd dimensional} applied to the bijection $\Phi|_{C_1(u) \cap \mathcal{U} (C)} : C_1(u) \cap \mathcal{U} (C)\to \tilde{C}_1(\Phi(u)) \cap \mathcal{U} (\tilde{C})$ proves the existence of a complex-linear triple isomorphism $T_1: C_1(u) \to \tilde{C}_1(\Phi(u))$ whose restriction to  $C_1(u) \cap \mathcal{U} (C)$ is precisely $\Phi|_{C_1(u) \cap \mathcal{U} (C)}$, that is \begin{equation}\label{eq Phi and T1 coincide on Peirce1 of u type 5} \Phi (w) = T_1(w), \hbox{ for all } w\in  C_1(u) \cap \mathcal{U} (C).
\end{equation}

The Peirce-0 subspace is treated next. It is well known that $C$ has rank--2 (cf. \cite{loos1977bounded} or \cite[Table 1 in page 210]{Ka97}). It follows from this fact that the JBW$^*$-subtriple $C_0(u)$ must have rank--1, and thus it is isometrically tripe isomorphic to complex Hilbert space. The same argument shows that every norm-one element in in the Hilbert space $C_0(u)$ is a minimal tripotent in $C$. Furthermore, since $C_2(u) = \mathbb{C} u$ and dim$(C_1(u)) = 10$, we can actually affirm that $C_0(u)$ is a triple isomorphic to a 5-dimensional complex Hilbert space. By Lemma \ref{c first consequences}$(c)$, $\Phi$ maps $C_0(u) \cap \mathcal{U} (C)$ onto the set $\tilde{C}_0(\Phi(u)) \cap \mathcal{U} (\tilde{C})$, however since the involved JBW$^*$-triples have rank--1 we do not have intrinsic tools to work with the mapping $\Phi|_{C_0(u) \cap \mathcal{U} (C)}$ (cf. Remark \ref{r rank 1 for op bijections}). We shall see how to avoid this difficulty. Let us first two orthogonal (in the Hilbert sense), norm-one elements $e,v$ in the Hilbert space $C_0(u)$. We claim that \begin{equation}\label{eq Hilbert combination in the sphere type 5} \Phi (\alpha e +\beta v) = \alpha \Phi(e) +\beta \Phi(v), \hbox{ for all } \alpha,\beta\in \mathbb{C} \hbox{ with } |\alpha|^2 + |\beta|^2 =1.
\end{equation}

Clearly, $e,$ $v$ and $\alpha e +\beta v$ are minimal tripotents in $C$. By \cite[Lemma 3.10]{FerPe18Adv} one of the following statements holds:\begin{enumerate}[$(i)$]\item There exist minimal tripotents $e_2,e_3,e_4$ in $C$ such that $(e,e_2,e_3,e_4)$ is a quadrangle and $v$ is a linear combination of $e,e_2,e_3,$ and $e_4$;
\item There exist a minimal tripotent $\tilde e\in C$ and a rank two tripotent $w\in C$ such that $(e, w,\tilde e)$ is a trangle and $v$ is a linear combination of $e, w,$ and $\tilde e$.
\end{enumerate}

We shall deal with both cases in parallel. If $(i)$ (respectively, $(ii)$) holds we set $\hat{e}=e+e_3$ (respectively, $\hat{e}=e+\tilde{e}$). In any of the cases, $\hat{e}$ is a ran-2 tripotent and $e,v\in C_2(\hat{e}).$ Lemma \ref{l Peirce 2 for fully non-complete tripotents improved} asserts that $\Phi|_{\mathcal{U} (C_2(\hat{e}))} : \mathcal{U} (C_2(\hat{e}))\to \mathcal{U} (\tilde{C}_2 (\Phi(\hat{e})))$ is a bijection preserving the local order in both directions and orthogonality between tripotents. The continuity of $\Phi$ forces the mapping $f$ given by Proposition \ref{p all maps fu coincide new} for $\Phi$ to be the identity or the conjugation on $\mathbb{T}$. Theorem \ref{t biorder preserving from a spin into a Cartan factor} proves the existence of a complex-linear or conjugate-linear triple isomorphism $\tilde{U}: C_2(\hat{e}) \to  \tilde{C}_2 (\Phi(\hat{e}))$ whose restriction to $\mathcal{U} (C_2(\hat{e}))$ coincides with $\Phi|_{\mathcal{U} (C_2(\hat{e}))}$. The mapping $\tilde{U}$ must be complex-linear because the mapping $f$ asociated with $\Phi$ has been assumed to be the identity on $\mathbb{T}$. Since $e,v, \alpha e +\beta v \in C_2(\hat{e}),$ we deduce that $$\Phi (\alpha e +\beta v ) = \tilde{U} (\alpha e +\beta v) = \alpha \tilde{U}(e) +\beta\tilde{U} (v) = \alpha \Phi(e) +\beta\Phi(v),$$ which proves the claim in \eqref{eq Hilbert combination in the sphere type 5}.\smallskip

Now let $\{e_1,e_2,e_3,e_4,e_5\}$ be an orthonormal basis of the Hilbert space $C_0(u)$. By applying \eqref{eq Hilbert combination in the sphere type 5}, we can deduce in a finite number of steps, like in the proofs of Theorem \ref{t spin order 2Good}\eqref{eq Hilbert sums in the sphere spin} or Theorem \ref{t BHK rectangular}\eqref{eq Phi is linear on families of mc min trip}, that \begin{equation}\label{eq Phi is linear on an ob of the Hilbert C0u} \Phi \left( \sum_{j=1}^5 \alpha_j e_j \right) =  \sum_{j=1}^5 \alpha_j \Phi (e_j),
\end{equation} whenever $\alpha_1,\ldots, \alpha_5\in \mathbb{C}$ with $\displaystyle \sum_{j=1}^5 |\alpha_j|^2=1.$  We consider the linear mapping $T_0 : C_0(u) \to \tilde{C}_0(\Phi(u))$ define on the orthonormal basis $\{e_1,e_2,e_3,e_4,e_5\}$ by $$T_0\left( \sum_{j=1}^5 \beta_j e_j\right) = \sum_{j=1}^5 \beta_j \Phi(e_j).$$ Since non-zero tripotents in the Hilbert space are precisely the elements in its sphere, the conclusion in \eqref{eq Phi is linear on an ob of the Hilbert C0u} is equivalent to say that \begin{equation}\label{eq Phi and T0 coincide on tripotents} \Phi (w ) = T_0 (w), \hbox{ for all tripotent } w\in C_0(u).
\end{equation}

We are in a position to employ a similar argument to that in the final paragraphs of the proof of Theorem \ref{t type 2 CF odd dimensional}. The mapping $T: C\to \tilde{C}$ given by \begin{equation}\label{eq def of the linear mapping T in the type 5} T (x) = T(\lambda u + P_1 (u) (x) + P_0(u) (x)) = \lambda \Phi(u) + T_1 (P_1 (u) (x)) + T_0 (P_0(u) (x)).
\end{equation} is linear and well defined because $C = \mathbb{C} u \oplus C_1(u) \oplus C_0(u)$. We shall finish as soon as we prove that $T(w) = \Phi(w)$ for all $w\in \mathcal{U} (C).$ By Proposition \ref{p identity principle}, it suffices to prove that $T(e) = \Phi(e)$ for every minimal tripotent $e$ in $C$. This can be done by a new application of \cite[Lemma 3.10]{FerPe18Adv} and repeating exactly the same arguments given at the end of the proof of Theorem \ref{t type 2 CF odd dimensional} with \eqref{eq Phi and T1 coincide on Peirce1 of u type 5}, \eqref{eq Phi and T0 coincide on tripotents} and \eqref{eq def of the linear mapping T in the type 5} in the roles of \eqref{eq def T1}, \eqref{eq def T2} and \eqref{eq def of the linear mapping T in the type 2}.
\end{proof}

\section{Main conclusions}\label{sec: conclusions}

This section is thought to reorganize the material obtained in previous results with the aim of presenting the main conclusion of this note which is  the next description of the bijections preserving the partial ordering in both directions and orthogonality between tripotents of two atomic JBW$^*$-triples.

\begin{theorem}\label{t main atomic JBWtriples} Let $\displaystyle M= \bigoplus_{i\in I}^{\ell_{\infty}} C_i$ and  $\displaystyle N = \bigoplus_{j\in J}^{\ell_{\infty}} \tilde{C}_j$ be atomic JBW$^*$-triples, where $C_i$ and $C_j$ are Cartan factors with rank $\geq 2$. Suppose that $\Phi : \mathcal{U}(M) \to \mathcal{U}(N)$ is a bijective transformation which preserves the partial ordering in both directions and orthogonality between tripotents. We shall additionally assume that $\Phi$ is continuous at a tripotent $u = (u_i)_i$ in $M$ with $u_i\neq 0$ for all $i$ {\rm(}or we shall simply assume that $\Phi|_{\mathbb{T} u}$ is continuous at a tripotent $(u_i)_i$ in $M$ with $u_i\neq 0$ for all $i${\rm)}. Then there exists a real linear triple isomorphism $T: M\to N$ such that $T(w) = \Phi(w)$ for all $w\in \mathcal{U} (M)$. Furthermore, $M$ decomposes as the direct sum of two orthogonal weak$^*$-closed ideals $M_1$ and $M_2$ such that $T|_{M_1}$ is complex-linear and $T|_{M_2}$ is conjugate-linear.
\end{theorem}

\begin{proof} Let us begin with an observation $\mathcal{U} (M) =\{ (u_i)_i : u_i\in \mathcal{U}(C_i)\}$. By Lemma \ref{l Peirce 2 for fully non-complete tripotents}$(a)$, for each $i\in I$ there exists a unique $\sigma(i)\in J$ such that $\Phi (\mathcal{U}(C_i)) = \mathcal{U}(\tilde{C}_{\sigma{i}})$. Applying the same argument to $\Phi^{-1}$ it can be deduced that $\sigma: I\to J$ is bijection. Clearly, $\Phi|_{\mathcal{U}(C_i)} : \mathcal{U}(C_i)\to \mathcal{U}(\tilde{C}_{\sigma(i)})$ is a bijection preserving the partial order in both directions and orthogonality between tripotents. Since $C_i$ and $\tilde{C}_{\sigma(i)}$ are Cartan factors we can apply the conclusions in Theorems \ref{t biorder preserving from a spin into a Cartan factor}, \ref{t Molnar thm rank 2 Cartan factors with a unitary}, \ref{t BHK rectangular}, \ref{t type 2 CF odd dimensional} and \ref{t exceptional type 5 CF} to deduce the existence of a complex-linear or conjugate-linear triple isomorphism $T_i: C_i\to \tilde{C}_{\sigma(i)}$ whose restriction to $\mathcal{U} (C_i)$ is precisely $\Phi|_{\mathcal{U}(C_i)}$. It is easy to see that the mapping $T : M\to N,$ $T((x_i)_i)  = (T_i(x_i))_i$ is a real-linear triple isomorphism whose restriction to $\mathcal{U} (M)$ is $\Phi$.\smallskip

Finally taking $I_1 :=\{i\in I : T_i \hbox{ is complex-linear}\},$  $I_2 = I\backslash I_1$ and $\displaystyle M_j := \bigoplus_{i\in I_j}^{\ell_{\infty}} C_i$ ($j=1,2$), it is clear that $M_1$ and $M_2$ are orthogonal weak$^*$ closed ideals with $M= M_1\oplus M_2$, $T|_{M_1}$ is complex-linear and $T|_{M_2}$ is conjugate-linear.
\end{proof}

The next corollary is a straight consequence of our main result, however, it could be also derived from Theorem \ref{t Molnar thm rank 2 Cartan factors with a unitary}. The result is interesting by itself for those researchers working on JB$^*$-algebras.

\begin{corollary}\label{t main atomic JBWalgebras} Let $M$ and $N$ be atomic JBW$^*$-algebras. Suppose that $\Phi : \mathcal{U}(M) \to \mathcal{U}(N)$ is a bijective transformation which preserves the partial ordering in both directions and orthogonality between tripotents. We shall additionally assume that $\Phi$ is continuous at a tripotent $u = (u_i)_i$ in $M$ with $u_i\neq 0$ for all $i$ {\rm(}or we shall simply assume that $\Phi|_{\mathbb{T} u}$ is continuous at a tripotent $(u_i)_i$ in $M$ with $u_i\neq 0$ for all $i${\rm)}. Then there exists a real linear triple isomorphism $T: M\to N$ such that $T(w) = \Phi(w)$ for all $w\in \mathcal{U} (M)$. Furthermore, $M$ decomposes as the direct sum of two orthogonal weak$^*$-closed ideals $M_1$ and $M_2$ such that $T|_{M_1}$ is complex-linear and $T|_{M_2}$ is conjugate-linear.
\end{corollary}

Wigner and Uhlhorn theorems, as well as the Bunce--Wright--Mackey--Gleason theorem, are milestone and influencing results with mathematics and physics. 
We would like to conclude this paper by contrasting our results with a generalizations of the Mackey--Gleason theorem for rectangular JBW$^*$-triples established by C.M. Edwards and G.T. R\"{u}ttimann in \cite{EdRutt99}.\smallskip

As it is well known, W$^*$-algebras are commonly used as a mathematical model for a statistical quantum-mechanical system, the bounded observables of which are represented by self-adjoint elements of a von Neumann algebra $W$, and the propositions are represented by elements of the complete orthomodular
lattice $\mathcal{P}(W)$ of projections in $W$. In this more traditional approach, states of the system are
represented by bounded finitely ortho-additive measures on $\mathcal{P}(W)$. The commented Bunce--Wright--Mackey--Gleason theorem affirms that, provided that $W$ does not contain $M_2(\mathbb{C})$ as a weak$^*$-closed ideal, the states of the system are the restrictions of bounded linear functionals on $W$.\smallskip

In an alternative approach, proposed by Gell-Mann and Hartle, states are represented by measures on the lattice $\mathcal{CP}(W)$ of centrally equivalent pairs of projections in a von Neumann algebra $W$ (\cite{Isham1994,IshamLinden1994,IshamLindenSchreckenberg1994,Wright1995}). A function $d: \mathcal{P}(W)\times \mathcal{P}(W)\to \mathbb{C}$ is called a \emph{decoherence functional} if it satisfies the following axioms: \begin{enumerate}[$(i)$]\item $d(p_1 + p_2, q) = d(p_1 , q) + d(p_2, q)$ whenever $p_1$ is orthogonal to $p_2$;
\item $d(p,q) = \overline{d(q,p)},$ for every $p$ and $q$;
\item $d(p, p)\geq 0,$ for each p;
\item $d(1, 1)=1$.
\end{enumerate} A detailed study of decoherence functionals for $B(H)$ for a finite-dimensional complex Hilbert space $H$ with dim$(H)\geq 3$, was conducted by Isham, Linden and Schreckenberg \cite{Isham1994,IshamLinden1994,IshamLindenSchreckenberg1994}. J.D.M. Wright proved in \cite{Wright1995} that when $d$ is a bounded decoherence functional associated with a von Neumann algebra $W$, then, provided $W$ has no direct summand of type $I_2$, $d$ can be represented as the difference between semi-innerproducts on $W$. C.M. Edwards and G.T. R\"{u}ttimann established in \cite[\S 6]{EdRutt98} that the bounded measures on $\mathcal{CP}(W)$ are the restrictions of bounded sesquilinear forms, or decoherence functionals, on $W\times W$, unless $W$ has a central projection $e$ such that $e W$ is isomorphic to $M_2(\mathbb{C})$. When $W$ is a type $I$ factor with dim$(H)\geq 3$, the second approach subsumes the first. For a system represented by a rectangular JBW$^*$-triple $M$ (i.e., a JBW$^*$-triple for which there exists a von Neumann algebra $W$ and a pair $p,q$ of projections in $A$ such that $M$ is isomorphic to the JBW$^*$-triple $pW q$), and we assume that $M\cong pW q$, where $p$ and $q$ are not equal, the first approach is not available. However, it is possible to appeal to the second approach. The properties of the complete lattice $\mathcal{CP}(W)_{(p,q)}$ of centrally equivalent pairs of projections in the W$^*$-algebra $W$ dominated by $(p, q)$ are examined in \cite{EdRutt98}, and in \cite{EdRutt99} it is raised the question of determining those bounded measures $m$ on $\mathcal{CP}(W)_{(p,q)}$ which have the property that, for each pair $(e_1, f_1)$, $(e_2, f_2)$ of either centrally orthogonal or rigidly collinear elements of $\mathcal{CP}(W)_{(p,q)}$ we have  $$m((e_1, f_1) \vee (e_2, f_2)) = m(e_1, f_1)+  m(e_2, f_2).$$ The main result in \cite{EdRutt99} proves that, provided that neither of the von Neumann algebras $p W p$ or $q W q$ has a direct summand of type $I_2$, such measures are the restrictions of a particular class of bounded sesquilinear functionals on $pW p \times q W q$. In the case in which $p$ and $q$ coincide, these are the decoherence functionals, mentioned above (cf. \cite[Lemma 5.1 and Theorems 5.2 and 5.3]{EdRutt99}).\smallskip

In contrast to the result by Edwards and Ruttimann, in our Theorem \ref{t main atomic JBWtriples} we work with the poset of all tripotents in a Cartan factor or in an atomic JBW$^*$-triple equipped with its intrinsic partial order and orthogonality without alluding to an external superstructure or von Neumann algebra containing it. Our result is then closer to the original aim in Wigner's and Uhlhorn's theorems.

\medskip\medskip

\textbf{Acknowledgements} Second author was partially supported by MCIN / AEI / 10. 13039 / 501100011033 / FEDER ``Una manera de hacer Europa'' project no. PGC2018-093332-B-I00, Junta de Andaluc\'{\i}a grants FQM375, A-FQM-242-UGR18 and PY20$\underline{\ }$ 00255, and by the IMAG--Mar{\'i}a de Maeztu grant CEX2020-001105-M / AEI / 10.13039 / 501100011033.\smallskip\smallskip

We deeply appreciate the useful and detailed suggestions and comments received from the anonymous referees of this paper. Special thanks are given to one of the reviewers for pointing out a mistake in one of the statements of an earlier version.\smallskip\smallskip

\textbf{Data availability} Data sharing is not applicable to this article as no new data were created or analyzed in this study. 


\smallskip



\begin{thebibliography}{0}

\bibitem{AlfShulStor78} E.M. Alfsen, F.W. Shultz, E. St{\o}rmer, A Gel'fand-Neumark theorem for Jordan algebras, \emph{Advances in Math.} \textbf{28}, no. 1, 11--56 (1978).

\bibitem{ArazFri78} J. Arazy, Y. Friedman, Contractive projections in $C_1$ and $C_\infty $, \emph{Memoirs of the American Mathematical Society} \textbf{200}, 165 pages (1978).

\bibitem{Ballentine} L.E. Ballentine, \emph{Quantum Mechanics. A Modern Development}, 2nd ed., World Scientific (2015).

\bibitem{Bargmann64} V. Bargmann, Note on Wigner's theorem on symmetry operations, \emph{J. Mathematical Phys.} \textbf{5}, 862--868 (1964).

\bibitem{BarDanHor88} T.J. Barton, T. Dang, G. Horn, Normal representations of Banach Jordan triple systems, \emph{Proc. Amer. Math. Soc.} \textbf{102}, no. 3, 551--555 (1988).

\bibitem{BarTi} T. Barton, R. M. Timoney, Weak$^*$-continuity of Jordan triple products and its applications, \emph{Math. Scand.} \textbf{59}, 177--191 (1986).

\bibitem{BarviHam2017} J. Barv{\'i}nek, J. Hamhalter, Linear algebraic proof of Wigner theorem and its consequences, \emph{Math. Slovaca} \textbf{67}, no. 2, 371--386 (2017).

\bibitem{Batt91} M. Battaglia, Order theoretic type decomposition of JBW$^*$-triples, \emph{Quart. J. Math. Oxford Ser.} (2) \textbf{42}, no. 166, 129--147 (1991).


\bibitem{BeLoPeRo} J. Becerra Guerrero, G. L{\'o}pez P{\'e}rez, A. M. Peralta, A. Rodr{\'\i}guez-Palacios, A. Relatively weakly open sets in closed balls of Banach spaces, and real ${\rm JB}^*$-triples of finite rank, \emph{Math. Ann.} \textbf{330},  no. 1, 45--58 (2004).

\bibitem{BraKaUp78} R. Braun, W. Kaup, H. Upmeier, A holomorphic characterisation of Jordan-C$^*$-algebras, \emph{Math. Z.} \textbf{161}, 277--290 (1978).

\bibitem{BuChu92} L.J. Bunce, Ch.-H. Chu,  Compact  operations, multipliers and Radon-Nikodym property in JB$^*$-triples, \emph{Pacific J. Math.} \textbf{153}, 249--265 (1992).

\bibitem{BuWri89} L.J. Bunce, J.D.M. Wright, Continuity and linear extensions of quantum measures on Jordan operator algebras, \emph{Math. Scand.} \textbf{64} 300--306 (1989).

\bibitem{BuWri92} L.J. Bunce, J.D.M. Wright, The Mackey--Gleason problem, \emph{Bull. Amer. Math. Soc.} \textbf{26}, 288--293 (1992).

\bibitem{BuWri94} L.J. Bunce, J.D.M. Wright, The Mackey--Gleason problem for vector measures on projections in von Neumann algebras, \emph{J. London Math. Soc.} \textbf{49}, 133--149 (1994).

\bibitem{BurFerGarMarPe} M. Burgos, F.J. Fern{\'a}ndez-Polo, J. Garc{\'e}s, J. Mart{\'\i}nez, A.M. Peralta, Orthogonality preservers in
C$^*$-algebras, JB$^*$-algebras and JB$^*$-triples, \emph{J. Math. Anal. Appl.} \textbf{348}, 220--233 (2008).

\bibitem{BurGarPe11} M. Burgos, J. Garc{\'e}s, A.M. Peralta, Automatic continuity of biorthogonality preservers between weakly compact JB$^*$-triples and atomic JBW$^*$-triples, \emph{Studia Math.} \textbf{204}, no. 2, 97--121 (2011).

\bibitem{CasdeVilahtiLevrero97} G. Cassinelli, E. de Vito, P. Lahti, A. Levrero, Symmetry groups in quantum mechanics and the theorem of Wigner on the symmetry transformations, \emph{Rev. Mat. Phys.} \textbf{8}, 921--941 (1997).

\bibitem{Chev2007} G. Chevalier, Wigner's theorem and its generalizations. In \emph{Handbook of quantum logic and quantum structures}, 429-475, Elsevier Sci. B.V., Amsterdam, 2007.


\bibitem{Cohen-Tannoudji} C. Cohen-Tannoudji, \emph{Quantum Mechanics}, Vol. 2 John Wiley \& sons, Hermann (2005).


\bibitem{DanFri87} T. Dang, Y. Friedman, Classification of JBW$^*$-triple factors and applications, \emph{Math. Scand.} \textbf{61}, no. 2, 292--330 (1987).



\bibitem{DymMckean72} H. Dym, H.P. McKean, \emph{Fourier series and integrals}. Probability and Mathematical Statistics, No. 14. Academic Press, New York-London, 1972.

\bibitem{EdRutt88} C.M. Edwards, G.T. R\"{u}ttimann, On the facial structure of the unit balls in a JBW$^*$-triple and its predual, \emph{J. Lond. Math. Soc.} \textbf{38}, 317--332 (1988).

\bibitem{EdRutt98} C.M. Edwards, G.T. R\"{u}ttimann, The lattice of weak*-closed inner ideals in a W$^*$-algebra, \emph{Comm. Math. Phys.} \textbf{197}, 131--166 (1998).

\bibitem{EdRutt99} C.M. Edwards, G.T. R\"{u}ttimann, Gleason's theorem for rectangular JBW$^*$-triples, \emph{Comm. Math. Phys.} \textbf{203}, no. 2, 269--295 (1999).



\bibitem{FerGarPe2012} F.J. Fern{\'a}ndez-Polo, J.J. Garc{\'e}s, A.M. Peralta, A Kaplansky theorem for JB$^*$-triples, \emph{Proc. Amer. Math. Soc.} \textbf{140} (9) 3179--3191 (2012).


\bibitem{FerMarPe04} F.J. Fern{\'a}ndez-Polo, J. Mart{\' i}nez, A.M. Peralta, Surjective isometries between real JB$^*$-triples, \emph{Math. Proc. Cambridge Phil. Soc.}, \textbf{137} 709--723 (2004).


\bibitem{FerPe18Adv} F.J. Fern\'andez-Polo, A.M. Peralta, Low rank compact operators and Tingley's problem, \emph{Adv. Math.} \textbf{338}, 1--40 (2018).

\bibitem{Fri2005} Y. Friedman, \emph{Physical applications of homogeneous balls}. With the assistance of Tzvi Scarr. Progress in Mathematical Physics, 40. Birkhäuser Boston, Inc., Boston, MA, 2005.

\bibitem{Fri2021}  Y. Friedman, A Physically Meaningful Relativistic Description of the Spin State of an Electron, \emph{Symmetry} 2021, \textbf{13}(10), 1853.

\bibitem{FriHak88} Y. Friedman, J. Hakeda, Additivity of quadratic maps, \emph{Publ. Res. Inst. Math. Sci.} \textbf{24}, no. 5, 707--722 (1988).

\bibitem{FriRu85} Y. Friedman, B. Russo, Structure of the predual of a JBW$^*$-triple, \emph{J. Reine u. Angew. Math.} \textbf{356}, 67--89 (1985).

\bibitem{FriRu86} Y. {Friedman}, B. {Russo}, {The Gelfand--Naimark theorem for JB$\sp*$-triples,} {\em {Duke Math. J.}}, \textbf{53}, 139--148 (1986).

\bibitem{FriRu89} Y. {Friedman}, B. {Russo}, Affine structure of facially symmetric spaces, \emph{Math. Proc. Cambridge Philos. Soc.} \textbf{106}, no. 1, 107--124 (1989).

\bibitem{FriRu93} Y. {Friedman}, B. {Russo}, Classification of atomic facially symmetric spaces, \emph{Canad. J. Math.} \textbf{45}, no. 1, 33--87 (1993).

\bibitem{FriRu01} Y. Friedman, B. Russo,  A new approach to spinors and some representation of the Lorentz group on them", \emph{Foundations of Physics}, \textbf{31}, no.12, 1733--1766 (2001).

\bibitem{Geher2014} G.P. Geh{\'e}r, An elementary proof for the non-bijective version of Wigner's theorem, \emph{Phys. Lett. A} \textbf{378}, no. 30-31, 2054--2057 (2014).

\bibitem{Geher2016} G.P. Geh{\'e}r, Wigner's theorem on Grassmann spaces, \emph{J. Funct. Anal.} \textbf{273}, no. 9, 2994--3001 (2017).

\bibitem{Gleason57} A.M. Gleason, Measures on the closed subspaces of a Hilbert space, \emph{J. Math. Mech.} \textbf{6}, 885--893 (1957).

\bibitem{Gyo2004} M. Gy\H{o}ry, A new proof of Wigner's theorem, \emph{Rep. Math. Phys.} \textbf{54}, no. 2, 159--167 (2004).

\bibitem{HamKalPe20} J. Hamhalter, O.F.K. Kalenda, A.M. Peralta, Finite tripotents and finite JBW$^*$-triples, \emph{J. Math. Anal. Appl.} \textbf{490} (2020), no. 1, 124217.

\bibitem{HamKalPePfi20} J. Hamhalter, O.F.K. Kalenda, A.M. Peralta, H. Pfitzner, Measures of weak non-compactness in preduals of von Neumann algebras and JBW$^\ast$-triples, \emph{J. Funct. Anal.} \textbf{278}, 1 (2020), 108300.

\bibitem{HamKalPePfi20Groth} J. Hamhalter, O.F.K. Kalenda, A.M. Peralta, H. Pfitzner, Grothendieck's inequalities for JB$^*$-triples: Proof of the
  Barton-Friedman conjecture, \emph{Trans. Amer. Math. Soc.} \textbf{374}, Number 2, 1327--1350 (2021).

\bibitem{HOS} H. Hanche-Olsen, E. St{\o}rmer, \emph{Jordan Operator Algebras}, Pitman, London, 1984.

\bibitem{Harris74} L.A. Harris, {Bounded symmetric homogeneous domains in infinite dimensional spaces}. In:
\emph{Proceedings on infinite dimensional Holomorphy (Kentucky 1973)}; pp. 13-40. Lecture Notes in
Math. 364. Berlin-Heidelberg-New York: Springer 1974.

\bibitem{Harris81} L.A. Harris, A generalization of C$^*$-algebras, \emph{Proc. Lond. Math. Soc.}, \textbf{42} no. 3 (1981) 331--361.

\bibitem{HervIs92} F.J. Herv{\'e}s, J.M. Isidro, Isometries and automorphisms of the spaces of spinors, \emph{Rev. Mat. Univ. Complut. Madrid} \textbf{5}, no. 2-3, 193--200 (1992).

\bibitem{HoMarPeRu} T. Ho, J. Martinez-Moreno, A.M. Peralta, B. Russo, Derivations on real and complex JB$^\ast$-triples, \emph{J. London Math. Soc.} (2) \textbf{65}, no. 1, 85--102 (2002).

\bibitem{Horn87} G. Horn, Characterization of the predual and ideal structure of a JBW$^*$-triple, \emph{Math. Scand.} \textbf{61}, no. 1, 117--133 (1987).

\bibitem{Horn87c} G. Horn, Coordinatization theorems for JBW$^*$-triples. \emph{Quart. J. Math. Oxford Ser.} (2) \textbf{38}, no. 151, 321--335 (1987).

\bibitem{Horn87b} G. Horn, Classification of JBW$^*$-triples of type I, \emph{Math. Z.} \textbf{196}, 271--291 (1987).

\bibitem{Isham1994} C.J. Isham, Quantum logic and histories approach to quantum theory, \emph{J. Math. Phys.} \textbf{35}, 2157--2185 (1994).

\bibitem{IshamLinden1994} C.J. Isham, N. Linden, Quantum temporal logic and decoherence functionals in the histories approach to generalised quantum theory, \emph{J. Math. Phys.} \textbf{35}, 5452--5476 (1994).

\bibitem{IshamLindenSchreckenberg1994} C.J. Isham, N. Linden, S. Schreckenberg, The classification of decoherence functionals: An analogue of Gleason's theorem, \emph{J. Math. Phys.} \textbf{35}, 6360--6370 (1994).

\bibitem{KalPe2019} O.F.K. Kalenda, A.M. Peralta, Extension of isometries from the unit sphere of a rank--2 Cartan factor, \emph{Anal. Math. Phys.} \textbf{11}, 15 (2021).

\bibitem{Ka83} W. Kaup, A Riemann Mapping Theorem for bounded symmentric domains in complex Banach spaces, \emph{Math. Z.} \textbf{183}, 503--529 (1983).

\bibitem{Ka97} W. Kaup, On real Cartan factors, \emph{Manuscripta Math.} \textbf{92}, 191--222 (1997).

\bibitem{KaUp77} W. Kaup, H. Upmeier, Jordan algebras and symmetric Siegel domains in Banach spaces, \emph{Math. Z.} \textbf{157}, 179--200 (1977).

\bibitem{LomontMendelson63} J.S. Lomont, P. Mendelson, The Wigner unitarity-antiunitarity theorem, \emph{Ann. of Math.} (2) \textbf{78}, 548--559 (1963).

\bibitem{loos1977bounded} O. Loos, \emph{Bounded symmetric domains and Jordan pairs}, Lecture Notes. Univ. California at Irvine, 1977.

\bibitem{McCrimMey81} K. McCrimmon, K. Meyberg, Coordinatization of Jordan triple systems, \emph{Comm. Algebra} \textbf{9}, no. 14, 1495--1542 (1981).

\bibitem{Molnar85} L. Moln{\'a}r, \emph{Selected Preserver Problems on Algebraic Structures of Linear Operators and on Function Spaces}, Lecture Notes in Math., vol. 1895, Springer-Verlag, Berlin, 2007.

\bibitem{Molnar96} L. Moln{\'a}r, Wigner's unitary-antiunitary theorem via Herstein's theorem on Jordan homomorphisms, \emph{J. Natur. Geom.} \textbf{10}, no. 2, 137--148 (1996).

\bibitem{Molnar98} L. Moln{\'a}r, An algebraic approach to Wigner's unitary-antiunitary theorem, \emph{J. Austral. Math. Soc.} \textbf{65}, 354--369 (1998).

\bibitem{Molnar99} L. Moln{\'a}r, A generalization of Wigner's unitary-antiunitary theorem to Hilbert modules, \emph{J. Math. Phys.} \textbf{40}, 5544--5554 (1999).

\bibitem{Molnar2000} L. Moln{\'a}r, Generalization of Wigner's unitary-antiunitary theorem for indefinite inner product spaces, \emph{Comm. Math. Phys.} \textbf{201}, 785--791 (2000).

\bibitem{Molnar2002} L. Moln{\'a}r, On certain automorphisms of sets of partial isometries, \emph{Arch. Math. (Basel)} \textbf{78}, no. 1, 43--50 (2002).

\bibitem{Neher87} E. Neher, \emph{Jordan triple systems by the grid approach}. Lecture Notes in Mathematics, 1280. Springer-Verlag, Berlin, 1987.

\bibitem{Pe2020} A.M. Peralta, On the extension of surjective isometries whose domain is the unit sphere of a space of compact operators, preprint 2020, arXiv:2005.11987v1



\bibitem{SharAlmeida90} C.S. Sharma, D.F. Almeida, A direct proof of Wigner's theorem on maps which preserve transition probabilities between pure states of quantum systems, \emph{Ann. Physics} \textbf{197}, no. 2, 300--309 (1990).

\bibitem{Uhlhorn63} U. Uhlhorn, Representation of symmetry transformations in quantum mechanics, \emph{Ark. Fysik} \textbf{23}, 307--340 (1963).

\bibitem{Wig31} E.P. Wigner, \emph{Gruppentheorie und ihre Anwendung auf die Quantenmechanik der Atomspektrum}, Fredrik Vieweg und Sohn, 1931.

\bibitem{Wig59} E.P. Wigner, \emph{Group theory: And its application to the quantum mechanics of atomic spectra}, Academic Press, New York-London 1959.

\bibitem{Wright1995} J.D.M. Wright, The structure of decoherence functionals for von Neumann quantum histories, \emph{J. Math. Phys.} \textbf{36}, 5409--5413 (1995).

\end{thebibliography}
\end{document}